\documentclass[twocolumn,prd,aps,superscriptaddress,preprintnumbers,tightenlines,showpacs,nofootinbib,eqsecnum,amsfonts,amsmath]{revtex4-1}
\usepackage[normalem]{ulem}
\usepackage{placeins}
\usepackage{color}
\usepackage{calc}
\usepackage{amsmath,amssymb,graphicx}
\usepackage{tensor}
\usepackage{bm}
\usepackage{times}
\usepackage[varg]{txfonts}
\usepackage[colorlinks, pdfborder={0 0 0}]{hyperref}
\usepackage{float}
\usepackage{dcolumn}
\usepackage[nolist,nohyperlinks]{acronym}
\usepackage{xspace}
\usepackage[abs]{overpic}
\usepackage{pict2e}
\usepackage{enumitem}
\usepackage[usenames,dvipsnames]{xcolor}
\usepackage[utf8]{inputenc}
\usepackage{acronym}
\usepackage{gensymb}
\usepackage{cleveref}
\usepackage[normalem]{ulem}
\usepackage{longtable}
\usepackage{verbatim}
\usepackage{multirow}
\usepackage{amsmath,amssymb}
\usepackage{csquotes}
\usepackage{braket}

\usepackage{subfigure}
\usepackage{placeins}
\usepackage{hyperref}
\usepackage{aas_macros}
\usepackage{soul}

 \newcommand{\at}[1]{\textcolor{blue}{Alex: #1}}

\def\addAEI{Max Planck Institute for Gravitationsphysik (Albert Einstein Institute), Am M\"{u}hlenberg 1, 14476 Potsdam, Germany}

\def\addMD{Department of Physics, University of Maryland, College Park, MD 20742, USA}
\def\addNASA{NASA Marshall Space Flight Center, Huntsville, Alabama 35811, USA}

\begin{document}

\title{Measuring source properties and quasi-normal-mode frequencies of \\ heavy massive black-hole binaries with LISA}

\author{Alexandre Toubiana}
\affiliation{\addAEI}

\author{Lorenzo Pompili}
\affiliation{\addAEI}

\author{Alessandra Buonanno}
\affiliation{\addAEI}
\affiliation{\addMD}

\author{Jonathan R. Gair}
\affiliation{\addAEI}

\author{Michael L. Katz}
\affiliation{\addAEI}
\affiliation{\addNASA}

\begin{abstract}
    The Laser Interferometer Space Antenna 
    (LISA) will be launched in the mid-2030s. It promises to observe the coalescence of massive black-hole (BH) binaries with signal-to-noise ratios (SNRs) reaching thousands. Crucially, it will detect some of these binaries with high SNR in both the inspiral and the merger-ringdown stages. Such signals are ideal for tests of general relativity (GR) using information from the whole waveform. Here, we consider astrophysically motivated binary systems at the high-mass end of the population observable by LISA, and simulate their 
    signals using the newly developed multipolar 
    effective-one-body model: \texttt{pSEOBNRv5HM}. 
    The merger-ringdown signal in this model depends on the binary properties (masses and spins), and also on parameters that describe fractional deviations from the GR quasinormal mode (complex) frequencies of the remnant BH. Performing full Bayesian analyses, we assess to which accuracy LISA will be able to constrain deviations from GR in the ringdown signal when using information from the whole signal. We find that these deviations can typically be constrained to within $10\%$ and in the best cases to within $1\%$. We also show that with this model we can measure the binary masses and spins with great accuracy even for very massive BH systems with low SNR in the inspiral.
    In particular, individual source-frame masses can typically be constrained to within $10\%$ and as precisely as $1\%$, and individual spins can typically be constrained to within $0.1$ and, in the best cases, to within $0.001$. We also probe the accuracy of the \texttt{SEOBNRv5HM} waveform family by performing synthetic injections of GR numerical-relativity waveforms. Using a novel method that we develop here to quantify the impact of systematic errors, we show that already for sources with SNR $\mathcal{O}(100)$, we would measure erroneous deviations from GR due to waveform model inaccuracies. One of the main sources of error is the mismodelling of the relative alignment between harmonics. 
    These results confirm the need for improving waveform models to perform tests of GR with binary BHs observed at high SNR by LISA.

\end{abstract}

\maketitle

\section{Introduction}

We are now well into the era of gravitational-wave (GW) astronomy, with 90 
observations of compact-object binaries~\cite{LIGOScientific:2021usb,LIGOScientific:2021djp} by the LIGO-Virgo-KAGRA (LVK) Collaboration~\cite{LIGOScientific:2014pky,VIRGO:2014yos,KAGRA:2020tym} and other claimed detections~\cite{Nitz:2021zwj,Olsen:2022pin}. The fourth observing run of the LVK Collaboration has just started, with the promise of many new detections thanks to improved sensitivity \cite{KAGRA:2013rdx}. In addition to unveiling an otherwise hardly detectable population of binary black holes (BBHs)~\cite{LIGOScientific:2016vpg,LIGOScientific:2018jsj,LIGOScientific:2020kqk,KAGRA:2021duu,LIGOScientific:2020ufj}, constraining the equation of state of neutron stars~\cite{LIGOScientific:2017vwq,LIGOScientific:2018cki} and inferring astrophysical and cosmological information~\cite{LIGOScientific:2017adf,LIGOScientific:2017ync}, GWs allow us to
test general relativity (GR) \cite{LIGOScientific:2016lio,LIGOScientific:2019fpa,LIGOScientific:2020tif,LIGOScientific:2021sio} in the strong-gravity and high-velocity regime, which is not accessible to other experiments. Indeed, by comparing predictions for the GW signal of a BBH within GR to the observed data we can constrain deviations from GR. 

One of the most promising approaches to probe deviations from GR with GWs are the so-called ``ringdown tests''. In the last stage of the coalescence of a BBH, after the two BHs have merged, the remnant BH is in a perturbed state and relaxes to a steady state configuration through GW emission. This stage is called the ringdown. In this final stage, the signal is a superposition of damped sinusoids with frequencies and damping times that depend exclusively on the properties of the remnant \cite{Vishveshwara:1970zz,Press:1973zz,Chandrasekhar:1975zza,Kokkotas:1999bd,Ferrari:2007dd,Berti:2009kk}. Within GR, the ``no-hair'' conjecture \cite{Carter:1971zc} tells us that those are the mass and the spin of the final BH, since astrophysical BHs are expected to carry no electric charge. Some gravity theories predict additional ``hairs'' for BHs, due, for instance, to cosmological boundary conditions or the presence of nearby matter \cite{Healy:2011ef,Horbatsch:2011ye,Berti:2013gfa} or to additional fields \cite{Sotiriou:2013qea,Yagi:2015oca,Barausse:2015wia,Silva:2017uqg,Herdeiro:2018wub,Julie:2018lfp,Julie:2019sab,Julie:2022huo}. In any case, the exact relation between the properties of the remnant and the spectrum of quasinormal modes (QNMs) (i.e., the sets of frequencies and damping times) is theory dependent \cite{Blazquez-Salcedo:2017txk,Blazquez-Salcedo:2016enn,Molina:2010fb,Cardoso:2009pk,Pani:2013ija,Pani:2013wsa,Mark:2014aja,Dias:2015wqa,Cardoso:2019mqo,McManus:2019ulj,Pierini:2021jxd,Pierini:2022eim,Cano:2023tmv,Cano:2023jbk,Wagle:2021tam}. By measuring two or more QNMs we can test if the signal agrees with GR. This is the basic idea behind BH spectroscopy \cite{Detweiler:1980gk,Dreyer:2003bv}. On the other hand, the amount to which each mode is excited (i.e., its amplitude) and the relative phases between them do depend on the properties of the BHs in the binary and the binary dynamics \cite{Flanagan:1997sx,Berti:2005ys,JimenezForteza:2020cve,Forteza:2022tgq}. Therefore, a consistent modelling of the merger-ringdown together with the inspiral can improve our ability to measure the QNMs, and to constrain deviations from GR during the ringdown. This is the approach followed in Refs.~\cite{Brito:2018rfr,Ghosh:2021mrv}, where the authors developed a parametrised model of the ringdown signal as part of the full inspiral-merger-ringdown (IMR) waveforms~\cite{Pan:2011gk,Cotesta:2018fcv} in the effective one-body (EOB) formalism~\cite{Buonanno:1998gg,Buonanno:2000ef}. Such a model can be used to perform parametrised (or theory-agnostic) ringdown tests of GR by allowing the QNMs to deviate from their GR prediction: A departure from the Kerr spectrum would be indicative of non-GR effects. The model has also been extended to parametrise the plunge-merger stages in Ref.~\cite{Maggio:2022hre}, and to carry out theory-specific tests of GR in the ringdown in Ref.~\cite{Silva:2022srr}. Here, we employ the parametrised ringdown test, which has already been applied to analyse the GW signals observed by the LVK Collaboration, showing so far consistency with GR \cite{Brito:2018rfr,Ghosh:2021mrv,LIGOScientific:2020tif,LIGOScientific:2021sio,Maggio:2022hre}. The precision of the test has so far been limited by the low signal-to-noise ratio (SNR) of the sources, which has led to measurement errors for the frequency and decay time of the dominant QNM on the order of $10\%$ and $20\%$, respectively, when combining events in a hierarchical way~\cite{LIGOScientific:2021sio}. More specifically, for LIGO-Virgo observations this test can be applied only 
when both the pre- and postinspiral regimes have at least SNR $\sim 8$, which has been the case for 12 binary systems~\cite{LIGOScientific:2021sio}. The best single-event measurement~\cite{LIGOScientific:2020tif} has been obtained with GW150914, which has a total SNR of 24~\cite{LIGOScientific:2016aoc}.
Other approaches have also been developed to do BH spectroscopy with LIGO-Virgo data~\cite{LIGOScientific:2020tif,LIGOScientific:2021sio}, using a superposition of damped sinusoids~\cite{Carullo:2019flw,Isi:2021iql}, in some cases augmented with QNM amplitudes calibrated to numerical-relativity (NR) simulations.

Scheduled for launch in the mid-2030s, the Laser Interferometer Space Antenna (LISA) \cite{Audley:2017drz} will detect massive BH binaries (MBHBs) with SNRs reaching thousands, sometimes both in the inspiral and in the merger-ringdown \cite{Klein:2015hvg,Bhagwat:2021kwv,Cotesta:2023}. MBHBs are therefore promising candidates for performing ringdown tests that use information from the full signal. Previous studies on ringdown analysis with LISA \cite{Berti:2005ys,Berti:2016lat,Bhagwat:2021kwv} focused on ``pure'' ringdown tests (i.e., using a superposition of damped sinusoids after the merger)
and employed simplified methods and criteria such as the Fisher matrix formalism~\cite{Finn:1992xs,Vallisneri:2007ev} to estimate the measurement accuracy and the distinguishability between QNMs. In this paper, we simulate LISA observations of MBHBs and run full Bayesian analyses on those in order to assess to which accuracy putative deviations from GR in the ringdown could be constrained from such observations when using information from the whole signal. Furthermore, we make use of parametrised EOB waveforms developed using the state-of-the-art multipolar aligned-spin model \texttt{SEOBNRv5HM}~\footnote{The generic name \texttt{SEOBNRvnEPHM} indicates that the version \texttt{vn} of the EOB model is calibrated to NR simulations (NR), includes spin (S) and precessional (P) effects, eccentricity (E) and higher modes (HM, i.e. higher harmonics).} developed in Refs.~\cite{Mihaylov:2023bkc,Khalil:2023kep,Pompili:2023tna,vandeMeent:2023ols}. Henceforth, we denote the parametrised model as \texttt{pSEOBNRv5HM}. Crucially, this model includes higher harmonics, which are expected to play an important role in LISA parameter estimation~\cite{Pitte:2023ltw}.

Since the expected population of MBHBs is highly uncertain \cite{Sesana:2007sh,Sesana:2010wy,Klein:2015hvg,Bonetti:2018tpf,Dayal:2018gwg,Barausse:2020mdt,Barausse:2020gbp}, here we focus on a few astrophysically realistic systems, compatible with the predictions of models where MBHs form from heavy seeds \cite{Latif:2016qau}. For simplicity, in this work we neglect the effect of spin precession and eccentricity. Our study is performed by using the same waveform model for generating mock injections and for estimating the parameters of the source. However, it is crucial to assess if such tests of GR could be spoiled by the limited accuracy of our theoretical models when performed on real data. Therefore, we assess the impact of systematics in waveform modelling on ringdown tests by simulating mock LISA injections with waveforms from NR and using \texttt{pSEOBNRv5HM} waveforms to perform the Bayesian analysis. Developing a new approach, we estimate from which SNR we expect to erroneously measure deviations from GR due to systematic effects. Finally, we assess to which extent a consistent modelling of the full signal allows us to measure the binary parameters in GR also for systems that are merger-ringdown dominated and have low SNR in the inspiral.


This paper is organised as follows. In Sec.~\ref{sec:wvf} we present the details of our parametrised EOB model, describe how the synthetic LISA observations are generated, and lay down the basis of our Bayesian analyses. In Sec.~\ref{sec:astro} we summarise the astrophysical systems that we simulate. We present our results when using the \texttt{pSEOBNRv5HM} model for both injection and parameter estimation in Sec.~\ref{sec:results}; then in Sec.~\ref{sec:syst} we discuss the impact of systematics and using a novel method, we show that our models are not yet accurate enough for the signals we expect with LISA. Finally, we present our conclusions in Sec.~\ref{sec:ccl}. In the appendices, 
we discuss how the settings of \texttt{pSEOBNRv5HM} waveforms impact the parameter estimation, and we show that, in GR, \texttt{SEOBNRv5HM} and \texttt{IMRPhenomTHM}~\cite{Estelles:2020twz}, a time-domain waveform model from the IMR phenomenological family ~\cite{Ajith:2007kx,Ajith:2009bn,Santamaria:2010yb},
predict similar measurement errors for the parameters of the source, that their measurement is little affected by adding the QNM deviation parameters in the \texttt{pSEOBNRv5HM} model, and show additional results of our study of systematic effects. 
Throughout this paper we will use natural units in which $c = G =1$. 

\section{Methods}\label{sec:wvf}

\subsection{Parametrised waveform model}\label{sec:wvf_eob}
We consider a binary with BH component (detector-frame) masses $m_1$ and $m_2$ and define the mass ratio $q = m_1/m_2 \geq 1$ and the total mass $M_t=m_1+m_2$. We limit to BHs moving on quasicircular orbits with aligned or anti-aligned spins (aligned spins for short) and define the (dimensionless) spin variables $\chi_1 = S_1/m_1^2$ and $\chi_2= S_2/m_2^2$, which range between $-1$ and $1$. We denote the luminosity distance of the source as $D_L$ and the cosmological redshift as $z$. We adopt the cosmology determined by the Planck mission (2018) \cite{Planck:2018vyg}. Masses, times, and frequencies are in the detector frame, unless they carry a subscript $s$. Source-frame masses $m_{1,s}$ and $m_{2,s}$ are related to the detector-frame ones by $m_i=(1+z)m_{i,s}$.

The GW polarisations can be expanded in the basis of spin-weight $-2$ spherical harmonics as
\begin{equation}
h_+(\mathbf{\Theta},\iota,\varphi_0;t ) - i h_\times(\mathbf{\Theta},\iota,\varphi_0;t) = \frac{1}{D_L} \sum_{\ell, m} {}_{-\!2}Y_{\ell m}(\iota,\varphi_0)\, h_{\ell m}(\mathbf{\Theta};t)\,,\label{eq:harmonic_decomp}
\end{equation}
where the parameters $(\iota, \varphi_0)$ denote the binary’s inclination angle with respect to the direction perpendicular to the
orbital plane and the azimuthal direction to the observer, respectively, and $\mathbf{\Theta}$ denotes the intrinsic parameters (masses and spins) of the binary. We build our parametrised model using the \texttt{SEOBNRv5HM} model~\cite{Pompili:2023tna} in GR, which includes several higher harmonics, notably the $(\ell, |m|)=(2,1)$, $(3,3)$, $(3,2)$, $(4,4)$, $(4,3)$ and $(5,5)$ harmonics, in addition to the dominant $(2,2)$ harmonic. For aligned-spin binaries, $h_{\ell m}=(-1)^{\ell} h_{\ell-m}^*$, therefore, we focus on $(\ell,m)$ harmonics with $m > 0$. 

In the EOB framework~\cite{Buonanno:2000ef}, the GW harmonics are decomposed as
\begin{equation}
\begin{aligned}
h_{\ell m}(\mathbf{\Theta}, t) &= h_{\ell m}(\mathbf{\Theta}, t)^\mathrm{insp-plunge}\, \theta(t_\mathrm{match}^{\ell m} - t) \\
&+ h_{\ell m}(\mathbf{\Theta}, t)^\mathrm{merger-RD}\,\theta(t-t_\mathrm{match}^{\ell m})\,,
\label{eq:EOBGW}
\end{aligned}
\end{equation}
where $\theta(t)$ is the Heaviside step function, $h_{\ell m}^\mathrm{insp-plunge}$ corresponds to the inspiral-plunge part of the waveform, while $h_{\ell m}^\mathrm{merger-RD}$ represents the merger-ringdown waveform. 
In particular, as explained in Ref.~\cite{Pompili:2023tna}, $t_\mathrm{match}^{\ell m}$ is chosen to be the peak of the $(2,2)$ harmonic amplitude for all $(\ell,m)$ harmonics except $(5,5)$, for which it is taken as the peak of the $(2,2)$ harmonic minus $10 M_t$. In the following, we suppress the $\mathbf{\Theta}$ dependence for ease of notation. 

For all harmonics, except for $(\ell, |m|)= (3,2)$ and $(4,3)$ which exhibit postmerger oscillations due to mode mixing~\cite{Buonanno:2006ui, Kelly:2012nd}, the merger-ringdown waveform employs the following ansatz~\citep{Bohe:2016gbl,Cotesta:2018fcv, Pompili:2023tna},
\begin{equation}
\label{RD}
h_{\ell m}^{\textrm{merger-RD}}(t) = \nu \ \tilde{\mathcal{A}}_{\ell m}(t)\ e^{i \tilde{\phi}_{\ell m}(t)} \ e^{-i \sigma_{\ell m 0}(t-t_{\textrm{match}}^{\ell m})},
\end{equation}
where $\nu = m_1 m_2 / (m_1+m_2)^2$ is the symmetric mass ratio of the binary and $\sigma_{\ell m0}$ is the complex frequency of the least-damped QNM, having overtone number zero, of the remnant BH. We define the corresponding oscillation frequency $f_{\ell m 0}$ and the damping time $\tau_{\ell m 0}$ respectively as
\begin{subequations}
\begin{align}
f_{\ell m 0}&=\frac{1}{2 \pi} \operatorname{Re}\left(\sigma_{\ell m 0}\right)=-\frac{1}{2 \pi} \sigma_{\ell m 0}^{\mathrm{I}},\\
\tau_{\ell m 0}&=-\frac{1}{\operatorname{Im}\left(\sigma_{\ell m 0}\right)}=-\frac{1}{\sigma_{\ell m 0}^{\mathrm{R}}}.
\end{align}
\end{subequations}

The functions $\tilde{\mathcal{A}}_{\ell m}(t)$ and $\tilde{\phi}_{\ell m}(t)$ are given by~\cite{Baker:2008mj,Damour:2014sva, Bohe:2016gbl,Cotesta:2018fcv,Pompili:2023tna}:
\begin{subequations}
\begin{align}
\label{eq:ansatz_amp}
\tilde{\mathcal{A}}_{\ell m}(t) &= c_{1,c}^{\ell m} \tanh[c_{1,f}^{\ell m}\ (t-t_{\textrm{match}}^{\ell m}) \ +\ c_{2,f}^{\ell m}] \ + \ c_{2,c}^{\ell m},\\
\label{eq:ansatz_phase}
\tilde{\phi}_{\ell m}(t) &= \phi_{\textrm{match}}^{\ell m} - d_{1,c}^{\ell m} \log\left[\frac{1+d_{2,f}^{\ell m} e^{-d_{1,f}^{\ell m}(t-t_{\textrm{match}}^{\ell m})}}{1+d_{2,f}^{\ell m}}\right],
\end{align}
\end{subequations}
where $ \phi_{\textrm{match}}^{\ell m}$ is the phase of the inspiral-plunge harmonic $(\ell, m)$ at $t = t_{\textrm{match}}^{\ell m}$. 

\begin{figure}
\centering
 \includegraphics[width=0.45\textwidth]{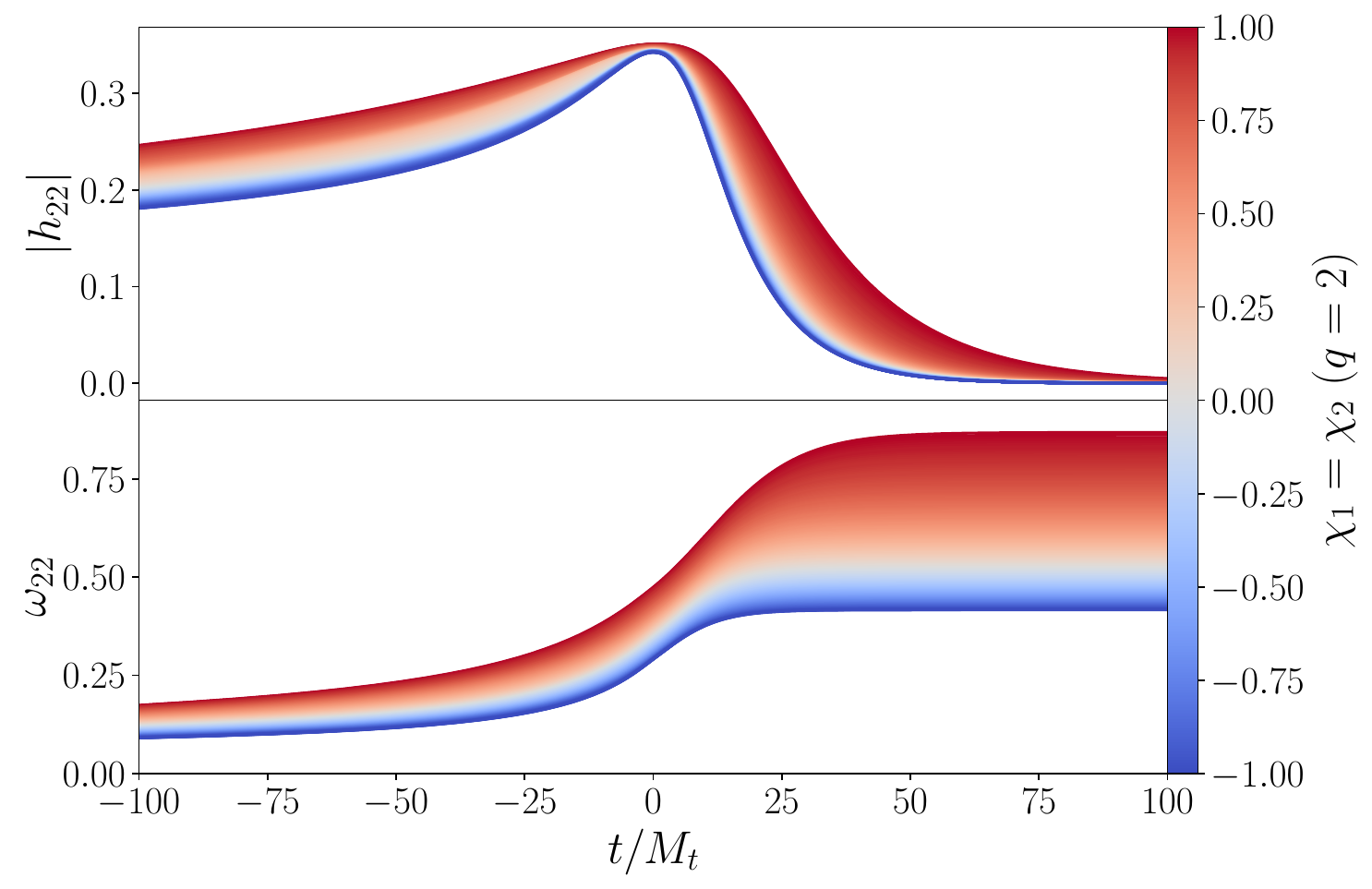}\\
 \centering
 \caption{Time evolution near the merger, which occurs at $t=0$, of the \texttt{SEOBNRv5HM} $(2,2)$ harmonic amplitude (top panel) and instantaneous frequency (bottom panel) when varying the spin components of a binary with mass ratio 2, assuming equal and aligned spins.
 }\label{fig:dep_spins}
\end{figure}

The coefficients $d_{1, c}^{\ell m}$ and $c_{i, c}^{\ell m}$ ($i=1,2$) are constrained by the requirement that 
the amplitude and phase of $h_{\ell m}(t)$ are continuously differentiable ($\mathcal{C}^1$) at $t=t_{\text {match}}^{\ell m}$. This allows us to write the coefficients $c_{i,c}^{\ell m}$ as~\cite{Cotesta:2018fcv,Pompili:2023tna}
\begin{subequations}
\begin{align}
\label{c1}
c_{1,c}^{\ell m} &= \frac{1}{c_{1,f}^{\ell
    m} \nu} \big[ \partial_t|h_{\ell
    m}^{\textrm{insp-plunge}}(t_{\textrm{match}}^{\ell m})| \nonumber \\
    &- \sigma^\textrm{R}_{\ell m} |h_{\ell
    m}^{\textrm{insp-plunge}}(t_{\textrm{match}}^{\ell
    m})|\big] \cosh^2{(c_{2,f}^{\ell m})}, \\
\label{c2}
c_{2,c}^{\ell m} &= \frac{ |h_{\ell
    m}^{\textrm{insp-plunge}}(t_{\textrm{match}}^{\ell
    m})|}{\nu} - \frac{1}{c_{1,f}^{\ell
    m} \nu} \big[ \partial_t|h_{\ell
    m}^{\textrm{insp-plunge}}(t_{\textrm{match}}^{\ell m})|  \nonumber \\
    &- \sigma^\textrm{R}_{\ell m} |h_{\ell
    m}^{\textrm{insp-plunge}}(t_{\textrm{match}}^{\ell
    m})|\big] \cosh{(c_{2,f}^{\ell m})}\sinh{(c_{2,f}^{\ell m})}, \\ \nonumber
\end{align}
\end{subequations}
and $d_{1,c}^{\ell m}$ as
\begin{align}
\label{d1}
d_{1,c}^{\ell m} &= \left[\omega_{\ell m}^{\textrm{insp-plunge}}(t_{\textrm{match}}^{\ell m}) -  \sigma^\textrm{I}_{\ell
      m}\right]\frac{1+ d_{2,f}^{\ell m}}{d_{1,f}^{\ell m}d_{2,f}^{\ell m}}\,,
\end{align}
where $\omega_{\ell m}^{\textrm{insp-plunge}}(t)$ is the frequency of the inspiral-plunge EOB harmonic. The coefficients $c_{i,f}^{\ell m}$ and $d_{i,f}^{\ell m}$ are obtained through fits to a large set of NR waveforms ($\sim 440$), spanning mass ratios up to 20 and spins up to 0.998, and BH perturbation-theory merger-ringdown waveforms for mass ratio 1000. Crucially, the fits depend on the binary's masses and spins $\mathbf{\Theta}$ and can be found in Appendix D in Ref.~\cite{Pompili:2023tna}. As an example, we illustrate in Fig.~\ref{fig:dep_spins} how the GW amplitude and frequency of the $(2,2)$ harmonic changes, during the late inspiral, merger and ringdown, as the component spins are varied, for a binary with mass ratio 2 and equal spins.

\begin{figure*}
\centering
 \includegraphics[width=\textwidth]{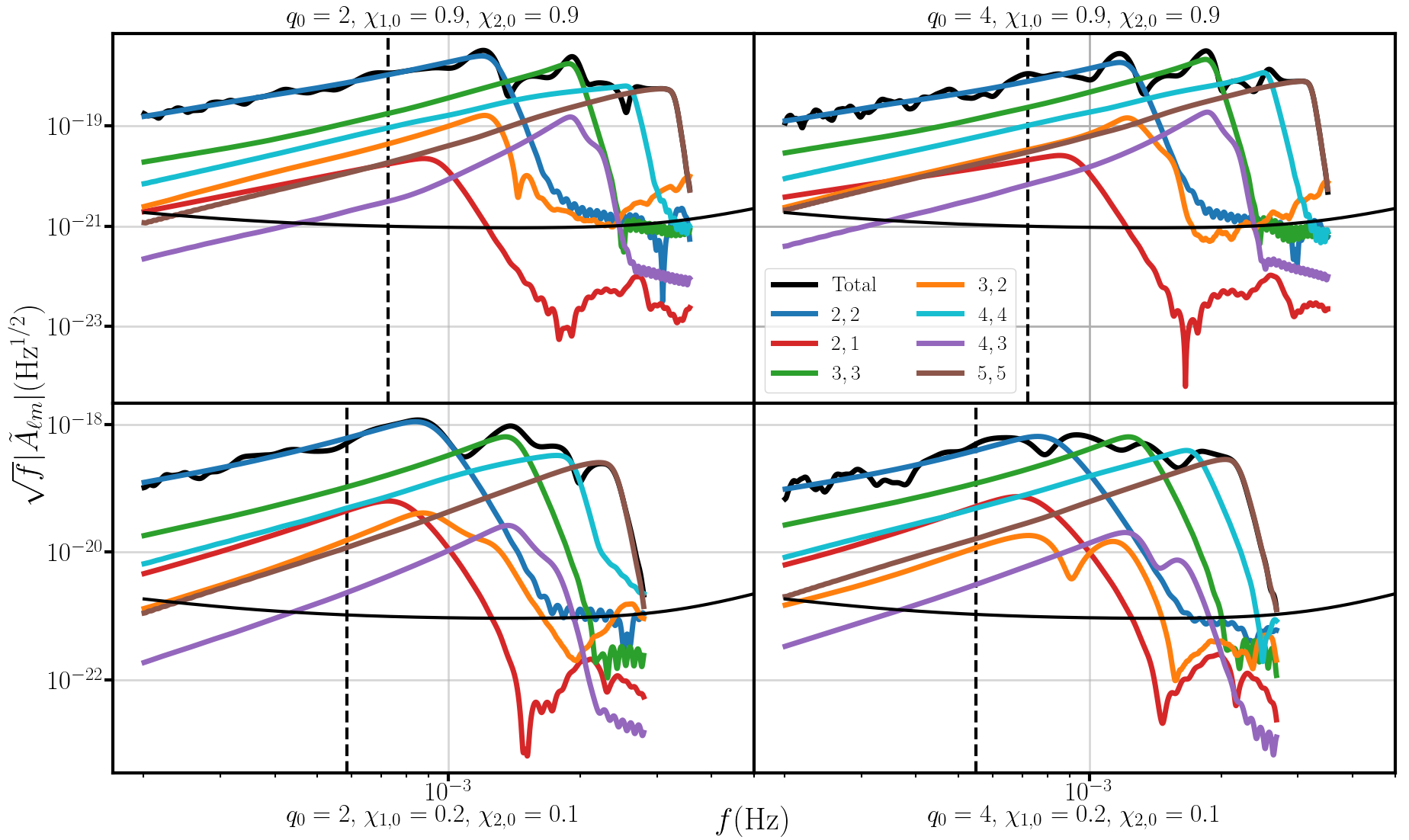}\\
 \centering
 \caption{Amplitude in frequency domain in the TDI channel $A$ broken into the contributions of each of the harmonics included in the \texttt{pSEOBNRv5HM} model, $\tilde{A}_{\ell m}$. The latter is computed using the harmonic decomposition of Eq.~(\ref{eq:harmonic_decomp}) in the TDI equations (\ref{eq:tdi_lw}). The thin black line shows the LISA PSD and the black dashed lines the GW frequency at $t_c$ (i.e., the separation between the inspiral and merger-ringdown regimes). We recall that $t_c$ is defined from the peak of $h_{22}$, and that the computation of the TDI variables involves second derivatives of the GW polarisation, leading to an offset between the maximum of $h_{22}$ and those of $A$ and $E$. Moreover, these quantities are defined in time-domain, whereas we show here the frequency-domain amplitudes, and the time-domain peak typically corresponds to lower frequencies than the frequency-domain one. We plot the amplitudes only for $M_{t,0}=2\times 10^7 \ M_{\odot}$ and $z_0=2.2$. For systems at $z_0=3.7$, one should simply rescale the amplitudes by $18,331/33,691 \simeq 0.54$, and for systems with $M_{t,0}=2 \times 10^8 
 \ M_{\odot}$, one should multiply the amplitudes by $10^2$ and the frequencies by $10^{-1}$. As expected, the $(2,2)$ harmonic is the loudest, followed by the $(3,3)$, $(4,4)$ and $(5,5)$ harmonics. For high-spins systems (upper row), the $(3,2)$ and $(4,3)$ harmonics are louder than the $(2,1)$ harmonic, whereas the opposite is true for low-spin systems (lower row).}\label{fig:ampls}
\end{figure*}

For the $(3,2)$ and $(4,3)$ harmonics, the mode-mixing behaviour is modelled by applying the previous construction to the spheroidal harmonics~\cite{Berti:2005gp} $(3,2,0)$ and $(4,3,0)$, which feature a monotonic amplitude and frequency evolution~\cite{KumarMehta:2019izs}. The spheroidal $(3,2,0)$ and $(4,3,0)$ harmonics can be related to the spherical harmonics by~\cite{Pompili:2023tna}
\begin{subequations}
\begin{align}
{}^{S}h_{320}(t)&\simeq \dfrac{ h_{32}(t)\,\mu_{2220}^{*} - h_{22}(t)\, \mu_{2320}^{*}}{\mu_{2330}^{*}\,\mu_{2220}^{*}},
\label{eq:unmixed_modes_1} \\
{}^{S}h_{430}(t)&\simeq \dfrac{h_{43}(t)\,\mu_{3440}^{*} - h_{33}(t)\,\mu_{3430}^{*}}{\mu_{3330}^{*}\,\mu_{3440}^{*}},
\label{eq:unmixed_modes_2}
\end{align}
\end{subequations}
where $\mu_{m\ell \ell^{\prime}n}$ are harmonic mixing coefficients, obtained using fits from Ref.~\cite{Berti:2014fga}.
Thus, the $(3,2)$ and $(4,3)$ harmonics are obtained by combining the $(3,2,0)$ and $(4,3,0)$ harmonics with the $(2,2)$ and $(3,3)$ harmonics, inverting Eqs.~(\ref{eq:unmixed_modes_1}) and (\ref{eq:unmixed_modes_2}).

Though overtones are not explicitly included in the merger-ringdown signal ansatz, unlike older versions of the model \cite{Pan:2011gk,Pan:2013rra}, their effect should be captured by the functions $\tilde{\mathcal{A}}_{\ell m}(t)$ and $\tilde{\phi}_{\ell m}(t)$. They contain free coefficients fitted against NR simulations and allow our ansatz to be more than a simple damped sinusoid with damping time and frequency given by the fundamental QNM. Moreover, we do not expect the linear perturbation description used in the ringdown to be valid right at the time used to transition from inspiral to merger-ringdown \cite{Cheung:2022rbm}, which for most modes is the peak of the $(2,2)$ harmonic. Thus, the choice of modelling in SEOBv5NRHM allows one to better capture such nonlinearities without having to include a transition phase.

In the \texttt{SEOBNRv5HM} model constructed in Ref.~\cite{Pompili:2023tna}, the complex QNM frequencies in GR are obtained for each $(\ell, m)$ harmonic as a function of the BH's final mass and spin using the \texttt{qnm} Python package \cite{Stein:2019mop}. The BH's mass and spin are, in turn, computed using the fitting formulas of Refs.~\cite{Jimenez-Forteza:2016oae,Hofmann:2016yih} respectively. 
In this work, following the strategy of Refs.~\cite{Gossan:2011ha,Meidam:2014jpa,Brito:2018rfr, Ghosh:2021mrv, Maggio:2022hre,Isi:2019aib,Isi:2021iql}, we introduce parametrised fractional deviations to the QNM frequencies, which are free parameters of the model (see Ref.~\cite{Silva:2022srr} where the deviations were mapped to specific gravity theories alternative to GR). More explicitly, we perform the substitutions
\begin{subequations}
\begin{align}
f_{\ell m 0} & \rightarrow f_{\ell m 0}\, (1 + \delta f_{\ell m }),\label{eq:nongr_freqs_a} \\ 
\tau _{\ell m 0} & \rightarrow \tau _{\ell m 0}\, (1 + \delta \tau_{\ell m }), \label{eq:nongr_freqs_b}
\end{align}
\end{subequations}
where for ease of notation we have dropped the zero overtone subscript in the deviation parameters. We shall denote this parametrised model as \texttt{pSEOBNRv5HM}. We note that allowing $\sigma_{\ell m 0}$ to vary freely also modifies the $c_{i,c}^{\ell m}$ and $d_{1,c}^{\ell m}$ coefficients in Eqs.~(\ref{c1}), (\ref{c2}), and (\ref{d1}), which enter the amplitude and phase functions $\tilde{\mathcal{A}}_{\ell m}(t)$ and $\tilde{\phi}_{\ell m}(t)$. As a consequence, such a modification can lead to deviations from the GR prediction in the ringdown signal starting soon after the merger. The plunge-merger stage of the waveform could be, in principle, also modified, as done, for example, in Ref.~\cite{Maggio:2022hre}, by introducing deviations with respect to the GR predictions to the time at which the amplitude peaks and to the value of the amplitude and frequency at this instant, for each waveform harmonic.

Finally, the inspiral-plunge EOB waveforms (\ref{eq:EOBGW}) are computed based on the two-body dynamics that are computed by solving Hamilton's equations with a suitable EOB Hamiltonian and radiation-reaction force (see Refs.~\cite{Khalil:2023kep, Pompili:2023tna} for details).

\subsection{Generation of LISA signals}\label{sec:lisa_sig}

We use the long-wavelength approximation \cite{Cutler:1997ta} to compute the response of LISA to an incoming GW, which is valid when the GW wavelength is much larger than the LISA arm length $L$ (i.e., in terms of the GW frequency, when ${\mbox 2\pi f L /c\ll 1}$). Given that $L=2.5 \times 10^9  \ {\rm m}$, this condition is satisfied for sources reaching maximum frequencies of $\sim 10^{-3}  \ {\rm Hz}$, such as the MBHBs we consider in this work. Under this approximation, LISA is somewhat similar to two LIGO or Virgo-type detectors rotated with respect to each other by $\pi/4$ and with angles of $\pi/3$ between the arms.

Transforming Eq.~(47) in Ref.~\cite{Marsat:2020rtl} to the time domain, we find that under the long-wavelength approximation the time-delay-interferometry (TDI) variables $A$, $E$, and $T$ \cite{Tinto:2004wu} (which, for an interferometer with equal arms and equal noise levels in each optical link, provide three noise-uncorrelated datasets) are given by
\begin{subequations}
\begin{align}
    &A=-3\sqrt{2} \left (\frac{L}{c} \right )^2 \left [ F_+(\lambda,\beta,\psi)\ddot{h}_{+}+F_{\times}(\lambda,\beta,\psi)\ddot{h}_{\times} \right ],  \nonumber \\
    &E=-3\sqrt{2} \left ( \frac{L}{c} \right )^2 \left [ F_+(\lambda+\pi/4,\beta,\psi)\ddot{h}_{+}+F_{\times}(\lambda+\pi/4,\beta,\psi)\ddot{h}_{\times} \right ], \nonumber \\
    &T=0, \label{eq:tdi_lw}
\end{align}
\end{subequations}
where $F_+$ and $F_{\times}$ are the antenna pattern functions:
\begin{subequations}
\begin{align}
    &F_{+}(\lambda,\beta,\psi)=\cos(2\psi) F_{+,0}(\lambda,\beta)+\sin(2\psi)F_{\times,0}(\lambda,\beta), \\
    &F_{\times}(\lambda,\beta,\psi)=-\sin(2\psi) F_{+,0}(\lambda,\beta)+\cos(2\psi)F_{\times,0}(\lambda,\beta), \\
    &F_{+,0}(\lambda,\beta)=\frac{1}{2}\left ( 1+\sin^2\beta \right ) \cos(2\lambda-\pi/3), \\
    &F_{\times,0}(\lambda,\beta)=\sin\beta\sin(2\lambda-\pi/3). 
\end{align}
\end{subequations}
In the above equations, $\lambda$, $\beta$, and $\psi$ are the longitude, latitude and polarisation in the LISA frame, respectively. We refer to Ref.~\cite{Marsat:2020rtl} for the relation between the angles in the LISA frame and those in the solar-system-barycentre frame. As anticipated, this is similar to the response of ground-based detectors. The main difference is that it is the second derivatives of the waveform polarisations that enter Eq.~(\ref{eq:tdi_lw}). This comes as a consequence of taking waveform differences in order to perform TDI, with a time step that goes to zero in the long-wavelength approximation. Finally, because most of the SNR of the signals we consider is accumulated in the last stages of the evolution (i.e., from a few hours to a few days), we do not take into account the motion of LISA about the Sun. Therefore, $\lambda$, $\beta$, and $\psi$ in Eq.~(\eqref{eq:tdi_lw}) are not varying with time. Henceforth, we will use $\tilde{A}$ and $\tilde{E}$ to denote the Fourier transform of $A$ and $E$.

We generate EOB waveforms from the frequency ${ \mbox f_{\rm gen}=5\times 10^{-5} [M_{t,0}/(2\times 10^7 M_\odot)]  \ {\rm Hz} }$ until the end of the signal. 
After transforming to the frequency domain, we keep the portion of the signal between $f_{\rm min}=2\times 10^{-4} [M_{t,0}/(2\times 10^7 M_\odot)]  \ {\rm Hz} $ and $f_{\rm max}$, to eliminate spurious features due to Fourier transform. The maximum frequency is chosen such that the frequency-domain amplitude is $1\%$ of its maximum value. We verified that our choice of $f_{\rm min}$ leads to a loss in SNR of less than $2\%$. 

The time and phase alignment of the signals is done in the following way. We define the time to coalescence $t_c$ as the moment the amplitude of the $(2,2)$ harmonic reaches its peak and define the phase of coalescence $\varphi_c$ as the phase of the $(2,2)$ harmonic contribution to the total waveform at $t_c$. In practice, the last step is done by choosing the azimuthal angle $\varphi_0$ such that the phase of ${}_{-\!2}Y_{22}(\iota,\varphi_0)\, h_{22}(t_c)$ is $\varphi_c$. This choice of $\varphi_0$ is then propagated consistently to other harmonics. We use $t_c$ to split between the inspiral and merger-ringdown regimes. We note that $t_c$ for EOB waveforms coincides with $t_\mathrm{match}^{\ell m}$ in Eq.~(\ref{eq:EOBGW}), for all $(\ell,m)$ harmonics except $(5,5)$.

\subsection{Bayesian analysis}\label{sec:bayesian}

We define the noise-weighted inner product between two data streams $d_1$ and $d_2$ as 
\begin{equation}
    (d_1|d_2)=4 {\mathcal Re} \left [ \int_0^{+\infty}\frac{d_1(f)d^*_2(f)}{S_n(f)}df \right ], \label{eq:inner_product}
\end{equation}
where $S_n(f)$ is the power spectral density (PSD). In this work, we use the SciRDv1 noise curve \cite{scirdv1}, which corresponds to the scientific requirement for the LISA mission and defines pessimistic noise levels compared to current predictions. For a given choice of the PSD, the SNR of a signal $h$ is defined as ${\rm SNR}=\sqrt{(h|h)}$.

\begin{table*}
 \begin{center}
   \begin{tabular}{c *{7}{c}}
   \cline{3-8}

  & & \multicolumn{3}{c|}{$M_{t,0}=2\times 10^7 M_{\odot}$} & \multicolumn{3}{c}{$M_{t,0}=2\times 10^8 M_{\odot}$} \\

\cline{3-8}

 & & \multicolumn{1}{c|}{IMR} &  \multicolumn{1}{c|}{Merger-Ringdown} & \multicolumn{1}{c|}{Inspiral} & \multicolumn{1}{c|}{IMR} &  \multicolumn{1}{c|}{Merger-Ringdown} & \multicolumn{1}{c}{Inspiral} \\

\hline

  \multicolumn{1}{c}{\multirow{2}{*}{$\chi_{1,0}=\chi_{2,0}=0.9$}} &  \multicolumn{1}{c}{$q_0=2$} & 
  \multicolumn{1}{c}{5505} & \multicolumn{1}{c}{5253} & \multicolumn{1}{c}{1645} & \multicolumn{1}{c}{623} &\multicolumn{1}{c}{616} & \multicolumn{1}{c}{98} \\

   \multicolumn{1}{c}{} & \multicolumn{1}{c}{$q_0=4$} & \multicolumn{1}{c}{5174} & \multicolumn{1}{c}{5017} &\multicolumn{1}{c}{1269} & \multicolumn{1}{c}{672} &\multicolumn{1}{c}{668} & \multicolumn{1}{c}{75}  \\
  
  \hline

  \multicolumn{1}{c}{\multirow{2}{*}{$\chi_{1,0}=0.2$, $\chi_{2,0}=0.1$}} & \multicolumn{1}{c}{$q_0=2$} & \multicolumn{1}{c}{2678} & \multicolumn{1}{c}{2558} & \multicolumn{1}{c}{795} & 
  \multicolumn{1}{c}{241} & \multicolumn{1}{c}{238} & \multicolumn{1}{c}{40} \\

  \multicolumn{1}{c}{} &  \multicolumn{1}{c}{$q_0=4$} & \multicolumn{1}{c}{1832} & \multicolumn{1}{c}{1741} & \multicolumn{1}{c}{571} & 
  \multicolumn{1}{c}{167} & \multicolumn{1}{c}{165} & \multicolumn{1}{c}{29} \\
  
  \hline

   \end{tabular}
\end{center}
 \caption{Full IMR SNR and its decomposition into the contribution of the merger-ringdown and inspiral stages for the systems considered in this work at $z_0=2.2$. Values at $z_0=3.7$ can be obtained by rescaling by $18,331/33,691 \simeq 0.54$. We recall that the inspiral and merger-ringdown SNRs add quadratically. All the systems we consider are merger-ringdown dominated, although some of them have high SNR in the inspiral as well, up to thousands.}\label{tab:snrs}
\end{table*}

To quantify the precision with which LISA observations will estimate the parameters of a source, we work in a Bayesian framework and compute the posterior distribution on the source parameters, $\theta$, given an observed dataset $d$, using Bayes' theorem:
\begin{equation}
    p(\theta|d)=\frac{p(d|\theta)p(\theta)}{p(d)},
\end{equation}
where $p(d|\theta)$ is the likelihood, $p(\theta)$ is the prior and $p(d)$ is the evidence. As long as we are not interested in model selection, the latter acts as a normalisation constant, and, thus, can be discarded. 
We take the prior to be flat in the (detector-frame) total mass $M_t$, the mass ratio $q$, the spins $-1 \leq \chi_1 \leq 1$ and $-1 \leq \chi_2 \leq 1$, the time to coalescence $t_c$, and the phase at coalescence $\varphi_c$. For the systems we consider here, the intrinsic parameters $M_t$, $q$, $\chi_1$, and $\chi_2$ are typically well measured, so that the actual priors have little importance. We take a flat prior on $\psi$, $\cos(\iota)$, and $\log_{10}(D_L)$ and fix the sky location ($\lambda$, $\beta$) to its true value to facilitate the convergence of the chains. Those parameters are not expected to correlate strongly with intrinsic parameters \cite{Marsat:2020rtl}, at least for aligned-spin binaries, and so this simplification should not significantly affect our conclusions. Finally, we take a flat prior between $-1$ and $1$ for the fractional deviations to the QNMs, $\delta f_{\ell m}$ and $\delta \tau_{\ell m}$. 
Assuming noise to be stationary and Gaussian, the likelihood reads
%
\begin{equation}
    p(d|\theta) \propto \prod_{c \in [A,E]}\exp \left [ -\frac{1}{2}(d_c-h_c(\theta)|d_c-h_c(\theta)) \right ].\label{eq:logl}
\end{equation}
The posterior distribution is then sampled via a Markov-chain Monte Carlo algorithm (MCMC). We use the Eryn sampler \cite{Karnesis:2023ras,eryn} for this purpose.

\section{Astrophysical binary systems}\label{sec:astro}

\begin{figure*}
\centering
 \includegraphics[width=\textwidth]{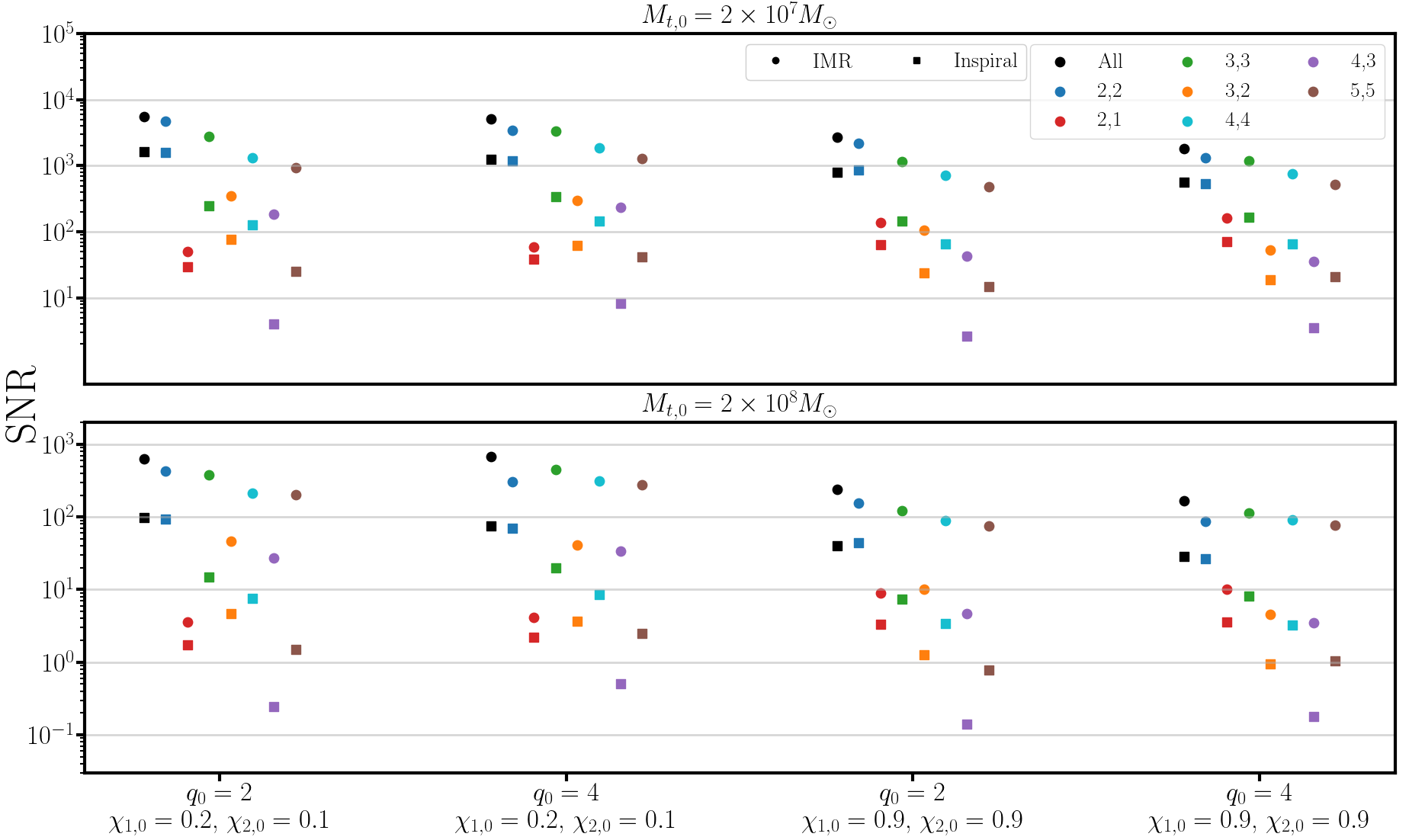}\\
 \centering
 \caption{Total SNR (black) and SNR in the individual harmonics (colours) for all the systems at $z_0=2.2$. The values at $z_0=3.7$ can be obtained by rescaling by $18,331/33,691 \simeq 0.54$. Circles show the full IMR SNR and squares the SNR in the inspiral stage.}\label{fig:snrs}
\end{figure*}

\begin{figure*}[btp]
\centering
 \includegraphics[width=\textwidth]{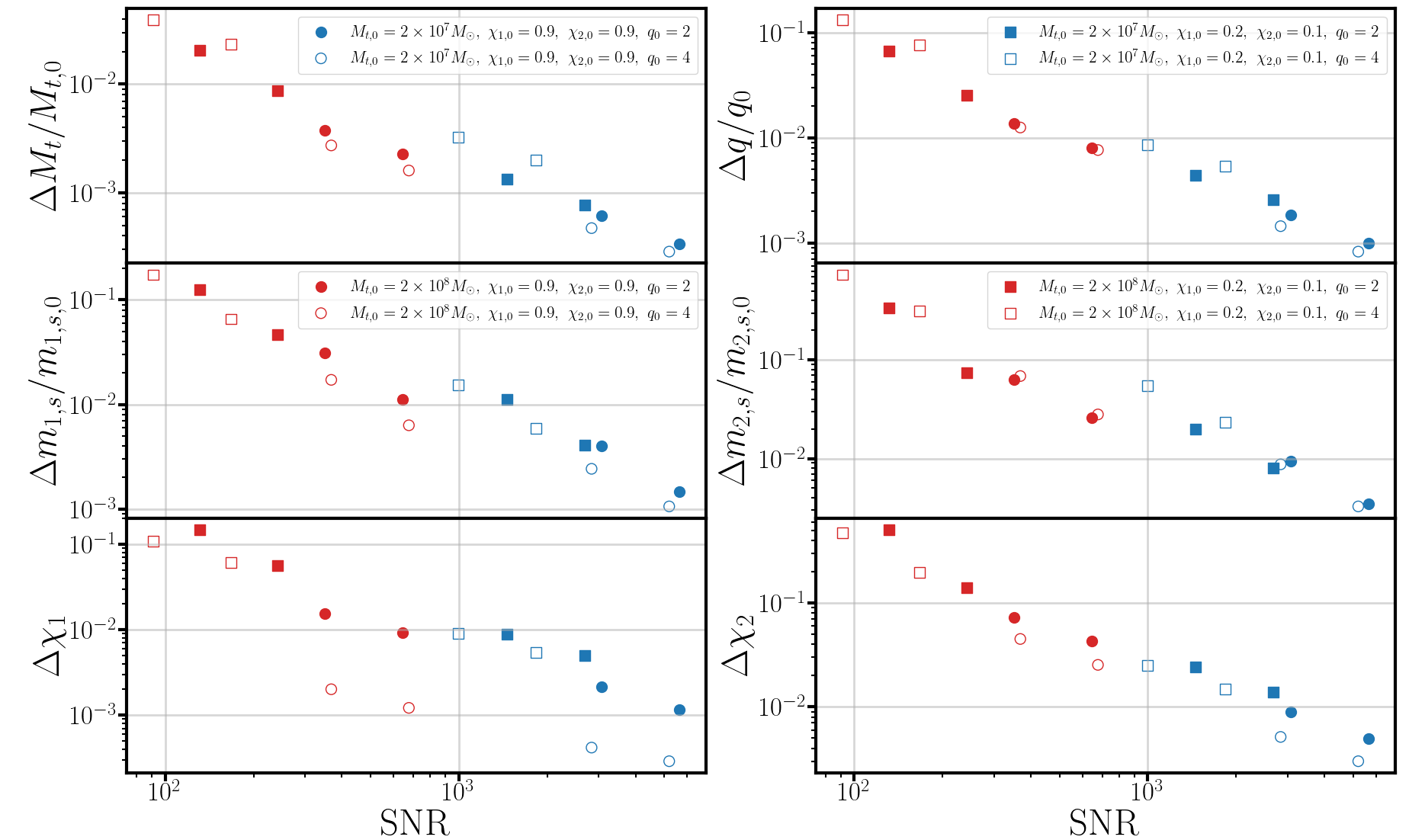}\\
 \centering
 \caption{Measurement error ($90\%$ confidence interval centred around the median) of intrinsic parameters. For mass parameters, we show the errors relative to the injection value. We inject signals within GR and recover them with GR templates (i.e., we employ \texttt{SEOBNRv5HM}). All points are doubled because of the two redshift possibilities; $z_0=2.2$ ($z_0=3.7$) yields larger (lower) SNR and lower (larger) errors. We show the detector-frame total mass and the mass ratio in the top row, the source-frame individual masses in the middle one, and the dimensionless spins in the bottom row. At leading order, measurement errors follow the $1/{\rm SNR}$ trend.}\label{fig:gr_errs}
\end{figure*}

We consider a set of 16 binary systems, defined from all possible combinations of the following choices of parameters:
\begin{itemize}
    \item $M_{t,0}=2 \times 10^7  \ M_{\odot}$ or $M_{t,0}=2 \times 10^8  \ M_{\odot},$
    \item $q_0=2$ or $q_0=4,$
    \item $\chi_{1,0}=\chi_{2,0}=0.9$ or $\chi_{1,0}=0.2, \ \chi_{2,0}=0.1, $ and
    \item $z_0=2.2$ ($D_{L,0}=18,331 \ {\rm Mpc}$) or $z_0=3.7$ ($D_{L,0}=33,691 \ {\rm Mpc}$).
\end{itemize}
Subscripts 0 indicate the true value of the parameter used to generate the synthetic injections.
This set of systems lies in the high-mass end of predictions for the population visible to LISA, as predicted from semianalytic models of MBH populations that use heavy seeds for the MBH progenitors \cite{Sesana:2007sh,Sesana:2010wy,Klein:2015hvg,Bonetti:2018tpf,Dayal:2018gwg,Barausse:2020mdt,Barausse:2020gbp}. Different heavy-seed scenarios have been proposed, such as the collapse of protogalactic disks as a result of bar instabilities \cite{Volonteri:2007ax}, the runaway collision of stars at the centre of galaxies \cite{1987gady.book.....B}, or the direct collapse of gas at the centre of galaxies \cite{Rees:1984si} (see Ref.~\cite{Latif:2016qau} for a review). Such heavy systems are the ones expected to have higher SNR in the merger-ringdown \cite{Bhagwat:2021kwv,Cotesta:2023}, and are, therefore, the most relevant ones to our analysis. In particular, very heavy systems ($M_{t,0}=2\times 10^8 \  M_{\odot}$) are expected to have very little SNR in the inspiral, and it is interesting to assess how well the parameters of such systems can be measured. Focusing on such massive systems is even more well motivated following the latest results from pulsar-timing-array observations~\cite{Antoniadis:2023rey,NANOGrav:2023gor,Reardon:2023gzh,Xu:2023wog}. If the apparent signal in the pulsar-timing-array data is generated by massive black hole inspirals, it indicates that MBHs might be more massive than originally expected. Semianalytical models predict a wide range of values for the mass ratio, but the vast majority of systems are predicted to have comparable masses. This is also the domain where our current IMR models are the most reliable.
The spin of a MBH typically depends on the amount of gas surrounding it and on how it has acquired mass and angular momentum through accretion \cite{Sesana:2014bea}. We consider two possibilities in order to cover both the case where MBHs are efficiently spun up and the case they are not. The relative alignment of the spins in a MBHB also depends on the presence of gas around the binary. Mergers happening in a gas-rich environment tend to have aligned spins due to the Bardeen-Peterson effect \cite{Bardeen:1975zz,Bogdanovic:2007hp}. Binaries formed through triplet interactions can also have misaligned spins, in addition to having high eccentricity \cite{Bonetti:2018tpf}. As discussed,  we neglect eccentricity and spin precession here for simplicity and focus on quasicircular binaries. 
Exploring different values for $q$, $\chi_1$, and $\chi_2$ is interesting, because they affect how much higher harmonics are excited and, therefore, how well it is possible to constrain the QNMs other than the fundamental one.
Finally, the redshifts at which MBHBs coalesce depend on when MBH seeds form and on the different timescales at play during the hardening of the binary \cite{Merritt:2004gc,Colpi:2014poa,Tremmel2018}. In heavy-seed models, MBHBs are expected to merge dominantly at late times (i.e., low redshift). Since $M_t$ is the detector-frame total mass, changing the redshift affects only the SNR of the system. Based on the predictions of semianalytical models \cite{Sesana:2007sh,Sesana:2010wy,Klein:2015hvg,Bonetti:2018tpf,Dayal:2018gwg,Barausse:2020mdt,Barausse:2020gbp}, we expect to observe up to a few tens of systems similar to the ones we defined during the nominal mission duration (four and a half years). For all binary systems we take $\iota_0=\pi/3$, $\psi_0=\pi/3$, $\lambda_0=\pi/3$, $\beta_0=\pi/3$, and $\varphi_{c,0}=0$.

In Fig.~\ref{fig:ampls}, we plot the frequency-domain amplitude of the TDI variable $A$ for the systems with $M_{t,0}=2\times 10^7 \ M_{\odot}$ and $z_0=2.2$ and the four combinations of mass ratio and spins. We show the contribution of each harmonic, $\tilde{A}_{\ell m}$, using 
Eq.~(\ref{eq:harmonic_decomp}) when computing the TDI variables [see Eq~(\ref{eq:tdi_lw})]. The black dashed lines indicate the GW frequency at $t_c$, which we choose as the separation between the inspiral and merger-ringdown regimes. As expected, the $(2,2)$ harmonic is the loudest, but higher harmonics are also important, in particular, the $(3,3)$, $(4,4)$ and $(5,5)$. This is better quantified in Fig.~\ref{fig:snrs}, where we show the total SNR and the contribution of each harmonic. We observe that the relative importance of the subdominant $(2,1)$, $(3,2)$ and $(4,3)$ harmonics depends primarily on the spins: $(2,1)$ is more dominant for low-spin systems, whereas $(3,2)$ and $(4,3)$ are more dominant for high-spin systems. We note the very high SNR of some of these systems, reaching a few thousands. Figure~\ref{fig:snrs} also shows the inspiral SNR, defined by using the GW frequency at $t_c$ as the upper limit in the integral in Eq.~(\ref{eq:inner_product}). As anticipated, the signals of the systems we consider are merger-ringdown dominated. In Table \ref{tab:snrs}, we give the IMR, merger-ringdown, and inspiral SNR of the systems at $z_0=2.2$. Although systems with $M_{t,0}=2\times 10^7 \  M_{\odot}$ (upper panel) still have high SNR also in the inspiral, $\sim 1000$, it is not the case for the very massive systems ($M_{t,0}=2\times 10^8 \  M_{\odot}$, lower panel), with inspiral SNRs as low as $\sim 30$. 

\section{Measuring source properties and QNMs with MBHB observations}\label{sec:results}

We work with zero-noise injections, as these are well suited to the goals of understanding systematics and measurement uncertainties \cite{Rodriguez:2013oaa}, and perform three types of analyses using the \texttt{SEOBNRv5HM} and \texttt{pSEOBNRv5HM} models as synthetic-signal injections and templates.

\begin{enumerate}
    \item We inject a synthetic signal \emph{without} deviations from GR (i.e., \texttt{SEOBNRv5HM}) and use templates in the Bayesian analyses \emph{not allowing} for deviations from GR (i.e., \texttt{SEOBNRv5HM});
    \item we inject a synthetic signal \emph{without} deviations from GR (i.e., \texttt{SEOBNRv5HM}) and use templates in the Bayesian analyses \emph{allowing} for deviations from GR (i.e., \texttt{pSEOBNRv5HM});
    \item we inject a synthetic signal \emph{with} deviations from GR (i.e., \texttt{pSEOBNRv5HM}) and use templates in the Bayesian analyses \emph{allowing} for deviations from GR (i.e., \texttt{pSEOBNRv5HM}).
\end{enumerate}

The first type of analysis will estimate how well the parameters of MBHBs can be constrained assuming GR is correct. It is the first study of this kind using EOB waveforms. The second will tell us how well the deviation parameters of QNMs can be constrained, and the third for which values of the deviation parameters we can detect non-GR effects in the ringdown. We perform these mock injections for all MBHBs described in Sec.~\ref{sec:astro}.

\subsection{Measurement of source parameters in GR}

We show in Fig.~\ref{fig:gr_errs} the width of the $90\%$ confidence interval centred around the median for the intrinsic parameters (i.e., the masses and spins) as a function of the SNR of the system. The colour, shape, and filling of the point indicate, respectively, the total mass, spin and mass ratio of the system, indicated in the legends. Each point is doubled because of the two redshifts used: $z_0=2.2$ ($z_0=3.7$) corresponds to the largest (smallest) SNR and the smallest (largest) measurement error. Note that we show the detector-frame total mass in the top row and the source-frame individual masses in the middle one. For all systems the parameters are well constrained, and we find that the error (or relative error for the mass parameters) goes as $1/{\rm SNR}$, as expected in the high SNR regime \cite{Finn:1992xs,Cutler:1994ys,Vallisneri:2007ev}. This relation is more scattered for the spin  parameters, especially for $\chi_1$: Systems with $q_0=4$ have better spin measurement, in agreement with \cite{Cotesta:2023}. This is because, for such systems (nonfilled points), higher harmonics become more dominant (see Figs.~\ref{fig:ampls} and \ref{fig:snrs}) and help improve the  measurement of the spins. It is remarkable that even for very massive systems (red points), which usually have low SNR in the inspiral, we get tight constraints on their parameters, similarly to what \cite{Baibhav:2020tma} found. This is the benefit of using a fully consistent modelling of the IMR signal, since in our model the merger-ringdown signal also informs us on the parameters of the component BHs in the binary.

\begin{figure}
\centering
 \includegraphics[width=0.49\textwidth]{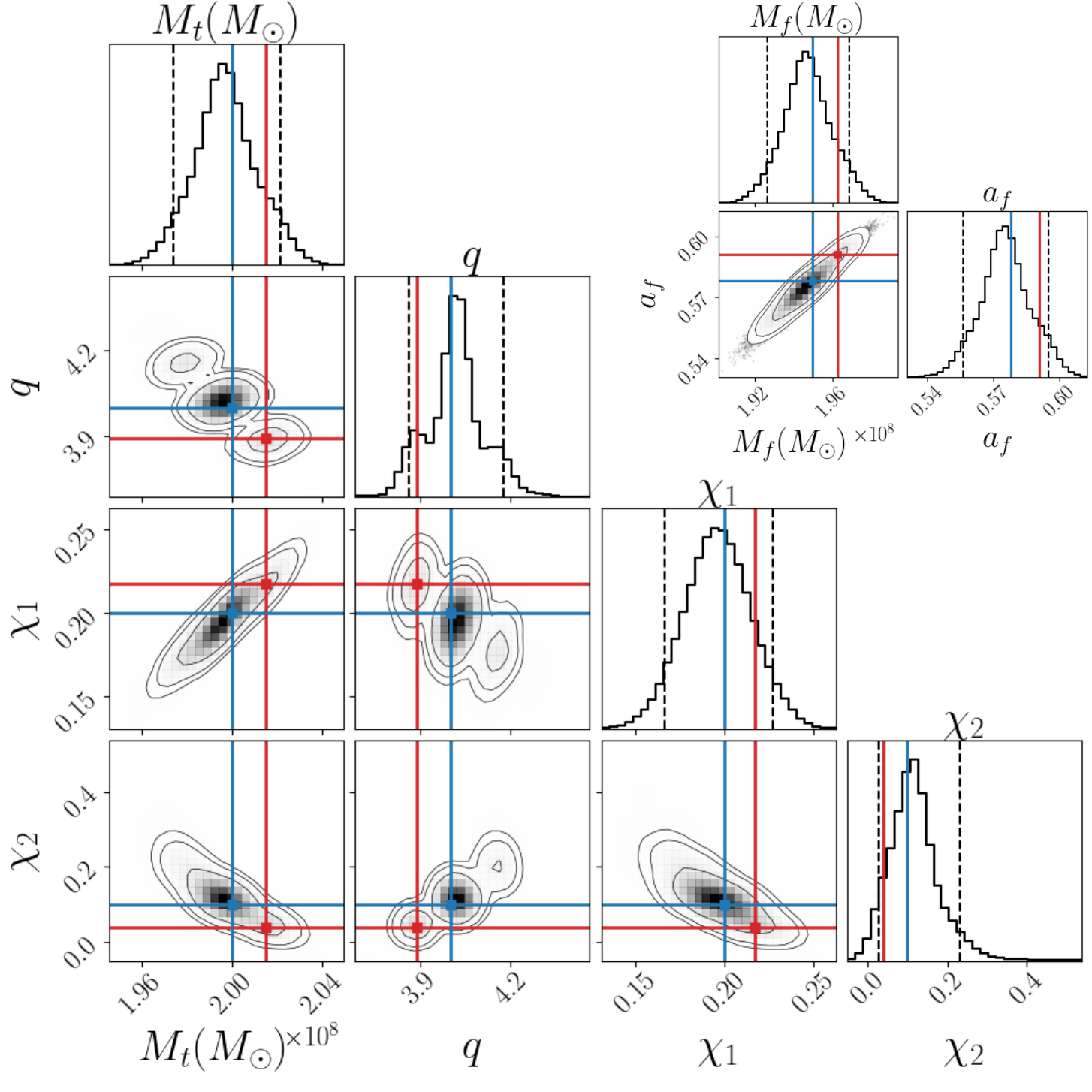}\\
 \centering
 \caption{Corner plot for the system with $M_{t,0}=2\times 10^8  \ M_\odot$, $\chi_{1,0}=0.2$, $\chi_{2,0}=0.1$, $q_0=4$, $z_0=2.2$, and SNR=167. The blues lines show the injection and the red ones a secondary mode. Contours show the $68\%, \ 90\%, \ {\rm and} \ 95\%$ confidence intervals, and dashed lines the 0.05 and 0.95 quantiles. On the upper-right part, we show the inferred properties of the remnant. The multimodality observed in the binary parameters is due to combinations of intrinsic parameters that yield remnant properties compatible with the true values within the measurement uncertainty, as can be seen from the absence of clear multimodality in $M_f$ and $a_f$. }\label{fig:corner_id15}
\end{figure}

\begin{figure}
\centering
 \includegraphics[width=0.45\textwidth]{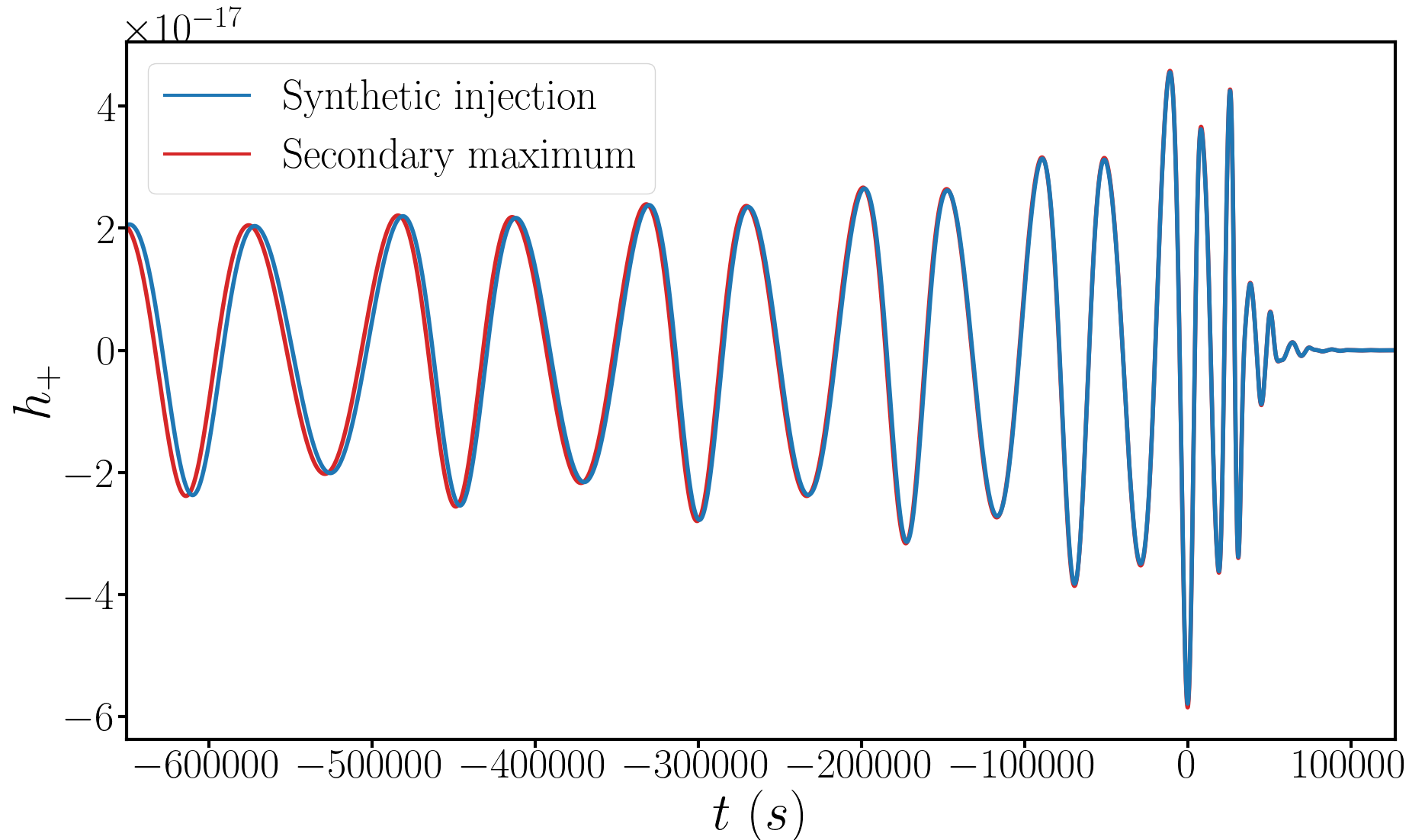}\\
 \centering
 \caption{Comparison between synthetic-injection waveform and the one at a secondary maximum (same colour code as in Fig.~\ref{fig:corner_id15}). The late inspiral and merger-ringdown signals match almost perfectly. The early inspiral signals are dephased, but the SNR in the inspiral is so low that it makes little difference.}\label{fig:wvf_sec_max_id15}
\end{figure}

For very massive systems with low SNR, we find a multimodality in intrinsic parameters, as illustrated in Fig.~\ref{fig:corner_id15}. Secondary modes arise from combinations of parameters that yield remnant parameters compatible with the true ones within the measurement uncertainty, as shown in the upper-right part in Fig.~\ref{fig:corner_id15}. As a consequence, the merger-ringdown signal remains quasi-identical to the synthetic injection. This can be seen in Fig.~\ref{fig:wvf_sec_max_id15}, where we compare the waveform of the injected parameters (in blue in Fig.~\ref{fig:corner_id15}) to the one with parameters from one of the secondary maxima (in red in Fig.~\ref{fig:corner_id15}). The early part of the inspiral signal is fairly different, but this has little importance because the system is merger-ringdown dominated, and the SNR in the inspiral is small compared to the total SNR ($\sim 30$, see Fig.~\ref{fig:snrs} and Table \ref{tab:snrs}).

In Appendix~\ref{app:res} we discuss the impact of the tolerance of the integrator used to solve the Hamilton equations and compute the EOB waveforms. In Appendix~\ref{app:imrt} we compare the measurement errors obtained with \texttt{pSEOBNRv5HM} to the ones obtained with the IMR phenomenological model \texttt{IMRPhenomTHM} \cite{Estelles:2020twz}, finding comparable results.

\begin{figure*}
\centering
 \includegraphics[width=\textwidth]{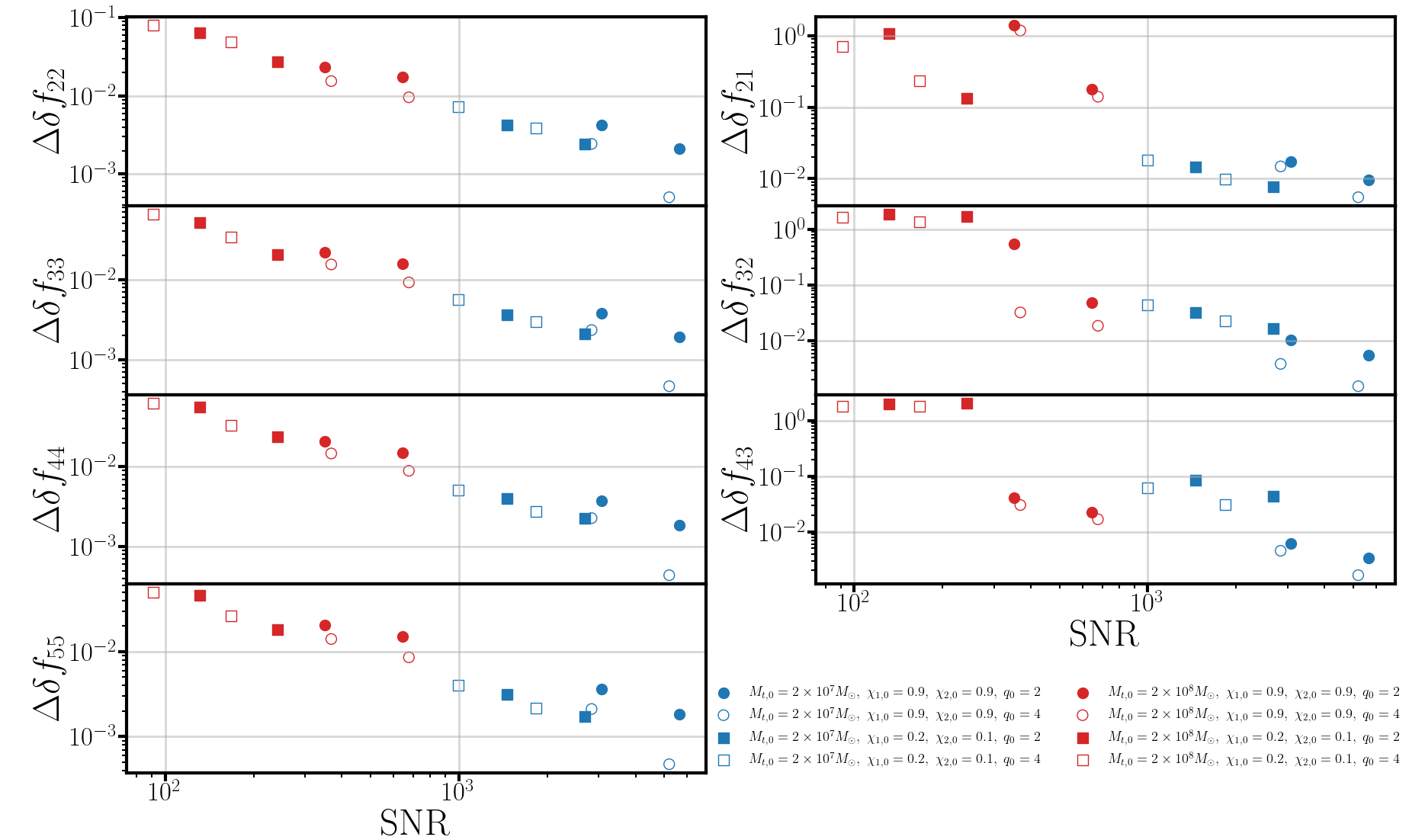}\\
 \centering
 \caption{Width of the $90\%$ confidence interval on $\delta f_{\ell m}$ centred around the median for GR injections. 
 }\label{fig:omega_errs}
\end{figure*}

\begin{figure*}
\centering
 \includegraphics[width=\textwidth]{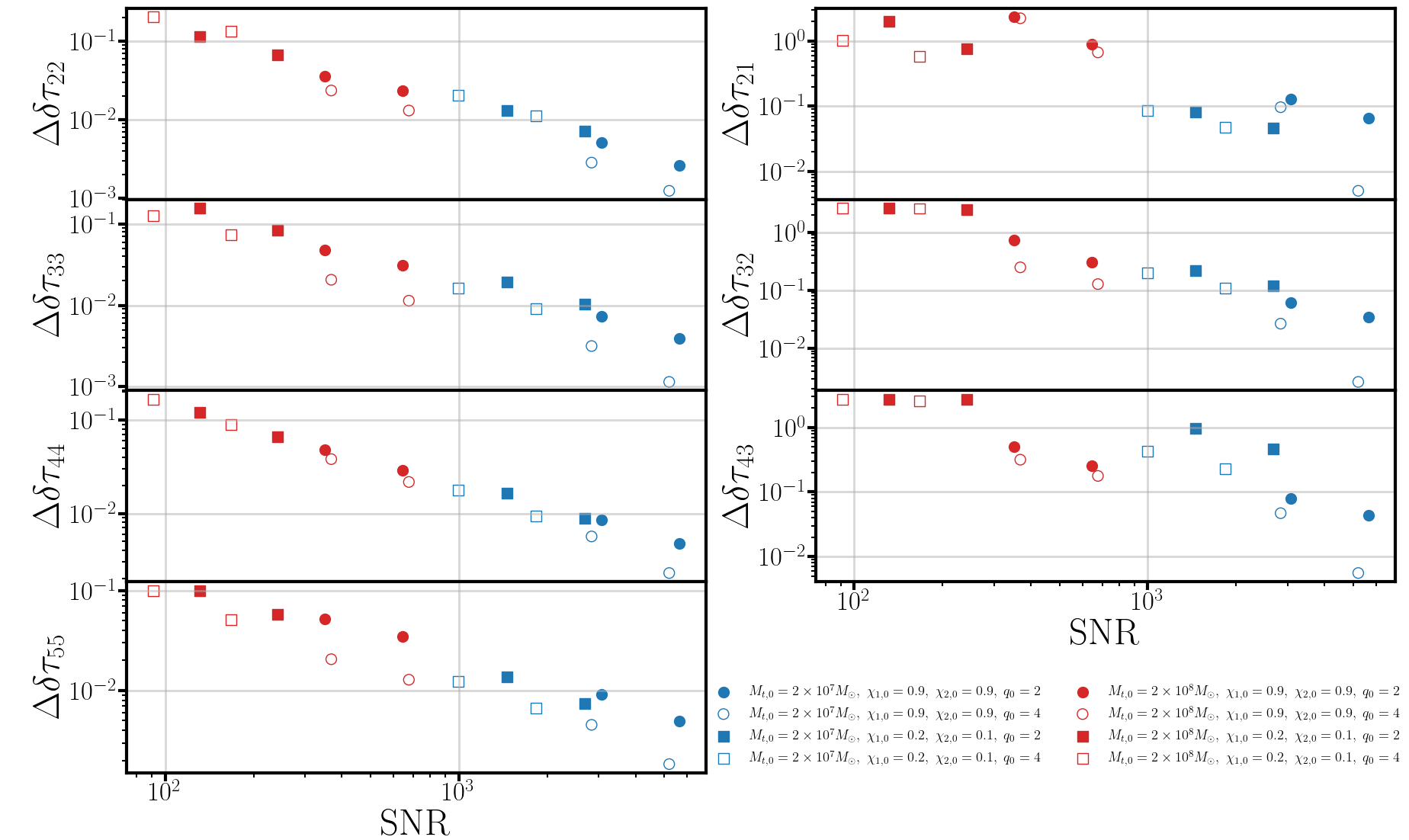}\\
 \centering
 \caption{Width of the $90\%$ confidence interval on $\delta \tau_{\ell m}$ centred around the median for GR injections. 
 }\label{fig:tau_errs}
\end{figure*}

\begin{figure*}
 \centering
\subfigure[\ $M_{t,0}=2\times 10^7 M_{\odot}$, $\chi_{1,0}=\chi_{2,0}=0.9$, $q_0=4$, $z_0=2.2$, SNR=5174.]{
    \centering \includegraphics[scale=0.17]{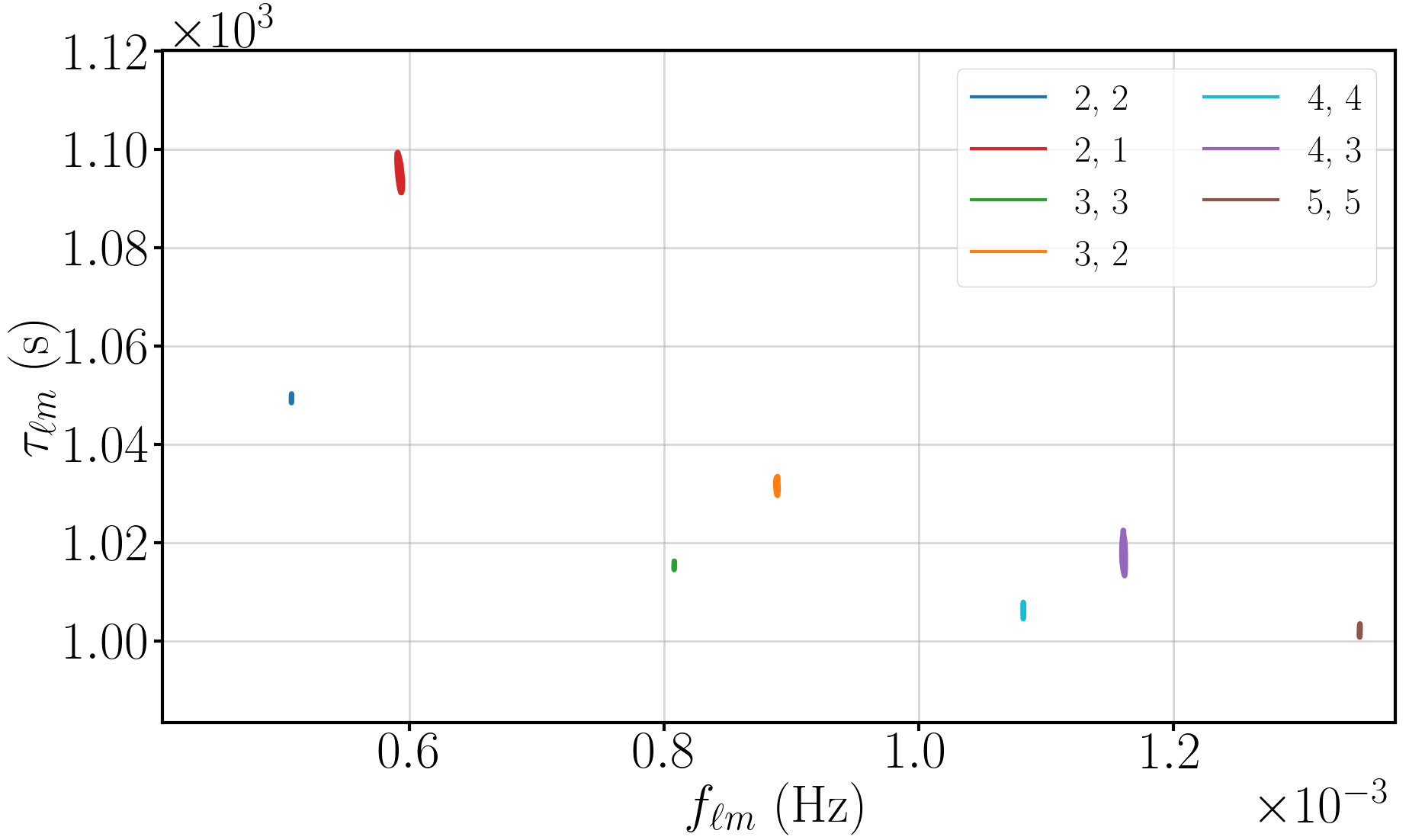}
    }
\centering
\subfigure[\ $M_{t,0}=2\times 10^8 M_{\odot}$, $\chi_{1,0}=\chi_{2,0}=0.9$, $q_0=4$, $z_0=2.2$, SNR=672.]{
    \centering \includegraphics[scale=0.17]{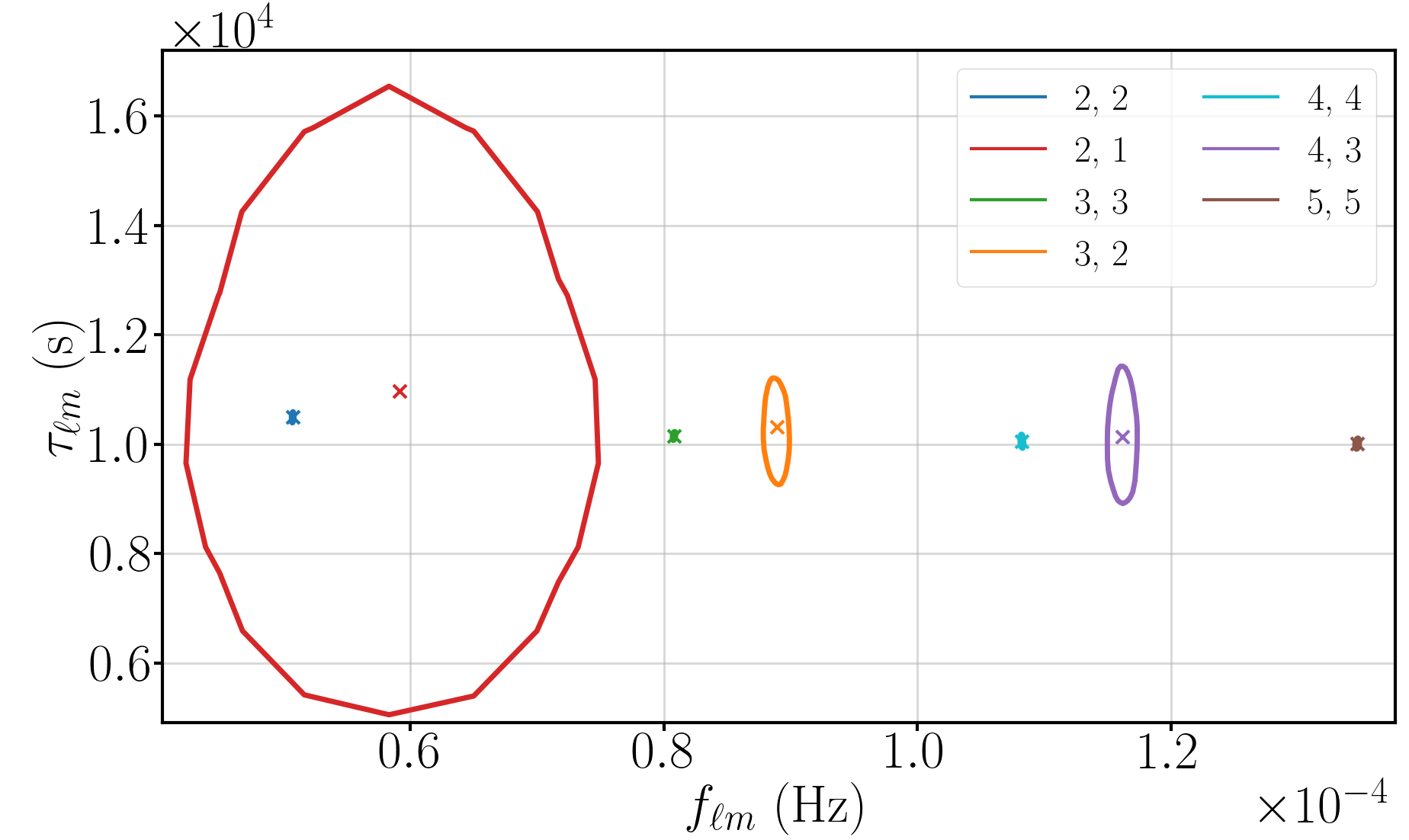}
   }
   \centering
   \centering
\subfigure[\ $M_{t,0}=2\times 10^7 M_{\odot}$, $\chi_{1,0}=0.2$, $\chi_{2,0}=0.1$, $q_0=2$, $z_0=3.7$, SNR=1457.]{
    \centering \includegraphics[scale=0.17]{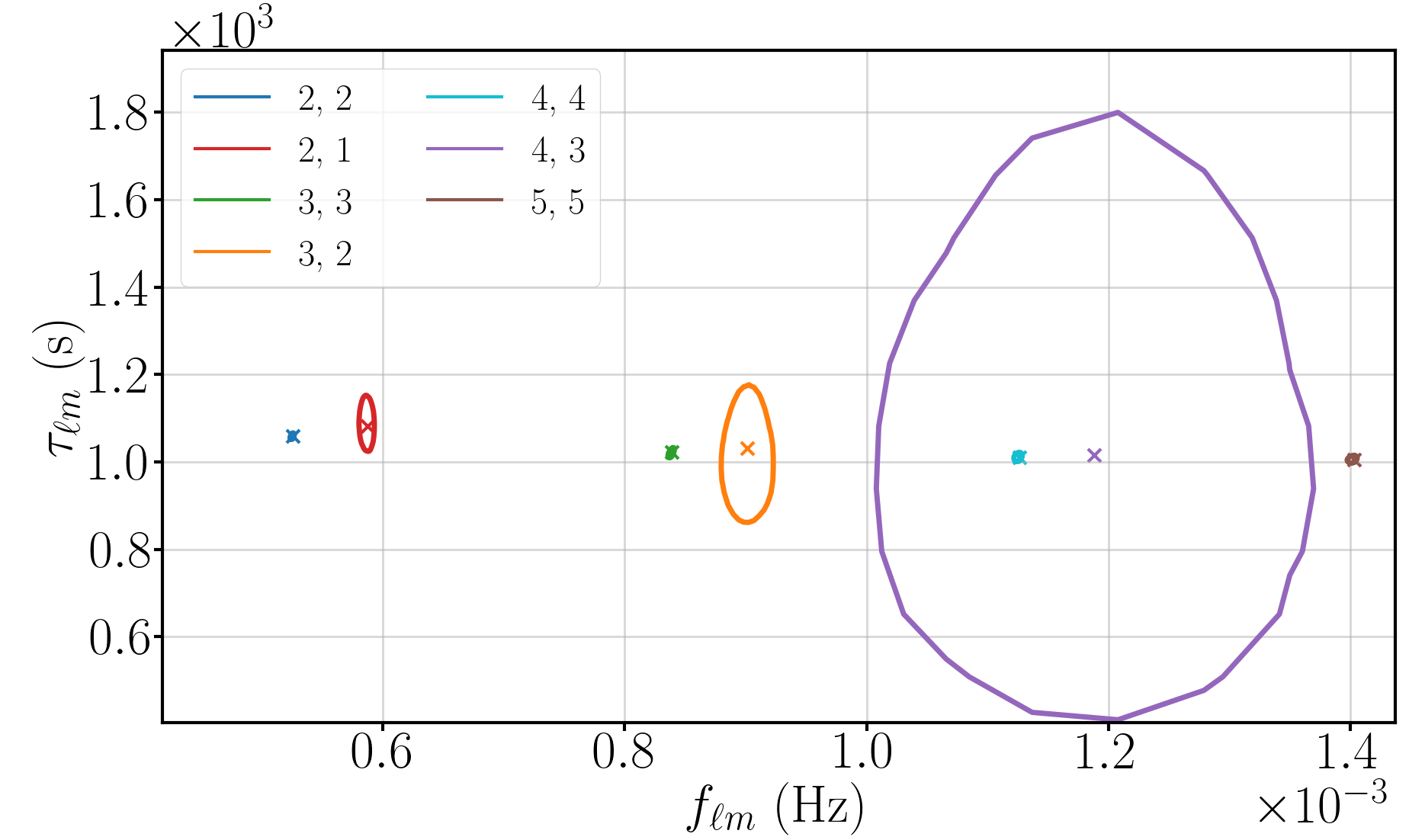}
   }
   \centering
     \centering
\subfigure[\ $M_{t,0}=2\times 10^8 M_{\odot}$, $\chi_{1,0}=0.2$, $\chi_{2,0}=0.1$, $q_0=2$, $z_0=3.7$, SNR=132.]{
    \centering \includegraphics[scale=0.17]{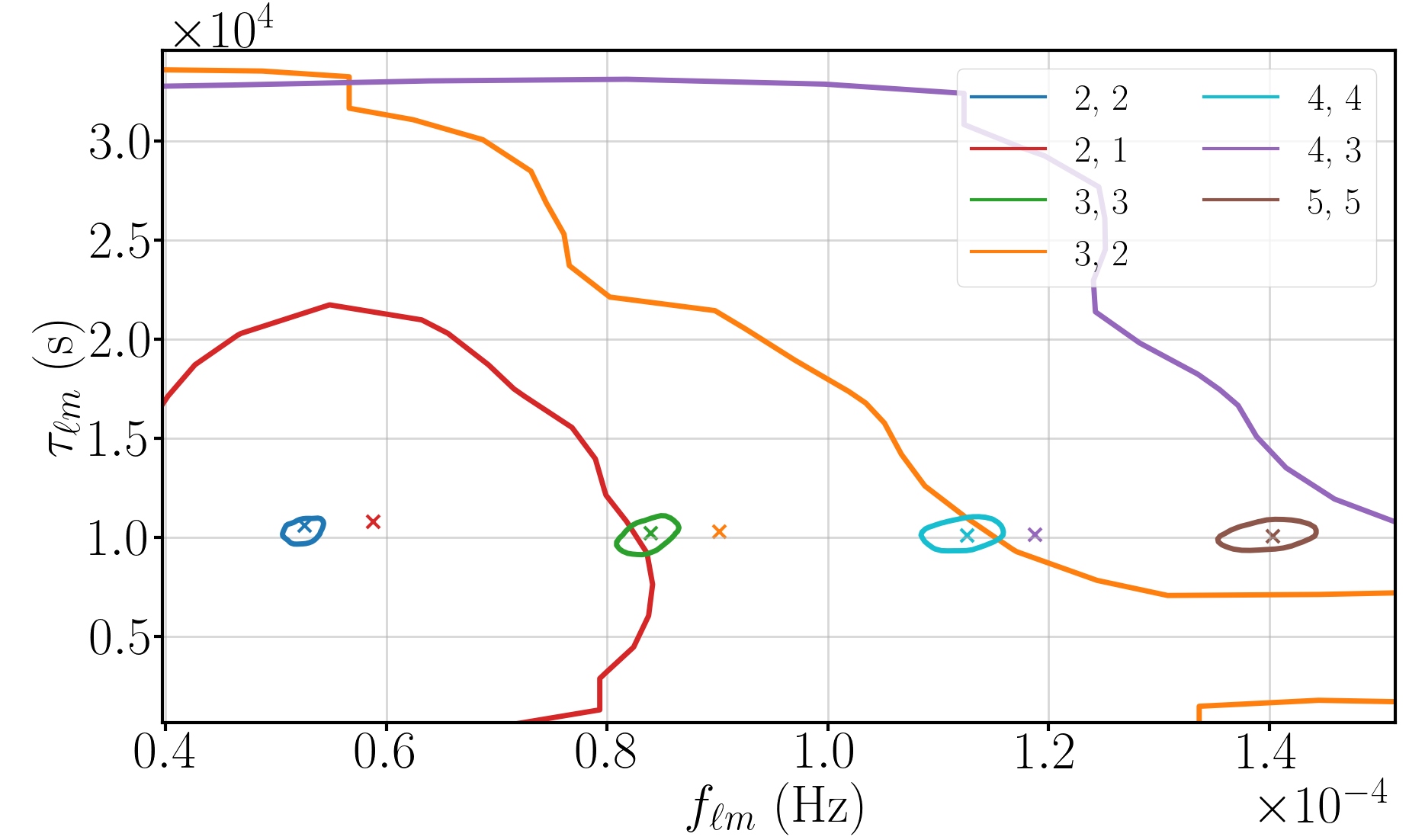}
   }
   \centering
   \caption{$90 \%$ confidence interval on the QNMs for four different systems. Crosses indicate the true values. Between columns only the total mass varies, and between rows mass ratio, spins, and redshift vary, keeping the same total mass. In every case, the synthetic injection is GR. The top-left panel illustrates why applying the Rayleigh criterion to both the damping time and the frequency to decide on the resolvability of QNMs is too stringent: Although the one-dimensional damping time posteriors overlap, the two-dimensional ones are well separated. That case corresponds to a ``best-case scenario.'' The other panels show cases where not all seven QNMs can be distinguished. The $(2,1)$, $(3,2)$, and $(4,3)$ modes are typically less-well constrained due to their lower SNR (see also Figs.~\ref{fig:omega_errs} and \ref{fig:tau_errs}). }\label{fig:qnms}
 \end{figure*}

\begin{figure*}
 \centering
\subfigure[\ Deviation parameters]{
    \centering \includegraphics[scale=0.35]{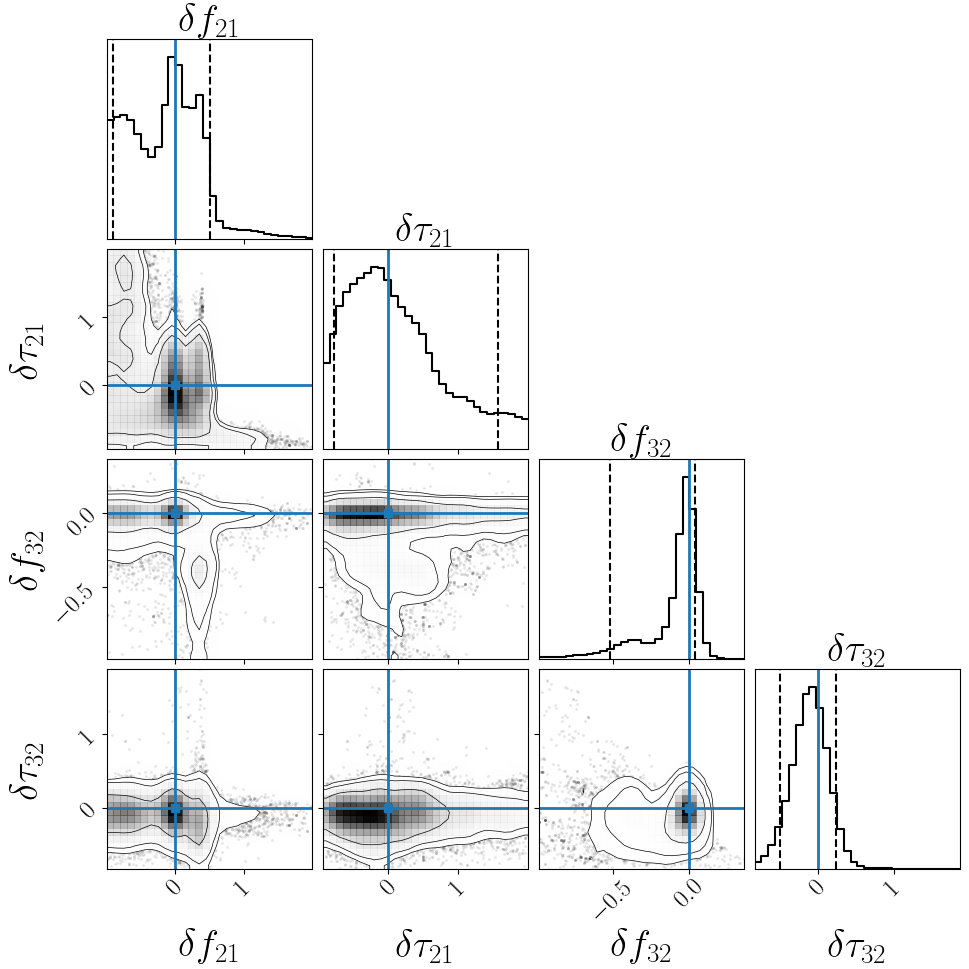}
    }
\centering
\subfigure[\ QNMs]{
    \centering \includegraphics[scale=0.35]{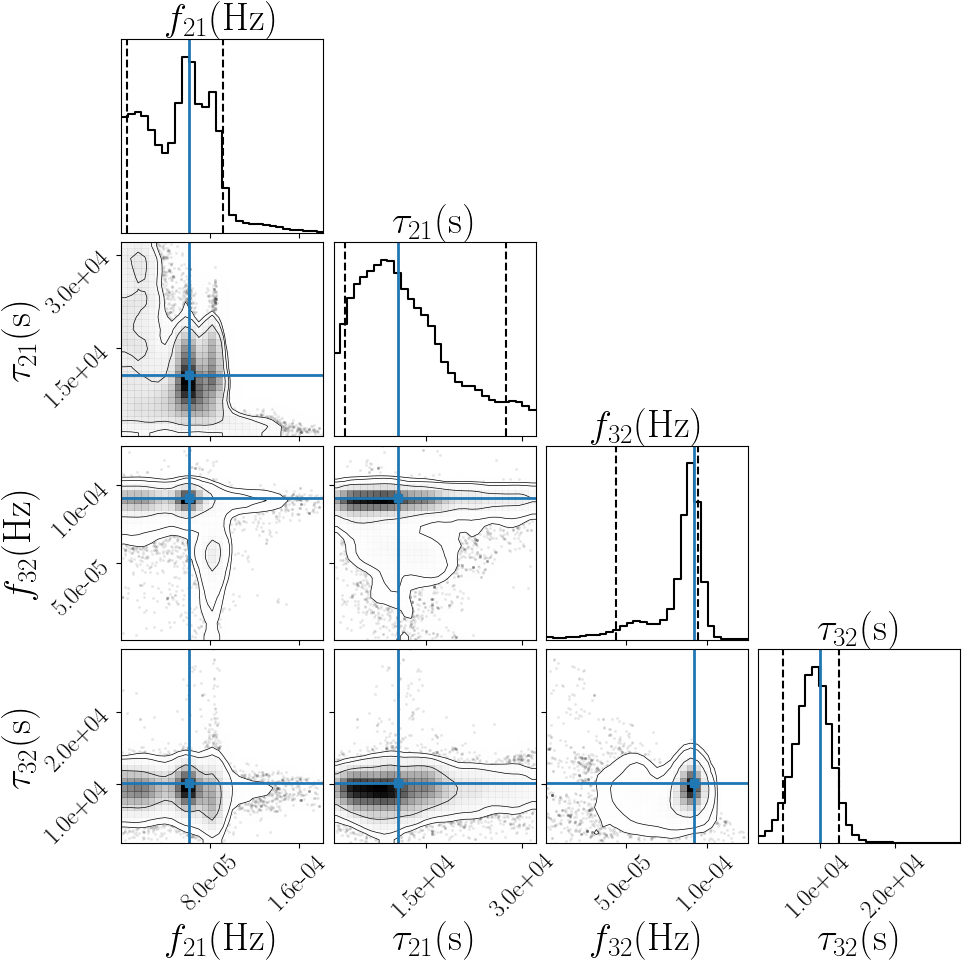}
   }
   \centering
   \centering
   \caption{Reduced corner plot of deviation parameters and QNMs for a GR injection of the system $M_{t,0}=2\times 10^8$, $\chi_{1,0}=\chi_{2,0}=0.9$, $q_0=2$, $z_0=3.7$, and SNR=339. We find a multimodality in the $\delta f_{21}$ posterior, which is the less-well measured harmonic for this system. This leads to a secondary mode in the $f_{21}$ posterior at frequencies that match $f_{32}$. Given that the damping times of the two harmonics are poorly measured and their posterior fairly compatible, this suggests that the $(2,1)$ harmonic is trying to match the $(3,2)$ harmonic and vice versa.}\label{fig:multimodality_qnm}
 \end{figure*}

\subsection{Measurement of QNMs and possible deviations from GR}

We now turn our attention to QNM measurements and constraints on deviations from GR using MBHB observations with LISA. 

\subsubsection{GR injections}\label{sec:gr_inj_non_gr_bayes}

First, we consider the case where the injected signal is compatible with GR, and allow for nonzero deviations when running the Bayesian analysis. In Figs.~\ref{fig:omega_errs} and \ref{fig:tau_errs}, we show the width of the $90\%$ confidence interval centred around the median on the deviation parameters (i.e., $\Delta \delta f_{\ell m}$ and $\Delta \delta \tau_{\ell m}$). We find that deviations to the frequency are generally better constrained than those to the damping time. As a consequence of the higher SNR of the dominant $(2,2)$, $(3,3)$, $(4,4)$, and $(5,5)$ harmonics, fractional deviations to their QNMs are better constrained than those to the subdominant $(2,1)$, $(3,2)$, and $(4,3)$ harmonics. For the former, $\delta f_{\ell m}$ and $\delta \tau_{\ell m}$ are typically constrained within 0.1 and even within 0.01 for the systems with $M_{t,0}=2\times 10^7  \ M_{\odot}$ (blue points) and follow the $1/{\rm SNR}$ trend, with some scatter for higher harmonics [especially the $(3,3)$ and $(5,5)$ harmonics] that depends on the mass ratio. Here again, the reason for this is that, for systems with $q_0=4$ (nonfilled points), higher harmonics are more excited (see Figs.~\ref{fig:ampls} and \ref{fig:snrs}), so we are able to better constrain deviations in their QNMs, in agreement with \cite{Bhagwat:2021kwv}. Deviations in subdominant harmonics are poorly constrained for very massive systems (red points), $\Delta \delta f_{\ell m}$ and $\Delta \delta \tau_{\ell m} \sim 1$, which given our prior range, translates into uninformative measurements. This is due to the low SNR of these harmonics. However, we get rather good constrains, within 0.1, for systems with $M_{t,0}=2 \times 10^7 \  M_{\odot}$ (blue points). We also find that for low-spin systems (circles) deviations to the $(2,1)$ harmonic are better constrained than the ones to the $(3,2)$ and $(4,3)$ harmonics, whereas the opposite happens for high-spin systems (squares), in agreement with our remark on their relative predominance in Sec.~\ref{sec:astro}. We show the impact of allowing for deviations from GR on the measurement of intrinsic parameters in Appendix~\ref{app:gr_errors}.

We translate these constraints on the fractional deviations into measurements of the QNMs in Fig.~\ref{fig:qnms}, where we show some representative examples of $90\%$ confidence regions on the QNMs of all the harmonics included in \texttt{pSEOBNRv5HM}. The upper-left panel shows a best-case scenario, where all QNMs can be perfectly distinguished. It illustrates that applying the Rayleigh criterion \cite{Berti:2005ys} to both the damping time and the frequency to decide on the distinguishability of QNMs is too stringent. Indeed, as can be seen from the upper-left panel in Fig.~\ref{fig:qnms}, the one-dimensional projection of the $90\%$ confidence regions onto the $y$ axis (damping time) can overlap [e.g., for the $(4,4)$ and $(4,3)$ QNMs], although the two-dimensional regions are well separated. Thus, one should really consider the two-dimensional regions in order to decide on the distinguishability of QNMs, as pointed out in Ref.~\cite{Isi:2021iql}. However, given the little correlation between $\tau_{\ell m}$ and $f_{\ell m}$, it is often enough to apply the Rayleigh criterion only to the damping time or to the frequency, as suggested in previous works~\cite{Berti:2005ys,Bhagwat:2019dtm,Ota:2019bzl,JimenezForteza:2020cve}. In the cases shown here, the frequency is enough to decide on the distinguishability. The upper-right and lower-left panels illustrate cases where not all QNMs can be resolved. As expected from our comments above, for a high-spin system such as the one shown in the upper-right panel, deviations to the $(2,1)$ QNM are poorly constrained, so its measurement uncertainty contour encloses that of the $(2,2)$ mode. Similarly, the lower-left panel shows a low-spin system, for which the $(4,3)$ QNM measurement uncertainty contour contains the $(4,4)$ one. Finally, the lower-left panel shows a worst-case scenario where the QNMs cannot be distinguished due to the large uncertainty on the subdominant harmonics. Note that the system in the lower-left panel has a total mass of $2\times 10^7  \ M_{\odot}$, illustrating that the confidence regions of QNMs are not always all well separated for systems with $M_{t,0}=2\times 10^7 
 \ M_{\odot}$, although they do tend to yield better results than for systems with $M_{t,0}=2\times 10^8  \ M_{\odot}$, as illustrated in Figs.~\ref{fig:omega_errs} and \ref{fig:tau_errs}. 

We find some cases of multimodality in the deviation parameters, as illustrated in Fig.~\ref{fig:multimodality_qnm}. They can be understood by looking at the corresponding values of QNMs. Indeed, the frequency of the secondary mode in the $(2,1)$ QNM matches the frequency of the $(3,2)$ QNM. Because their damping times are poorly constrained, they are fairly compatible. Thus, this multimodality can be understood as the subdominant harmonics trying to ``match'' each other. 

\begin{figure*}
\centering
 \includegraphics[width=0.98\textwidth]{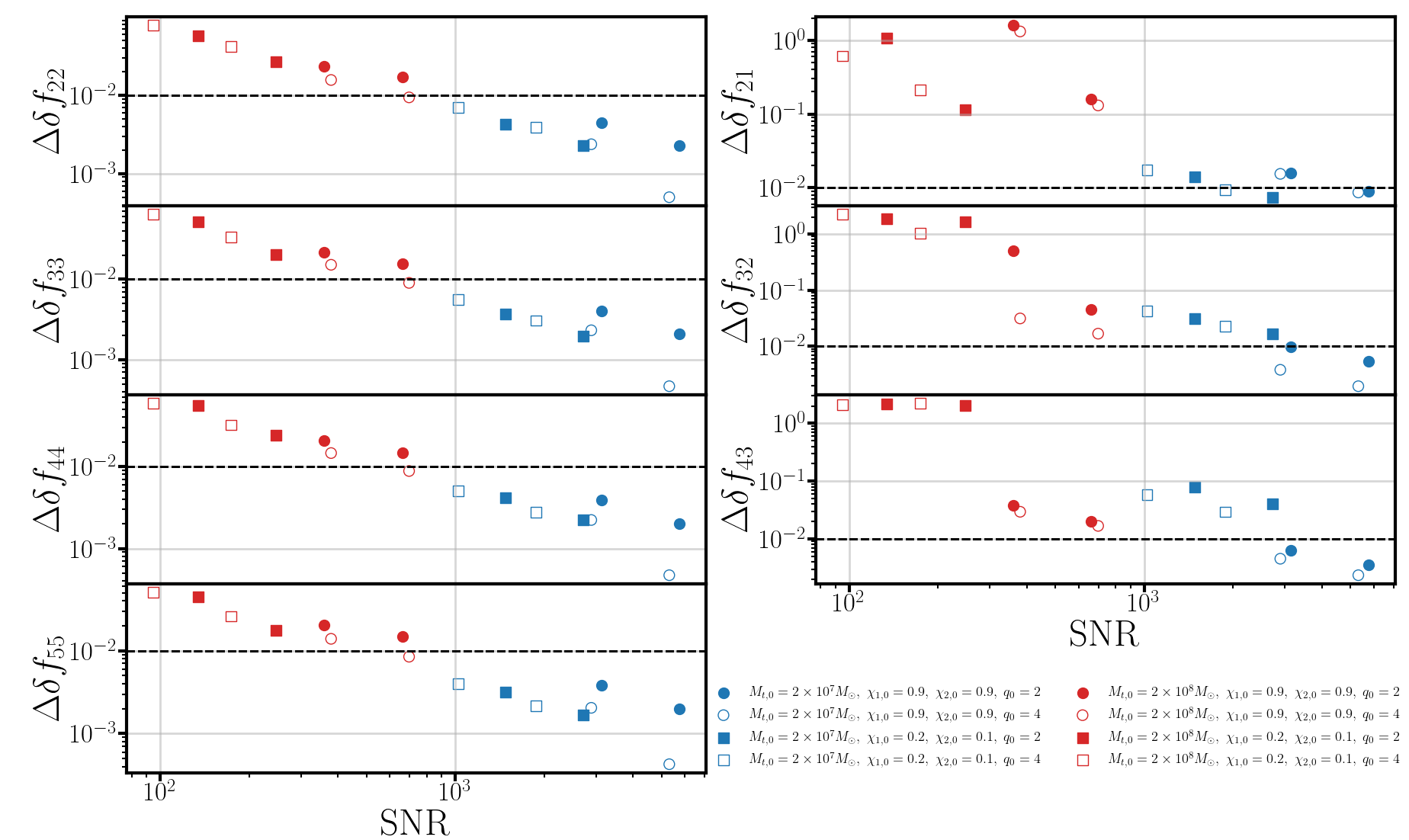}\\
 \centering
 \caption{Width of the $90\%$ confidence interval centred around the median for $\delta f_{\ell m}$ for injections with $\delta f_{\ell m}=\delta \tau_{\ell m}=0.01$, as indicated by the black-dashed lines.}\label{fig:omega_errs_ngr}
\end{figure*}

\begin{figure*}
\centering
 \includegraphics[width=0.98\textwidth]{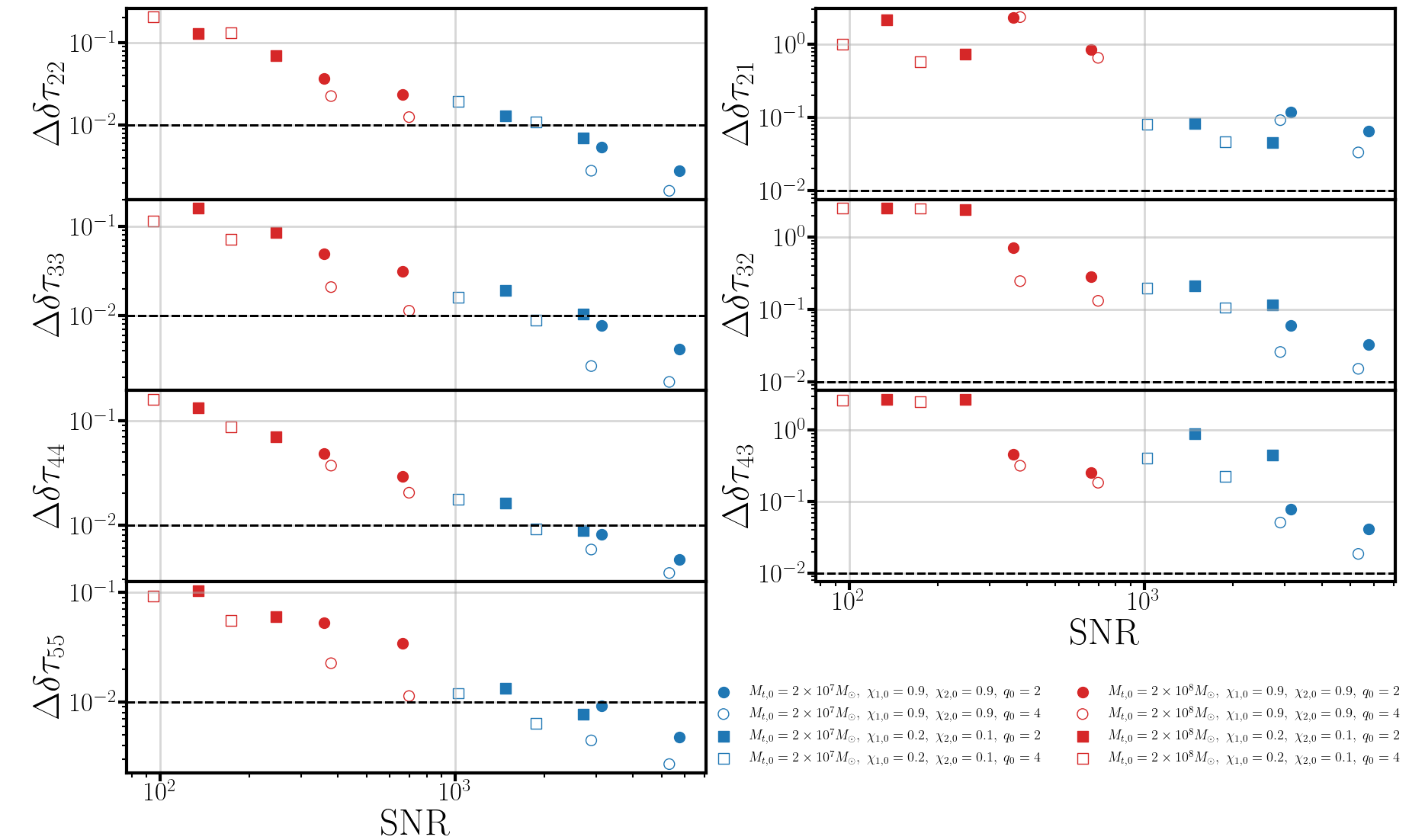}\\
 \centering
 \caption{Width of the $90\%$ confidence interval centred around the median for $\delta \tau_{\ell m}$ for injections with $\delta f_{\ell m}=\delta \tau_{\ell m}=0.01$, as indicated by the black-dashed lines.}\label{fig:tau_errs_ngr}
\end{figure*}

\subsubsection{Non-GR injections}

We now consider non-GR injections, and we generate synthetic signals with nonzero deviations to the QNMs. Deviations to GR in the QNM frequencies have been derived in non-GR theories and typically vary in the range $0.01$--$0.1$ or even smaller. In the spherically
symmetric case (i.e., nonspinning BHs), they were computed in theories such as
Einstein-Maxwell-dilaton~\cite{Ferrari:2000ep}, dynamical
Chern-Simons gravity~\cite{Molina:2010fb}, Einstein-dilaton-Gauss-Bonnet
gravity~\cite{Pani:2009wy,Blazquez-Salcedo:2016enn,Blazquez-Salcedo:2017txk}, higher-curvature gravity theories~\cite{Cano:2023jbk}, and for some solutions in massive
(bi)gravity~\cite{Brito:2013wya,Brito:2013yxa,Babichev:2015zub}. Recently, the computation of 
QNMs for spinning BHs in non-GR theories has received much attention, since the remnant BHs we are observing with LIGO and Virgo have typically spins of about 0.7. They include the Kerr-Newman case in Einstein-Maxwell theory~\cite{Pani:2013ija,Pani:2013wsa,Mark:2014aja,Dias:2015wqa}, Einstein scalar Gauss-Bonnet gravity~\cite{Pierini:2021jxd,Pierini:2022eim}, higher-curvature gravity theories~\cite{Cano:2023tmv,Cano:2023jbk}, and dynamical Chern-Simons theory~\cite{Wagle:2021tam}. Estimates for QNMs of spinning BHs in non-GR theories have also used the connection between the light ring and QNMs~\cite{Blazquez-Salcedo:2016enn,Glampedakis:2017dvb,Jai-akson:2017ldo,Glampedakis:2017cgd}, which is formally valid only in the eikonal $\ell \to \infty$ limit,
and are known to fail to describe some families of QNMs when additional
degrees of freedom are present~\cite{Blazquez-Salcedo:2016enn}.

Here, we assume the fractional deviation to GR to be 0.01 for all harmonics, for both frequencies and damping times. In Figs.~\ref{fig:omega_errs_ngr} and \ref{fig:tau_errs_ngr}, we show the width of the $90\%$ confidence interval centred around the median for $\delta f_{\ell m}$ and $\delta \tau_{\ell m}$. As a rule of thumb, we consider that a deviation can be measured when it is larger than the measurement error. Graphically, this corresponds to the points that are below the black dashed lines in Figs.~\ref{fig:omega_errs_ngr} and \ref{fig:tau_errs_ngr}. For this value of the deviation (i.e., $0.01$), we find that it could be detected in the frequency of the $(2,2)$, $(3,3)$, $(4,4)$, and $(5,5)$ harmonics of systems with $M_{t,0}=2\times 10^7  \ M_{\odot}$, and for the higher SNR ones we could also detect this deviation in their damping time. We note that the errors shown in Figs.~\ref{fig:omega_errs_ngr} and \ref{fig:tau_errs_ngr} are very similar to the ones we find when injecting GR signals (see Figs.~\ref{fig:omega_errs} and \ref{fig:tau_errs}). We also perform injections with deviations of 0.1 and 0.001 (not shown here), and find again similar errors. Therefore, we can extrapolate the results presented here and read from those figures which values of the deviations would be needed to detect them. For instance, a deviation of 0.1 could be detected for almost all systems presented here, in both the frequency and the damping time of the dominant harmonics (left column), and for the higher SNR systems, even of the subdominant harmonics (right column). Detecting a deviation in several harmonics, preferably in both the frequency and the damping time, would reinforce our confidence that we are truly observing effects in gravity theories alternative to GR. 


\section{Impact of systematics}\label{sec:syst}

\begin{figure}
\centering
 \includegraphics[width=0.49\textwidth]{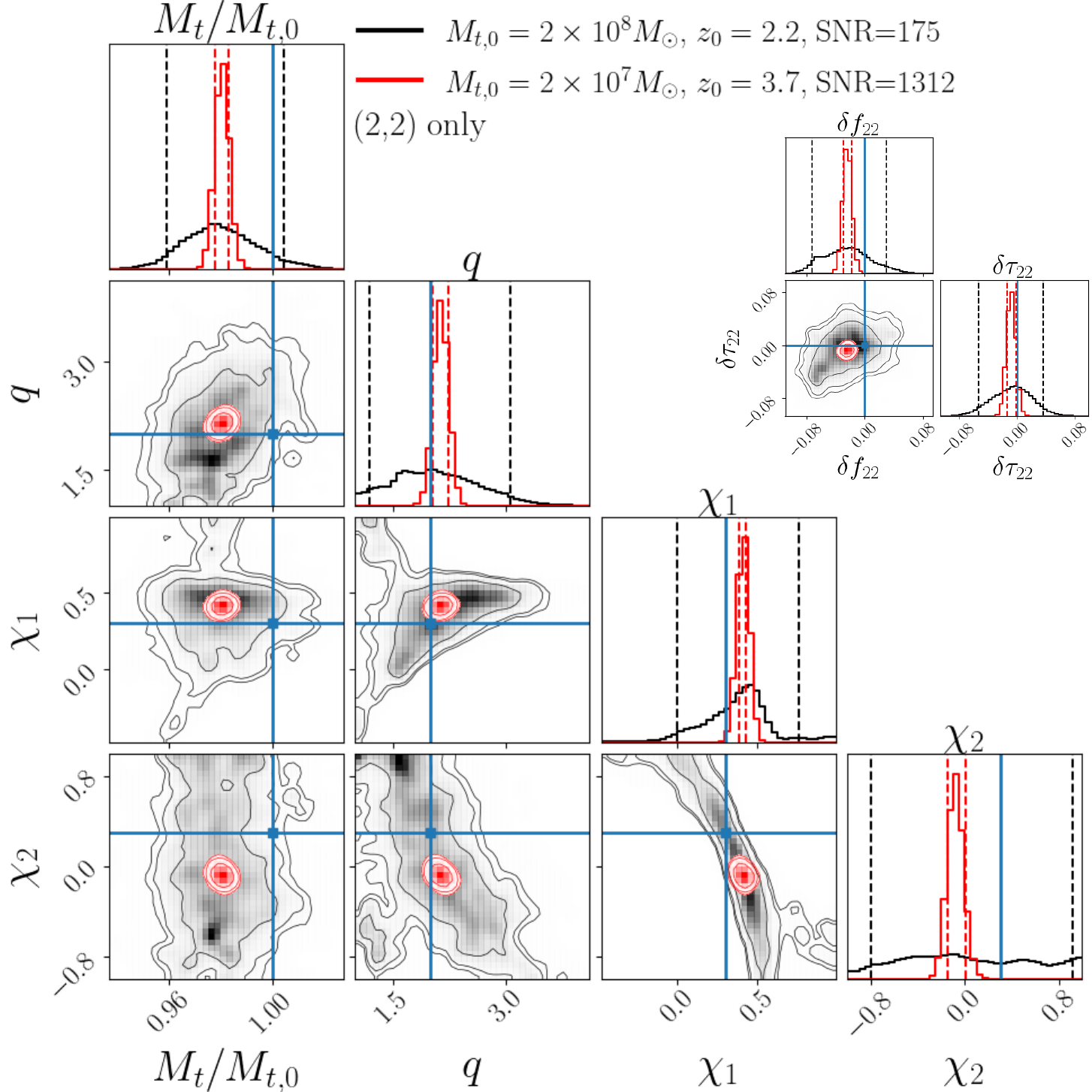}\\
 \centering
 \caption{Corner plot of intrinsic and deviation parameters for NR injections recovered with \texttt{pSEOBNRv5HM} templates. Contours show the $68\%$, $90\%$, and $95\%$ confidence intervals, and dashed lines the 0.05 and 0.95 quantiles. We compare the results for $M_{t,0}=2\times 10^8  \ M_{\odot}$, $z_0=2.2$ (black) and $M_{t,0}=2\times 10^7 M_{\odot}$, $z_0=3.7$ (red); both have $q_0=2$ and $\chi_{1,0}=\chi_{2,0}=0.3$. To ease comparison, we have rescaled the total mass to the injected value. In each case, both injection and templates contain only the $(2,2)$ harmonic. The heavier system has lower SNR, and the impact of mismodelling is small when including only the $(2,2)$ harmonic, so that the injected values (in blue) are well within the $90\%$ confidence regions. It is nonetheless relevant for the lighter system, due to its much higher SNR. In the latter case, one would be misled into thinking that the frequency of the $(2,2)$ QNM departs from the Kerr prediction. The spins are poorly measured for the heavy system due to the absence of higher harmonics. }\label{fig:corner_eob_nr_22}
\end{figure}

\begin{figure}
\centering
 \includegraphics[width=0.49\textwidth]{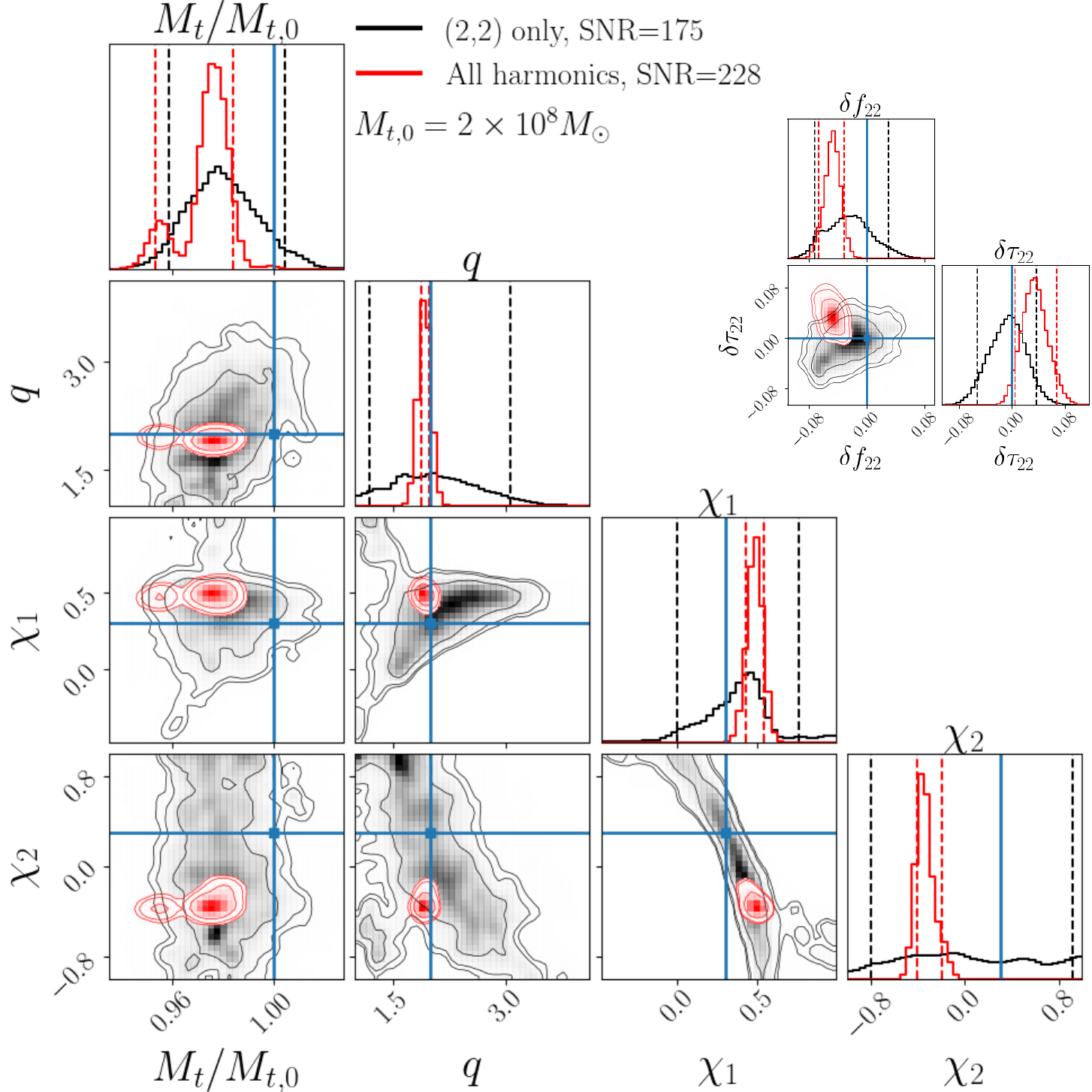}\\
 \centering
 \caption{The same as Fig.~\ref{fig:corner_eob_nr_22}, but now comparing results when including only the $(2,2)$ harmonic (black) versus when including all harmonics of \texttt{pSEOBNRv5HM} (red), for $M_{t,0}=2\times 10^8  \ M_{\odot}$ and $z_0=2.2$. The inclusion of higher harmonics worsens the match between NR and \texttt{pSEOBNRv5HM} waveforms, leading to significant biases in all parameters, in particular, the GR deviation ones. Note that, when including all harmonics, the deviation parameters for all QNMs are allowed to vary, and we find similar biases for all of them, but we show only the posterior for deviations in the $(2,2)$ harmonic because of space limitation.} \label{fig:corner_eob_nr_hm_22}
\end{figure}

In order to assess if the lack of accuracy in waveform modelling could spoil ringdown tests of GR, we generate synthetic injections with NR waveforms and recover them with \texttt{pSEOBNRv5HM} templates. We use the waveform SXS:BBH:2125 from the Simulating eXtreme Spacetimes Collaboration \cite{Boyle:2019kee} at the highest available resolution. It provides the signal of a BBH with mass ratio $2$ and aligned spins of magnitude $0.3$. First, we perform injections for the four total mass and redshift combinations detailed in Sec.~\ref{sec:astro}, allowing for deviations from GR in the QNMs when running our Bayesian analysis. Next, we investigate more methodically how systematic effects come into play and propose a novel method to assess their impact.

\subsection{Results on astrophysical systems}\label{sec:syst_astro}

First, we consider the case where the injected signal contains only the dominant (2,2) harmonic and include in the \texttt{pSEOBNRv5HM} templates the same harmonic content. We compare in Fig.~\ref{fig:corner_eob_nr_22} the posteriors obtained for injections with different total masses, giving   
SNR = 175 and 1312, respectively. The injected values (blue lines) are well within the $90\%$ confidence regions for the heavier system (black), having total mass $M_{t,0} = 2 \times 10^8  \ M_\odot$. However, due to its much higher SNR, the parameter estimation of the lighter system (red), having total mass $M_{t,0} = 2 \times 10^7  \ M_\odot$, is strongly biased. In particular, a deviation from GR in the frequency of the $(2,2)$ QNM is erroneously detected with high confidence. We recall that our model does not explicitly include overtones beyond $n=0$, as discussed in Sec. \ref{sec:wvf}. Including them in the merger-ringdown signal could, in principle, reduce this systematic bias, but would require fitting for the amplitude, phase and starting time of each overtone, which has proven very difficult so far \cite{Baibhav:2023clw}, and could end up having the opposite effect on the accuracy of the waveform. Next, we inject NR signals containing all the harmonics included in \texttt{pSEOBNRv5HM} and use templates with full harmonic content in the Bayesian analysis. As can be seen in Fig.~\ref{fig:corner_eob_nr_hm_22}, once we include higher harmonics (red), even for the more massive system the posterior shifts further from the true values while getting narrower, making it incompatible with the true parameters at more than $95\%$ confidence, in particular, for the GR deviation ones. We stress that the worsening for the system shown in Fig.~\ref{fig:corner_eob_nr_hm_22} is less due to the moderate increase in SNR than to the inclusion of higher harmonics, as we demonstrate in Sec.~\ref{sec:syst_meth}. This is not surprising since the accuracy of the \texttt{SEOBNRv5} model degrades when including higher harmonics, 
as comprehensive comparisons to $\sim 440$ NR waveforms and NR surrogate models have shown~\cite{Pompili:2023tna}.
One of the difficulties lies in the relative alignment between harmonics. When considering harmonics individually, we have freedom in the alignment (in phase and time) of the waveforms. When including several harmonics, we have a single phase shift and a single time shift that can be varied for all harmonics simultaneously. The relative alignments between them are fixed, and might not agree with the ones of NR waveforms. In the next section, we further illustrate how such tests of GR become less reliable when including higher harmonics.

\subsection{Exploring systematic effects}\label{sec:syst_meth}

\begin{table*}
 \begin{center}
   \begin{tabular}{c *{7}{c}}
   \cline{3-8}

   &  & \multicolumn{2}{c|}{$M_{t,0}=2\times 10^6 M_{\odot}$} & \multicolumn{2}{c|}{$M_{t,0}=2\times 10^7 M_{\odot}$} & \multicolumn{2}{c}{$M_{t,0}=2\times 10^8 M_{\odot}$} \\

\cline{2-8}

 & \multicolumn{1}{c|}{IMR} & \multicolumn{1}{c|}{Merger-Ringdown} & \multicolumn{1}{c|}{Inspiral}  &  \multicolumn{1}{c|}{Merger-Ringdown} & \multicolumn{1}{c|}{Inspiral} &  \multicolumn{1}{c|}{Merger-Ringdown} & \multicolumn{1}{c}{Inspiral}  \\

\hline

  \multicolumn{1}{c}{(2,2) only} &  \multicolumn{1}{c}{175} & 
  \multicolumn{1}{c}{102} & \multicolumn{1}{c}{101} & \multicolumn{1}{c}{161} & \multicolumn{1}{c}{68} &\multicolumn{1}{c}{168} & \multicolumn{1}{c}{49} \\
  
  \hline

  \multicolumn{1}{c}{All harmonics} & \multicolumn{1}{c}{175} &
  \multicolumn{1}{c}{106} & \multicolumn{1}{c}{140} & \multicolumn{1}{c}{165} &  \multicolumn{1}{c}{58} & \multicolumn{1}{c}{171} & \multicolumn{1}{c}{35} \\

  \hline

   \end{tabular}
\end{center}
 \caption{Full IMR SNR and its decomposition into the contribution of the merger-ringdown and inspiral stages for the NR waveform SXS:BBH:2125 \cite{Boyle:2019kee}, assuming a total SNR of 175. We show results both for when only the $(2,2)$ mode is considered and when all harmonics modelled in \texttt{SEOBNRv5HM} are included. }\label{tab:snrs_syst}
\end{table*}

Our goal here is to understand how systematic effects come into play, and, in particular, how do they depend on which portion of the signal  (inspiral or merger-ringdown) dominates. 
In addition to total masses of $2\times 10^7 \  M_{\odot}$ and $2\times 10^8  \ M_{\odot}$, we also consider $M_t=2\times10^6 \  M_{\odot} $. The minimum frequency used to analyse the signal remains $f_{\rm min}=2\times 10^{-4} [M_{t,0}/(2\times 10^7 M_\odot)]  \ {\rm Hz} $. This is a pessimistic choice for signals with $M_t=2\times10^6  \ M_{\odot} $, as they can accumulate significant SNR in the early inspiral, but with this choice, the length of the signal that is analysed (in geometric units) is kept fixed for different total masses. Moreover, we fix the SNR of the sources rather than their distance, to allow for a fair comparison between systems. Unless specified, we take the SNR to be 175 [as for the system with $M_t=2\times10^8  \ M_{\odot} $ at $z_0=2.2$ when including only the $(2,2)$ harmonic shown in Sec.~\ref{sec:syst}]. We stress that the long-wavelength approximation for the LISA response is expected to break for $M_t=2\times10^6  \ M_{\odot}$. Overall, the signals considered in this section might not be astrophysically realistic (e.g., we expect systems with $M_t=2\times10^7  \ M_{\odot}$ to have much larger SNR than 175), but we seek to have a methodic understanding of how systematic effects appear for different signal morphologies. We start by considering injections with the $(2,2)$ harmonic only, and then with all other harmonics included in \texttt{SEOBNRv5HM}. In both cases, the injected signal and templates have the same harmonic content. In Table \ref{tab:snrs_syst}, we give the merger-ringdown and inspiral SNR of the systems we consider, assuming an IMR SNR of 175, for both harmonic content options.

\subsubsection{$(2,2)$ only}\label{sec:syst_22}

\begin{figure}
\centering
 \includegraphics[width=0.49\textwidth]{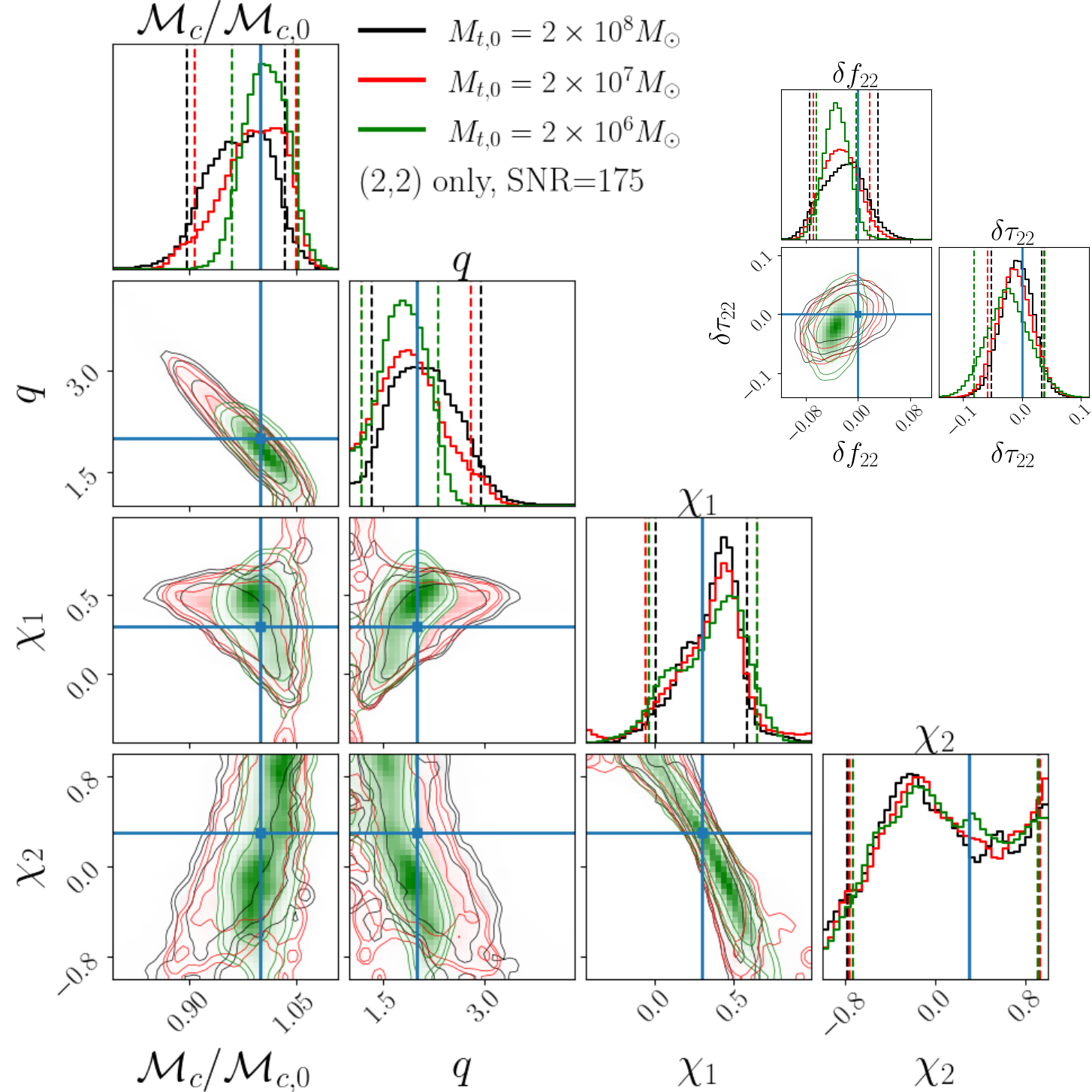}\\
 \centering
 \caption{Corner plot for the parameters estimated using \texttt{pSEOBNRv5HM} to analyse a mock NR injection for different total masses, considering only the $(2,2)$ harmonic and fixing the SNR of the injection to be 175. Contours show the $68\%$, $90\%$, and $95\%$ confidence intervals, and dashed lines the 0.05 and 0.95 quantiles. In the upper-right part, we show the posterior on the fractional deviations to GR QNMs.}\label{fig:corner_comp_om_124}
\end{figure}

\begin{figure}
 \centering
\subfigure[Inspiral.]{
    \centering \includegraphics[scale=0.18]{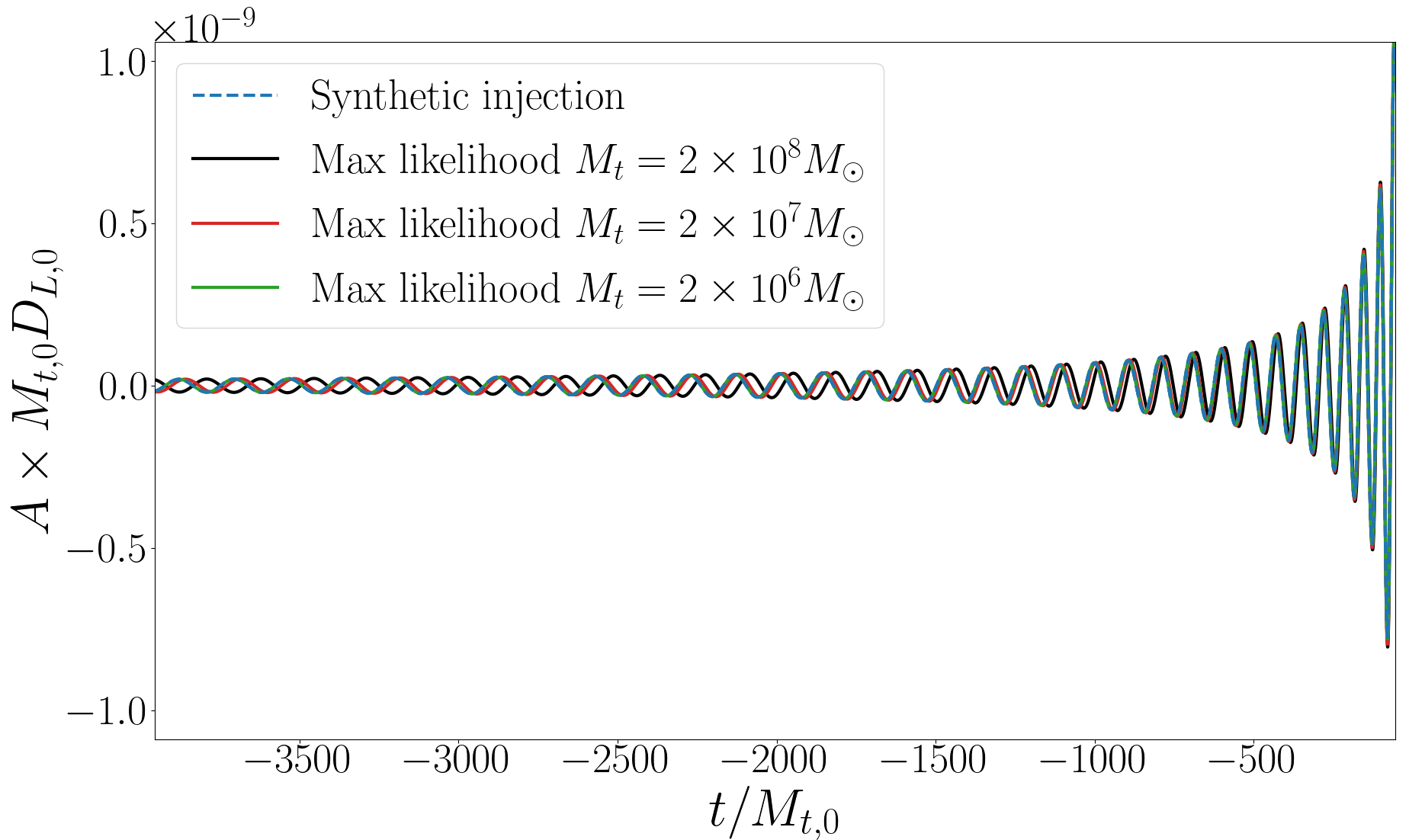}
    }
\centering
\subfigure[Merger-ringdown.]{
    \centering \includegraphics[scale=0.18]{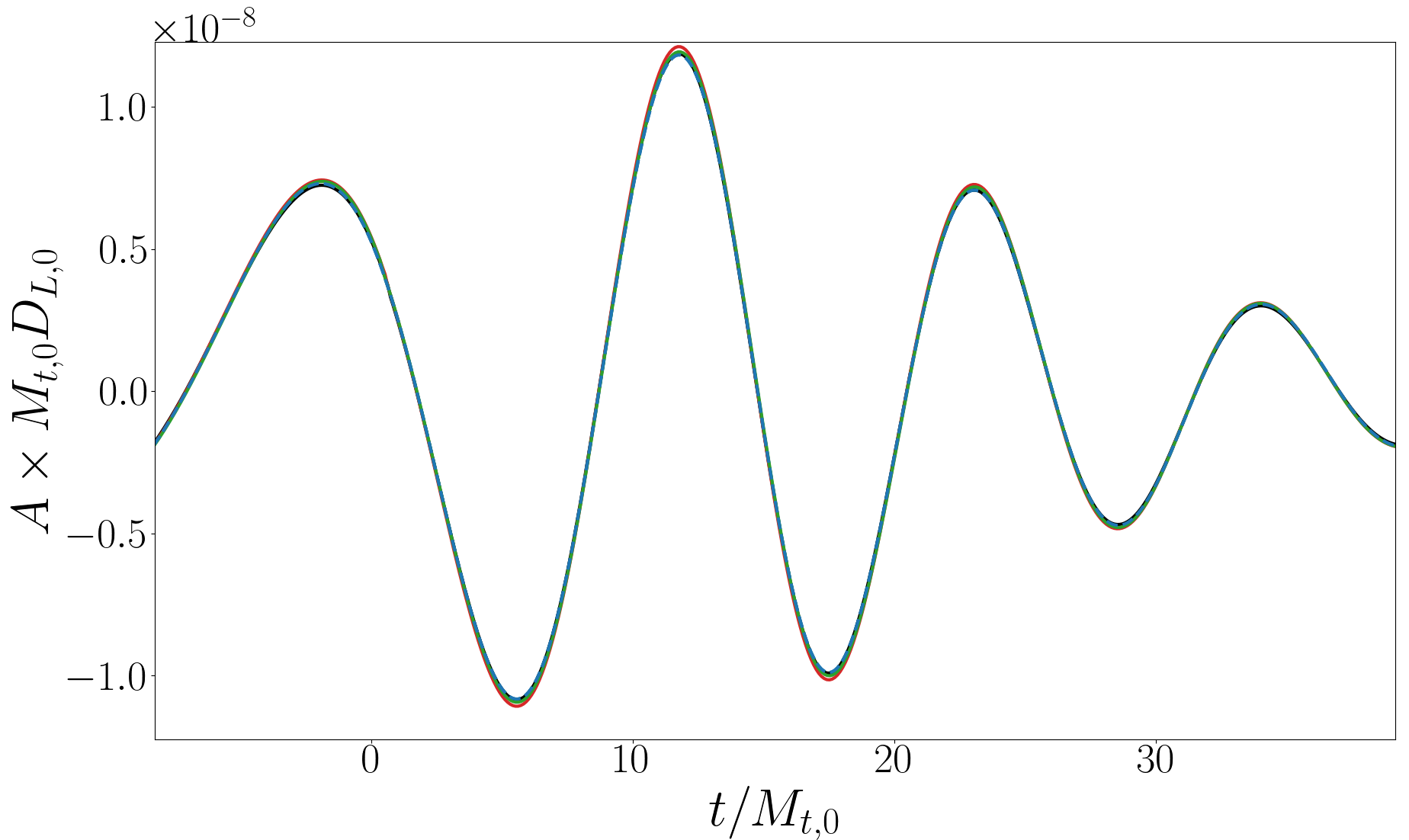}
   }
   \centering
   \caption{Comparison between the waveforms of the maximum likelihood points (from the analyses shown in Fig.~\ref{fig:corner_comp_om_124}) and the injection waveform. We plot the adimensionalised TDI strain $A$ as a function of time in geometric units to allow for the comparison between different total masses.}\label{fig:comp_wvfs_om}
 \end{figure}

In Fig.~\ref{fig:corner_comp_om_124}, we compare the posterior distributions obtained for different total masses. Note that, unlike in previous plots, we show the chirp mass, defined as $\mathcal{M}_c=(m_1^3m_2^3/(m_1+m_2))^{1/5}$, rather than the total mass. This is the mass combination that is best measured during the inspiral. We find that, as the total mass decreases, the chirp mass and the mass ratio are better measured and in better agreement with the true value. This is because the fraction of the SNR in the inspiral is larger for lighter systems, and this is the portion of the signal where our templates are in better agreement with NR waveforms. In Fig.~\ref{fig:comp_wvfs_om}, we compare the adimensionalised TDI strain $A$ for the highest likelihood point in each case. The upper panel focuses on the inspiral, and we can see that the agreement is better for lighter mass systems. The lower panel focuses on the merger-ringdown portion [we recall that $t/M_t=0$ is the peak of the $(2,2)$ amplitude], and we can see that the system with $M_t=2\times10^8  \ M_{\odot}$ is the one in the best agreement with the injection, as expected from the fact that it is the system for which the fraction of SNR in the merger-ringdown is the largest.

\begin{figure}
\centering
 \includegraphics[width=0.49\textwidth]{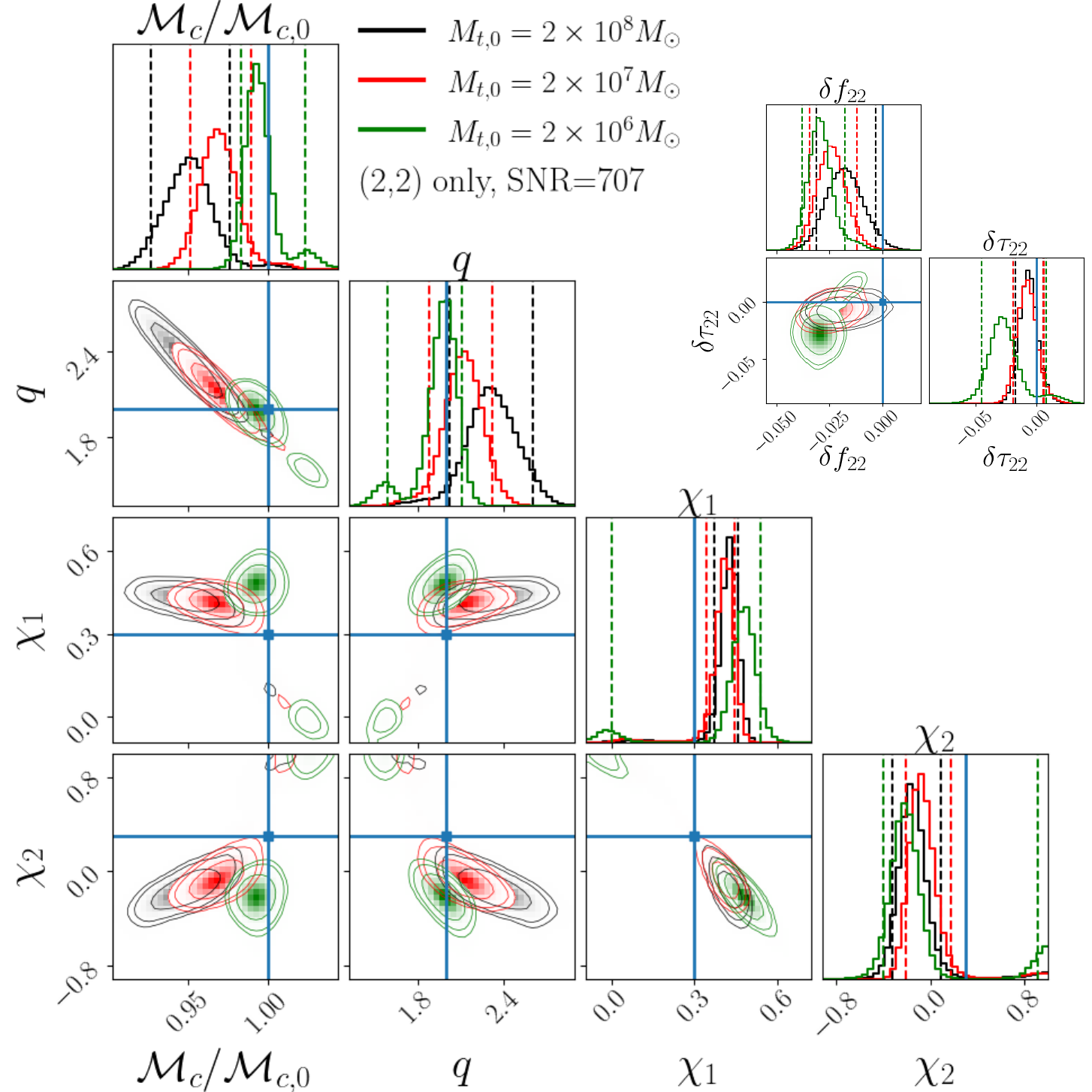}\\
 \centering
 \caption{The same as Fig.~\ref{fig:corner_comp_om_124} for an IMR SNR of 707. }\label{fig:corner_comp_om_500}
\end{figure}

Looking at the caption in Fig.~\ref{fig:corner_comp_om_124}, it might appear surprising that the system with $M_t=2\times10^6 \ M_{\odot}$ is the one for which the GR deviation parameters are the largest (while still being very much consistent with 0). The reason behind it is that the determination of intrinsic parameters for this system really comes from the inspiral, and the GR deviation parameters are then estimated to maximise the match in the merger-ringdown portion of the signal. In contrast, the intrinsic parameters of the $M_t=2\times10^8  \ M_{\odot}$ system are chosen to maximise the match in the merger-ringdown. This becomes clearer if we increase the SNR, as in Fig.~\ref{fig:corner_comp_om_500}, where we take the IMR SNR to be 707. The posteriors become thinner and the intrinsic parameters in the $M_t=2\times10^8  \ M_{\odot}$ case are no longer compatible with the true value. Moreover, for $M_t=2\times10^6  \ M_{\odot}$, we definitely favour nonzero deviations from GR. The fact that this is needed to maximise the match, even for an SNR of 175, is likely a consequence of the model used for the merger-ringdown [see Eq.~(\ref{RD}) and below]. When changing the QNMs, the amplitude and phase coefficients in Eqs.~(\ref{c1}) and (\ref{c2}) change consistently, possibly providing a better match, in particular, around the merger, where we use a purely phenomenological description, and discrepancies between our templates and NR waveforms are larger. Moreover, our model might not be accounting for higher overtones accurately enough. Therefore, favouring a nonzero value for the GR deviation parameters does not necessarily reflect that the physical QNMs are different from the Kerr ones, but rather that, given the functional form assumed, such QNM values provide overall a better fit to data. It would be interesting to explore if we still find such deviations from GR when allowing for the additional modifications around the merger proposed in \cite{Maggio:2022hre}.

\begin{figure}
\centering
 \includegraphics[width=0.49\textwidth]{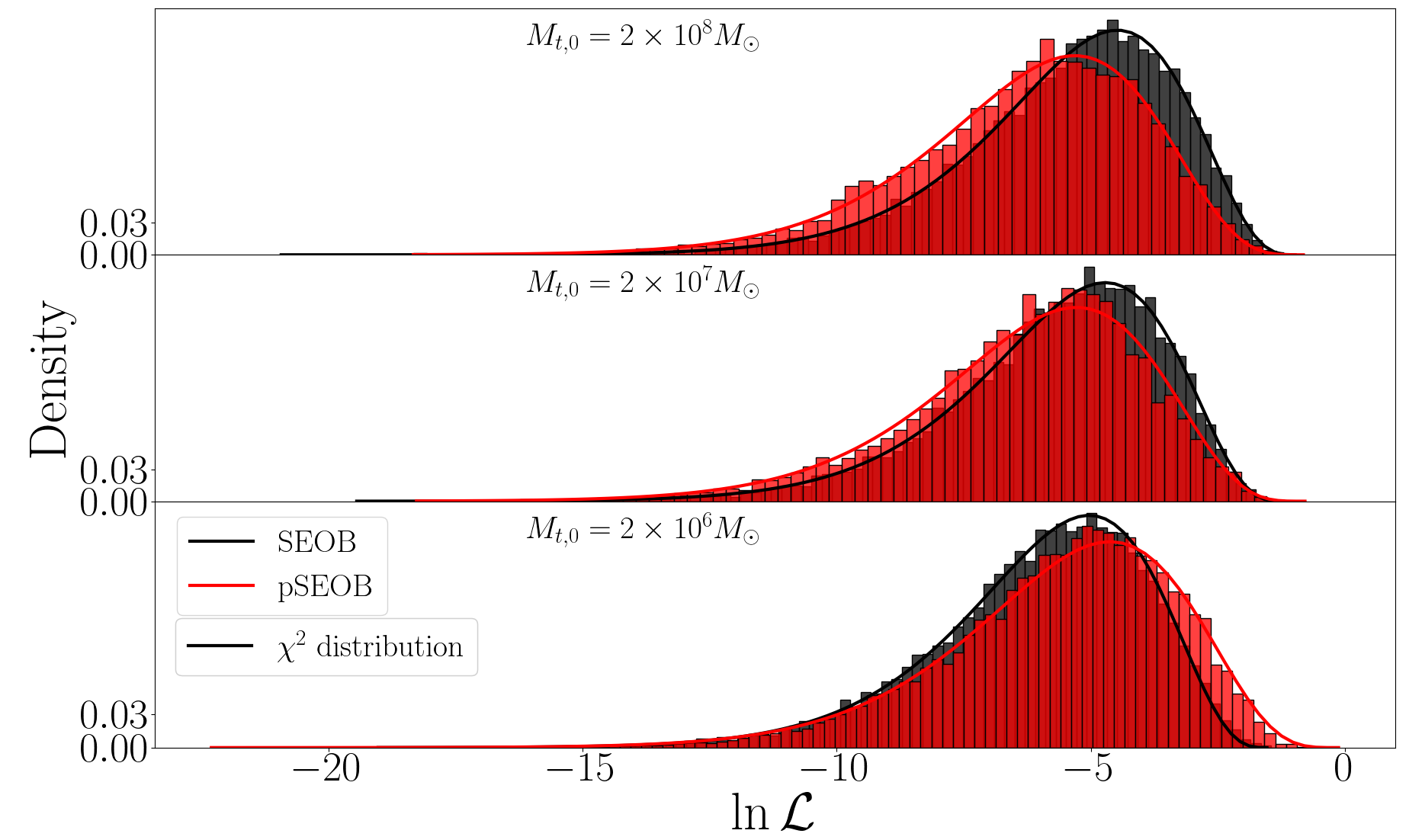}\\
 \centering
 \caption{Distribution of log-likelihood values when analysing with \texttt{SEOBNRv5HM} and \texttt{pSEOBNRv5HM} for different total masses.}\label{fig:logl_comp_nonmodgr}
\end{figure}

Finally, in Fig.~\ref{fig:logl_comp_nonmodgr}, we show the distribution of log-likelihood values obtained from running parameter estimation allowing or not for deviations from GR (\texttt{pSEOBNRv5HM} versus \texttt{SEOBNRv5HM}) for the three values of the total mass (fixing the SNR to 175). We find that, as the total mass decreases, the likelihood values increase, and allowing for deviations from GR improves more the fit, in agreement with Fig.~\ref{fig:corner_comp_om_124}. We can use the results shown in this plot to estimate the SNR from which it will be favoured to allow for deviations from GR. For this purpose, we introduce the Akaike information criterion (AIC) \cite{1100705}, defined as
\begin{equation}
    {\rm AIC}=2n_p-2\ln\hat{\mathcal{L}}, \label{eq:aic}
\end{equation}
where $\hat{\mathcal{L}}$ is the maximum likelihood and $n_p$ is the number of free parameters. The latter accounts for the dimensionality penalty. When choosing between different models to describe observed data, the one with minimum AIC is favoured. Moreover, we can estimate the log-Bayes factor between model 1 and model 2 as
\begin{equation}
    \ln \mathcal{B}=-\frac{1}{2}({\rm AIC}_1-{\rm AIC}_2). \label{eq:log_bayes}
\end{equation}
In \cite{Toubiana:2024car}, we justify the use of the expression above in the case of Gaussian posteriors and discuss how it relates to the standard indistinguishability criterion \cite{Flanagan:1997kp, Lindblom:2008cm, McWilliams:2010eq, Chatziioannou:2017tdw, Purrer:2019jcp}.
Here, $n_p$ is 9 when using \texttt{SEOBNRv5HM} and 11 when using \texttt{pSEOBNRv5HM}. From the definition of the likelihood [Eq.~(\ref{eq:logl})], we can write the log-likelihood (up to an additional constant that depends only on the noise properties of the detector) as
%
\begin{align}
    \ln\mathcal{L}&= \sum_{c \in [A,E]} (d_c|s_c(\theta)) -\frac{1}{2} (d_c|d_c)-\frac{1}{2} (s_c(\theta)|s_c(\theta))  \\
    &= \sum_{c \in [A,E]} (d_c|d_c) \sqrt{\frac{(s_c(\theta)|s_c(\theta))}{(d_c|d_c)}} \left ( \frac{(d_c|s_c(\theta))}{\sqrt{(d_c|d_c) (s_c(\theta)|s_c(\theta))}} \right . \nonumber \\ 
    & \left . -\frac{1}{2} \sqrt{\frac{(d_c|d_c)}{(s_c(\theta)|s_c(\theta))}} -\frac{1}{2} \sqrt{\frac{(s_c(\theta)|s_c(\theta))}{(d_c|d_c)}} \right ).\label{eq:logl_exp}
\end{align}
For a zero-noise injection, $d_c=s_{0,c}$, and $(d_c|d_c)={\rm SNR}_c^2$ is the SNR of the signal in the TDI channel $c$. If we vary the loudness of the injection by changing the distance (as we do here) and rescale the template by the same factor, which corresponds to rescaling the distance, then $(d_c|d_c)$ is the only term that depends on the SNR. Thus, we get the simple scaling for the log-likelihood:
\begin{equation}
    \ln\mathcal{L}=\ln\mathcal{L}_{175}\left (\frac{{\rm SNR}}{175} \right)^2. \label{eq:scaling_logl}
\end{equation}
Note that this scaling does not rely on approximating the likelihood as a Gaussian on the event parameters. Combining Eqs.~\ref{eq:aic}, \ref{eq:log_bayes}, and \ref{eq:scaling_logl}, we find that the log-Bayes factor for \texttt{pSEOBNRv5HM} versus \texttt{SEOBNRv5HM} is
\begin{align}
    \ln \mathcal{B}&=\left(\ln\hat{\mathcal{L}}_{ \texttt{pSEOBNRv5HM},175}-\ln\hat{\mathcal{L}}_{\texttt{SEOBNRv5HM},175} \right) \left (\frac{{\rm SNR}}{175} \right)^2 \nonumber \\ &- (n_{p,\texttt{pSEOBNRv5HM}}-n_{p,\texttt{SEOBNRv5HM}}).
\end{align}
In general, it can be difficult to accurately estimate $\ln \hat{\mathcal{L}}$ from the MCMC alone, because it lies in the higher-end tail of the log-likelihood distribution (see Fig.~\ref{fig:logl_comp_nonmodgr}). It becomes harder as the dimensionality of the parameter space increases. To get a better estimate of the maximum likelihood, we start from the Gaussian approximation. Under this hypothesis,
\begin{align}
    \ln \mathcal{L}&=\ln \hat{\mathcal{L}}-\frac{1}{2} \theta^tC^{-1}\theta \\
    &=\ln \hat{\mathcal{L}}-\frac{1}{2} \sum_i^{n_p} \frac{\theta^{\prime 2}_i}{\sigma^{\prime 2}_i}, \label{eq:gauss_logl}
\end{align}
where $\theta^{\prime}$ are the coordinates of $\theta$ in the basis of eigenvectors of the covariance matrix $C$ and $\sigma^{\prime 2}_i$ are the eigenvalues of $C$. 
From Eq.~(\ref{eq:gauss_logl}), we can see that $2(\ln \hat{\mathcal{L}}-\ln \mathcal{L})$ follows a $\chi^2$ distribution with $n_p$ degrees of freedom. Since the mean of a $\chi^2$ distribution with $n_p$ degrees of freedom is $n_p$, denoting the mean by $\braket{\cdot}$, we have
\begin{equation}
    \ln \hat{\mathcal{L}}=\braket{\ln \mathcal{L}}+\frac{n_p}{2}.\label{eq:lnlmax}
\end{equation}
The mean log-likelihood is less dependent on sampling the tails of the distribution than the maximum log-likelihood. Thus, we can use the log-likelihood samples we get from MCMC to estimate it and then use Eq.~(\ref{eq:lnlmax}) to estimate $\hat{\mathcal{L}}$. We recall that our prior on $\theta$ is flat, so the likelihood values obtained with MCMC are fair draws of the distribution followed by the log-likelihood. In Fig.~\ref{fig:logl_comp_nonmodgr}, we overplot in full lines the probability density function of the theoretical distribution of log-likelihood values assuming that $\ln \mathcal{L} \sim \ln \hat{\mathcal{L}}- \frac{1}{2}\chi^2(n_p) $, and $ \hat{\mathcal{L}}$ was estimated with the procedure described above. We can see that the agreement between this prediction and the distribution we obtain with MCMC is remarkable, even though the likelihood is not actually Gaussian in the event parameters. Moreover, we have verified that the integrated weight of the theoretical probability density function above the maximum likelihood we find with our MCMC is typically below $1/N_s$, where $N_s$ is the number of MCMC samples. This indicates that our sampling of the log-likelihood function is compatible with its theoretical estimate.  

Using this method to estimate $\hat{\mathcal{L}}$, we find that, for SNR=175, we have $\ln \mathcal{B}<0$ for all three total masses. This is in agreement with Fig.~\ref{fig:corner_comp_om_124}, where the posteriors are compatible with GR at least at $\sim 68\%$ confidence. 
We now estimate the SNR for which \texttt{pSEOBNRv5HM} would be definitely favoured with respect to \texttt{SEOBNRv5HM}. Following the Kass-Raftery scale \cite{doi:10.1080/01621459.1995.10476572}, we adopt the criterion $\ln \mathcal{B}>3$ to estimate that a model is favoured with respect to another. 
For $M_t=2\times 10^6 \ M_{\odot}$, we estimate ${\rm SNR}\gtrsim 330$, and for $M_t=2\times 10^7 \ M_{\odot}$ and $M_t=2\times 10^8 \ M_{\odot}$, ${\rm SNR}\gtrsim 598$ and ${\rm SNR}\gtrsim 977$, respectively. These values are in agreement with Fig.~\ref{fig:corner_comp_om_500}, where having a zero value for the GR modifications is supported for $M_t=2\times 10^8 \ M_{\odot}$, but it is not in the other cases.

\subsubsection{All harmonics}

\begin{figure}
\centering
 \includegraphics[width=0.49\textwidth]{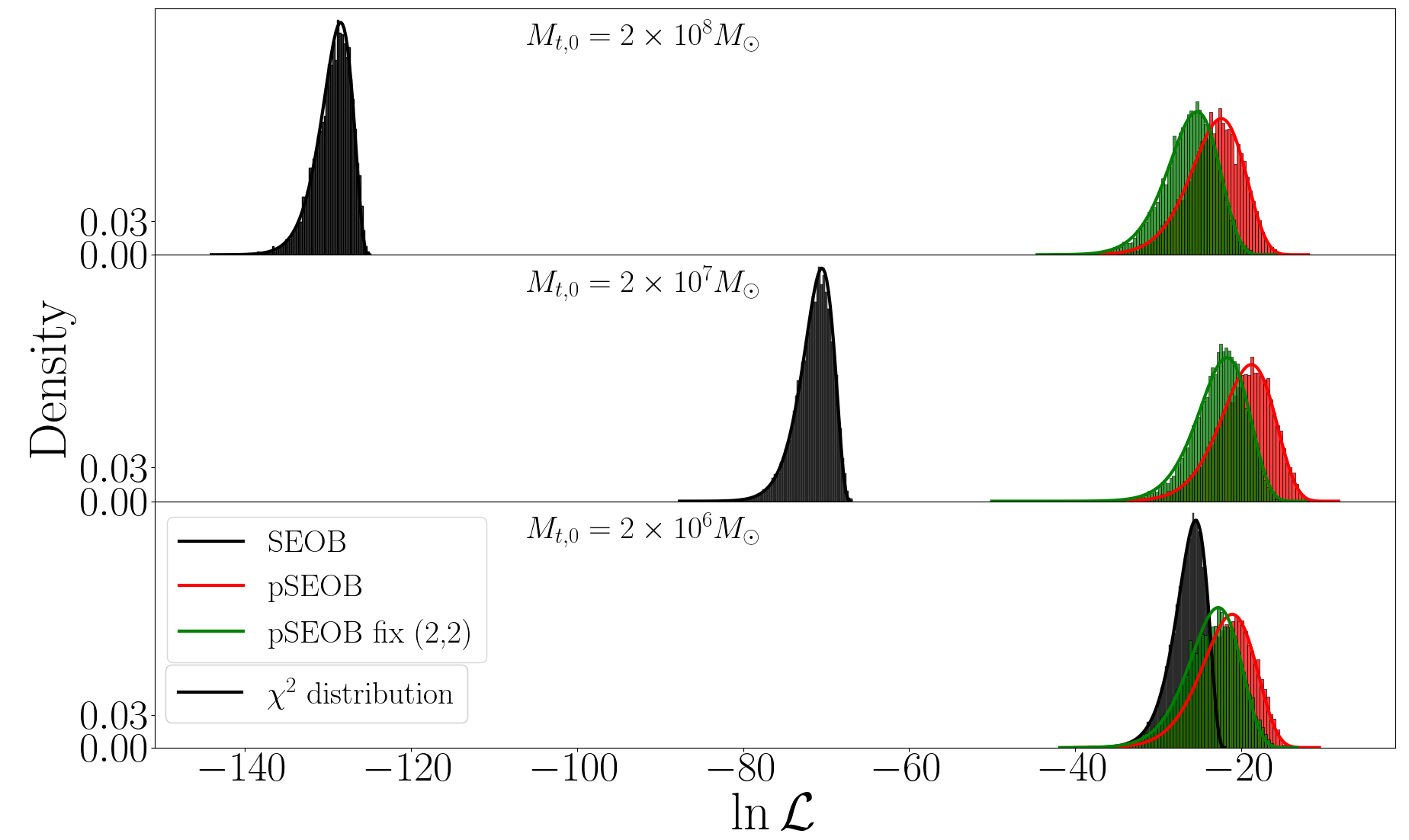}\\
 \centering
 \caption{ The same as Fig.~\ref{fig:logl_comp_nonmodgr}, now including all harmonics and considering an additional case where $\delta f_{22}$ and $\delta \tau_{22}$ are fixed to 0.}\label{fig:logl_comp_opts}
\end{figure}

In Fig.~\ref{fig:logl_comp_opts}, we show the equivalent of Fig.~\ref{fig:logl_comp_nonmodgr} when including all harmonics, fixing the total IMR SNR to 175. We consider one extra case: using for the template the \texttt{pSEOBNRv5HM} model with the $(2,2)$ QNM fixed to its Kerr value (i.e., $\delta f_{22}=\delta \tau_{22}=0$). The motivation for this is the following: The rationale for this test of GR is that the parameters of the binary are measured from the inspiral; from there, we can estimate the final mass and spin and, therefore, the QNM spectrum. We then seek to measure deviations from this spectrum. However, for heavier systems, most of the information comes from the merger-ringdown and the measurements are no longer "independent."  Thus, we want to investigate how the parameter estimation changes when not allowing for deviations in the dominant mode. 

First, we note that the likelihood values we get when including all harmonics are lower than when including only the $(2,2)$ mode, despite the significant increase in the number of free parameters in the \texttt{pSEOBNRv5HM} case (23 versus 11), showing that the agreement with the injected NR waveform worsens. Focusing on the \texttt{SEOBNRv5HM} case first (in black), we find that the hierarchy of likelihood values between the different total masses is compatible with the values reported in Table \ref{tab:snrs_syst}: The best fit is for the $M_t=2\times 10^6 \ M_{\odot}$ system, because most of its SNR comes from the inspiral, where our templates are more reliable. In the \texttt{pSEOBNRv5HM} case, we find that the quality of the fit improves significantly, with a larger improvement for more massive systems. Surprisingly, we find higher likelihood values for the $M_t=2\times 10^7 \ M_{\odot}$ system than for the $M_t=2\times 10^6 \ M_{\odot}$ one. This happens because, in the \texttt{pSEOBNRv5HM} case, the additional parameters allow a substantial improvement in the match in the merger-ringdown portion of the signal but not in the inspiral. To better understand this, we start from Eq.~(\ref{eq:logl_exp}), and make the assumption that $(d_c|d_c)=(s_{0,c}|s_{0,c}) \simeq (s_c(\hat{\theta})| s_c(\hat{\theta}))$, where $\hat{\theta}$ is the maximum likelihood point. In other words, we assume that the loudness of the recovered signal is virtually the same as the one of the injection. If we further assume that the contribution from each channel is roughly the same, we can write 
\begin{align}
  \ln\hat{\mathcal{L}}&= -2 {\rm SNR}^2 (1-{\rm FF}(s_0,h)), \label{eq:loglike_mismatch}
\end{align}
where ${\rm FF}$ is the fitting factor between the true signal and our templates defined as the maximised overlap:
\begin{align}
{\rm FF}(s_0,h)&={\rm max}_{\theta}  \ \mathcal{O}(s_0|h(\theta)) \label{eq:fitting_factor} \\
\mathcal{O}(h_1,h_2)&=\frac{\left( h_1 | h_2\right)}{\sqrt{\left( h_1 |h_1\right)\left( h_2 | h_2\right)}}.\label{eq:overlap}
\end{align}
%
In the previous equations, we can split between the contribution coming from the inspiral (I) and the one coming from the merger-ringdown (MRD): 
\begin{align}
  \ln\hat{\mathcal{L}}&= -2  \left(  \left ({\rm SNR}^2 (1-{\rm FF}) \right)_{\rm I}+\left ( {\rm SNR}^2(1-{\rm FF}) \right )_{\rm MRD} \right).
\end{align}
From Table \ref{tab:snrs_syst}, we see that the fraction of SNR coming from the merger-ringdown for the $M_t=2\times 10^7 \ M_{\odot}$ system is larger than the one coming from the inspiral for the $M_t=2\times 10^6 \ M_{\odot}$ one. Thus, if introducing the QNM deviations allows a sufficient improvement in the match in the merger-ringdown for the $M_t=2\times 10^7 \ M_{\odot}$ system, we can obtain higher likelihood values for it than for the $M_t=2\times 10^6 \ M_{\odot}$ system, unlike in the \texttt{SEOBNRv5HM} case. As in the case of the $(2,2)$ harmonic only, we can estimate the SNRs for which \texttt{pSEOBNRv5HM} is favoured with respect to \texttt{SEOBNRv5HM}, now including all harmonics. Note that now $n_p=23$ for \texttt{pSEOBNRv5HM}. We find ${\rm SNR} \gtrsim 68$, ${\rm SNR} \gtrsim 93$ and ${\rm SNR} \gtrsim 214$ for $M_t=2\times 10^8 \ M_{\odot}$, $M_t=2\times 10^7 \ M_{\odot}$ and $M_t=2\times 10^6 \ M_{\odot}$, respectively. We stress that these values are below the typical SNRs we expect for MBHBs, in particular, for sources with $M_t=2\times 10^6 \ M_{\odot}$ and $M_t=2\times 10^7 \ M_{\odot}$. Systems with $M_t=2\times 10^8  \ M_{\odot}$ could also have SNR above the respective limit, as is the case for the system shown in Fig.~\ref{fig:corner_eob_nr_hm_22}, for which GR is indeed excluded at more than $95\%$ confidence. Moreover, our estimates are in agreement with the corner plots shown in Figs.~\ref{fig:corner_comp_opts_om8}, \ref{fig:corner_comp_opts_om7} and \ref{fig:corner_comp_opts_om6} in Appendix ~\ref{app:pe_nr}, where we show the posterior for the different total masses in all three scenarios. For an IMR SNR of 175, only the lighter system is compatible with GR in the \texttt{pSEOBNRv5HM} case. We note that, in Ref.~\cite{Ghosh:2021mrv}, the authors performed a similar study for LIGO detectors at design sensitivity and found the GR value of QNMs to be well within the $90\%$ confidence interval even when injecting an NR waveform with an SNR of 75. 
This result is in qualitative agreement with our estimate of ${\rm SNR} \gtrsim 214$ for \texttt{pSEOBNRv5HM} to be favoured with respect to \texttt{SEOBNRv5HM} for the $M_t=2\times 10^6  \ M_{\odot}$ system, which is the closest one among our systems (in terms of relative contribution of the inspiral and merger-ringdown) to the system considered in Ref.~\cite{Ghosh:2021mrv}.    

From Figs.~\ref{fig:corner_comp_opts_om8}, \ref{fig:corner_comp_opts_om7} and \ref{fig:corner_comp_opts_om6}, it is possible to see that the posteriors in the \texttt{pSEOBNRv5HM} with fixed $(2,2)$ QNM case are slices of the \texttt{pSEOBNRv5HM} posteriors on the $\delta f_{22}=\delta \tau_{22}=0$ hypersurfaces, as they should be. It is harder to see that the posteriors in the \texttt{SEOBNRv5HM} correspond indeed to slices of the \texttt{pSEOBNRv5HM} posteriors on the $\delta f_{\ell m}=\delta \tau_{\ell m}=0$ hypersurfaces, because this slice of the posterior yields much lower likelihood values than explored by the sampler in the \texttt{pSEOBNRv5HM} case, as can be seen from Fig.~\ref{fig:logl_comp_opts}, except in the $M_t=2\times 10^6  \ M_{\odot}$ case (Fig.~\ref{fig:corner_comp_opts_om6}).  
We find that the chirp mass and the mass ratio tend to be in ``better'' agreement with their true values in the \texttt{pSEOBNRv5HM} fixed $(2,2)$ case than in the full \texttt{pSEOBNRv5HM} one, but we have no conclusive evidence of this. The spins are usually wrongly measured, with only a secondary mode in the \texttt{SEOBNRv5HM} case for $M_t=2\times 10^6  \ M_{\odot}$ containing the true values. This suggests that the inclusion of spins in our waveforms is one of the bottlenecks for performing accurate parameter estimation. 

It might seem surprising that the likelihood values in the \texttt{pSEOBNRv5HM} with fixed $(2,2)$ case are only a little lower than in the full case, as illustrated in Fig.~\ref{fig:logl_comp_opts}, even more given that $(2,2)$ is the dominant mode. A likely explanation for this is as follows. Deviations in the QNMs can partially compensate for the misalignment in harmonics between NR and \texttt{pSEOBNRv5HM} waveforms, as they change the amplitude and phase of the different harmonics, and improve the match, reinforcing our claim in the discussion around Fig.~\ref{fig:corner_comp_om_124}: Measuring nonzero GR deviations does not necessarily mean that the physical QNMs are different from their predicted Kerr values.  However, the phase shift and the time shift that we fit when doing parameter estimation are defined with respect to the peak of the $(2,2)$ amplitude. Therefore, it is less crucial to allow for deviations in this harmonic. It would be interesting to investigate how the match worsens when fixing other QNMs, in particular, the next-to-dominant $(3,3)$, or when fixing several but not all of them. We leave this for future investigation.

These studies show that, because of the high SNRs that we will reach with LISA, the accuracy requirement for waveforms is much more stringent than for current ground-based detectors. Moreover, our results suggest that particular attention should be paid to the modelling of higher harmonics, as their inclusion increases biases, and of spins, as their measurement tends to be very biased. Let us stress that even the NR waveforms we currently use are not accurate enough for SNRs of thousands~\cite{Purrer:2019jcp,Pompili:2023tna} and would need to be improved by at least one order of magnitude. 

\section{Conclusion}\label{sec:ccl}

Gravitational-wave observations have provided us with brand-new opportunities to test GR. In particular, ringdown tests are one of the most promising possibilities to detect deviations from GR. In this work, we have assessed how a fully consistent modelling of the IMR signal will allow us to perform high-precision ringdown tests with MBHB observations by LISA. To do so, we have performed synthetic injections of astrophysically realistic systems and analysed them with templates in GR (\texttt{SEOBNRv5HM}) and with parameterised deviations from GR (\texttt{pSEOBNRv5HM}), using the newly released \texttt{SEOBNRv5HM} waveform family~\cite{Mihaylov:2023bkc,Khalil:2023kep,Pompili:2023tna,vandeMeent:2023ols,Ramos-Buades:2023ehm}. More specifically, the \texttt{pSEOBNRv5HM} templates allow for deviations to the QNMs (frequency and damping time) of all the harmonics included in the model. All our analyses have been done in a fully Bayesian framework. 

First, we have considered the case where we use the GR \texttt{SEOBNRv5HM} templates for both the synthetic injection and the Bayesian analysis. We find that having a consistent modelling of the whole signal allows us to measure the parameters of the binary accurately, even for signals with very little SNR in the inspiral (e.g., very massive MBHBs with total mass $\sim 10^8 \ M_{\odot}$). Source-frame masses can typically be measured within $10\%$ and even within $1\%$ for systems with total mass $\sim  \ 10^7 M_{\odot}$. Spins can be measured within 0.1 and down to 0.001 for the primary BH. Second, we have shown that deviations to the QNMs of the dominant harmonics [i.e., the $(2,2)$, $(3,3)$, $(4,4)$ and $(5,5)$ harmonics], can be constrained within $10\%$ and down to $1\%$ for systems with total mass of the order $10^7 \ M_{\odot}$. Those are also the magnitude of deviations that we could measure in a non-GR signal. Converting the measurement of fractional deviations into measurements of QNM frequencies, we find that for most systems we could accurately measure and distinguish the QNMs of several harmonics, up to all seven included in the \texttt{pSEOBNRv5HM} model in the most favourable cases (i.e., total mass $10^7  \ M_{\odot}$, mass ratio $\sim 2-4$, highly spinning, and at $z\sim 2$).

Then, we have assessed the impact of systematics on ringdown tests by using NR waveforms to perform synthetic injections and analysing them with \texttt{pSEOBNRv5HM}. In order to estimate at which SNRs we would expect to erroneously measure deviations from GR, we have developed a novel approach based on the Akaike information criterion
We have found that, in particular, when higher harmonics are included, parameter estimation is significantly biased already for SNRs of $\mathcal{O}(100)$, leading to the erroneous detection of deviations from GR in high SNR signals.
The results we have obtained when using \texttt{pSEOBNRv5HM} for the injection and the Bayesian analysis give us a sense of the incredibly high precision to which we will be able to perform ringdown tests with LISA. However, in order not to jeopardise those tests, the accuracy of our waveform models needs to be improved far beyond current standards, which is one of the major challenges facing the GW community over the next few years. 

In this work, we have focused on MBHB systems that are in the high-mass end of predictions for LISA observations, typically produced in astrophysical models where MBHs form from the evolution of heavy seeds \cite{Latif:2016qau}. This is because those are the ones for which we expect the highest SNR in the merger-ringdown~\cite{Bhagwat:2021kwv,Cotesta:2023}. However, it would be interesting to assess how observations of lighter systems, which might be more numerous, could be used to detect deviations from GR in the ringdown. Also, we have neglected the effect of spin precession and eccentricity, which might not be appropriate, in particular, if MBHBs harden through triplet interactions \cite{Bonetti:2018tpf}. Finally, the waveform model we used in this work allows for deviations from GR only in the ringdown, whereas, if deviations are present, we should expect them to affect the whole signal. Different theory-agnostic formalisms have been proposed to account for deviations in the inspiral~\cite{Yunes:2009ke,Agathos:2013upa,Mehta:2022pcn}, typically by modifying the post-Newtonian expansion of the GW phase \cite{Blanchet:2013haa}, and progress has recently been made to account for deviations in the plunge-merger stage~\cite{Maggio:2022hre}. It would be interesting to assess how to link modifications in different parts of the signal or at least to assess how the constraints change when accounting for all possible modifications. We leave these studies for future work.

After the first submission of this work we identified an error in our parameter estimation pipeline: we were missing a factor of $\frac{1}{2}$ inside the exponential entering the likelihood [Eq.~\ref{eq:logl}]. The SNR values cited in the original version of the paper were not affected by this error. Because we consider zero-noise injections, this can be interpreted as, when performing parameter estimation, either (i) placing the sources a factor $\sqrt{2}$ closer, or (ii) dividing the LISA PSD by 2, i.e. considering a level of noise $\sqrt{2}$ lower. In both cases the SNR of the systems for which we performed parameter estimation need to be multiplied by $\sqrt{2}$. We have decided to adopt the first possibility, and changed the redshifts of the fiducial sources we consider from 3 and 5 to 2.2 and 3.7, and applied the corresponding $\sqrt{2}$ factor to all SNR values quoted in the paper and shown in the figures. We stress that, given the uncertainty on the exact LISA noise, and that we use the SciRdv1 noise curve, which is considered to be a pessimistic estimate of the LISA noise, the second possibility would also have been a reasonable choice. This updated version matches the one published in Phys. Rev. D.

\section*{Acknowledgments}

It is our pleasure to thank Serguei Ossokine for providing assistance in the usage of publicly available LALSimulation \cite{lalsuite} codes and getting access to NR waveforms, as well as Danny Laghi and Vasco Gennari for comments on the Bayes' factor scale. We express our gratitude to the anonymous referee for their valuable comments on our manuscript.

The computational work for this manuscript was carried out on the \texttt{Hypatia} compute cluster at the Max Planck Institute for Gravitational Physics in Potsdam.

The waveform model \texttt{SEOBNRv5HM} is publicly available through the Python package \texttt{pySEOBNR} \href{https://git.ligo.org/waveforms/software/pyseobnr}{\texttt{git.ligo.org/waveforms/software/pyseobnr}}. Stable versions of \texttt{pySEOBNR} are published through the Python package Index (PyPI) and can be installed via \texttt{pip install pyseobnr}.

\appendix

\section{Settings for the \texttt{SEOBNR} waveforms}\label{app:res}

\begin{figure}[h!]
\centering
 \includegraphics[width=0.49\textwidth]{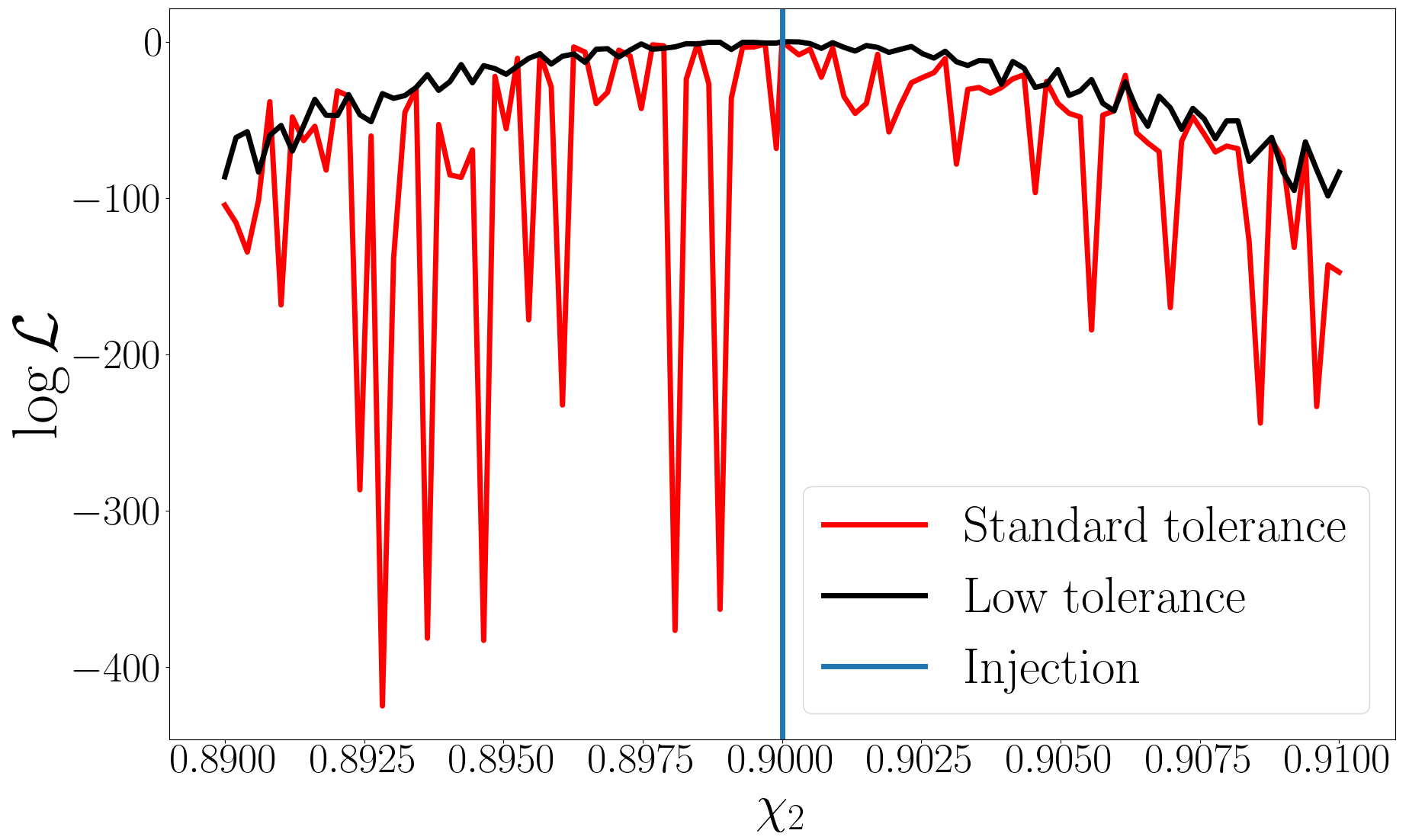}\\
 \centering
 \caption{Log-likelihood as a function of $\chi_2$ when using two different tolerances, when integrating the EOB equations of motion, for a system with $M_{t,0}=2\times  \ 10^8 M_{\odot}$, $\chi_{1,0}=\chi_{2,0}=0.9$, $q_0=2$, $z_0=2.2$, and SNR=623. When using waveforms computed with standard tolerance, the likelihood can be a very nonsmooth function of the binary parameters, inducing secondary maxima, and yielding very small values even close to the injected value. }\label{fig:plot_logl_res}
\end{figure}

\begin{figure}[h!]
\centering
 \includegraphics[width=0.5\textwidth]{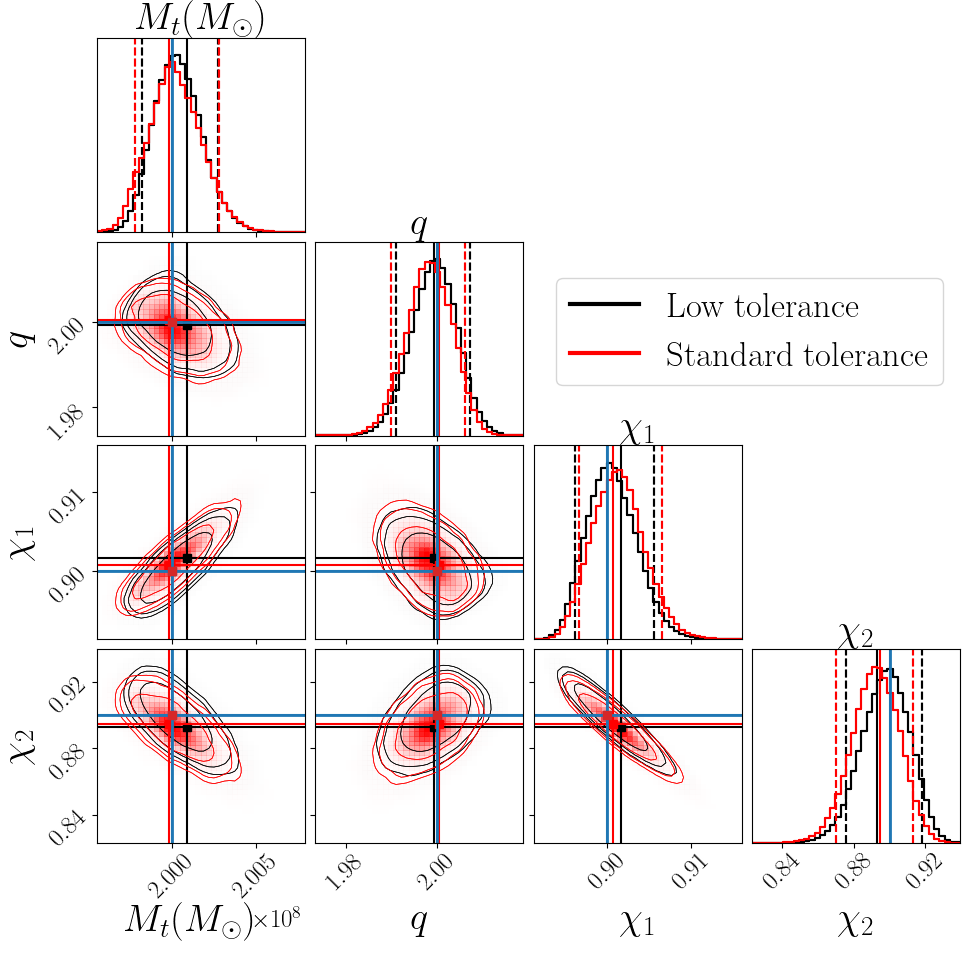}\\
 \centering
 \caption{Comparison between parameter estimation results when using low-tolerance waveforms (black) and standard ones (red) for the same system as in Fig.~\ref{fig:plot_logl_res}. Blue lines indicate the injection point; black and red symbols indicate the maximum-posterior point of the low-tolerance and standard-tolerance runs, respectively. The nonsmoothness of the likelihood function illustrated in Fig.~\ref{fig:plot_logl_res} induces secondary maxima over the parameter space. Those lead to very non-Gaussian distributions and induce an apparent bias when projecting onto one-dimensional posteriors. However, as indicated by the black and red lines, the maximum-posterior point is close to the synthetic injection, as expected when using flat priors. }\label{fig:corner_comp_res}
\end{figure}

\texttt{SEOBNR} waveforms are generated by numerically integrating the EOB equations of motion (e.g., see Ref.~\cite{Pompili:2023tna}). They are computed up to a given accuracy that depends on the tolerance used in the integration. As a consequence, at fixed tolerance, the waveform function is not smooth on the manifold of waveform parameters, and the inner product between waveforms [Eq.~(\ref{eq:inner_product})] is an oscillating function on that manifold. For low-SNR systems, these oscillations are negligible, since they correspond to very small changes in the likelihood between neighbouring points, but for high SNRs these oscillations become important. This is illustrated in Fig.~\ref{fig:plot_logl_res}, where we plot the log-likelihood as a function of $\chi_2$ for ``standard-tolerance'' and ``low-tolerance'' waveforms, keeping all the other parameters at their true value. When using the standard tolerance, the likelihood shows many local extrema and can reach very small values even close to the synthetic injection (indicated by the blue line). For comparison, the minimum log-likelihood in the parameter estimation runs done for this paper are typically $\sim -15$. The smoothness of the log-likelihood improves significantly when using low-tolerance waveforms. The counterpart of this improvement is a slowing down of the waveform computation. Our low-tolerance waveforms are $\sim 5$ times slower to compute than the ones with the standard \texttt{SEOBNRv5HM} configuration. 
We stress that the effect of these oscillations is exaggerated by looking at one slice of the parameter space (i.e., keeping all the other parameters fixed). Variations in other parameters compensate for these oscillations and make these local extrema less pronounced. However, as we show in Fig.~\ref{fig:corner_comp_res}, the existence of several local extrema makes the posterior non-Gaussian and the marginal one-dimensional distributions can peak away from the true value. This is similar to the effect discussed in Sec.~V.C in Ref.~\cite{Marsat:2020rtl} in the context of sky localisation with LISA. For the system shown in Fig.~\ref{fig:corner_comp_res}, this apparent bias disappears when using low-tolerance waveforms. Let us stress that this apparent bias is an effect of projecting a non-Gaussian posterior onto one-dimensional posteriors. As indicated by the black and red lines, in each case, the maximum-posterior points found by our sampler are close to the injection point, as expected when using flat priors. All the results presented in the main body of this paper were obtained using low-tolerance waveforms.

\begin{figure}[!h]
\centering
 \includegraphics[width=0.5\textwidth]{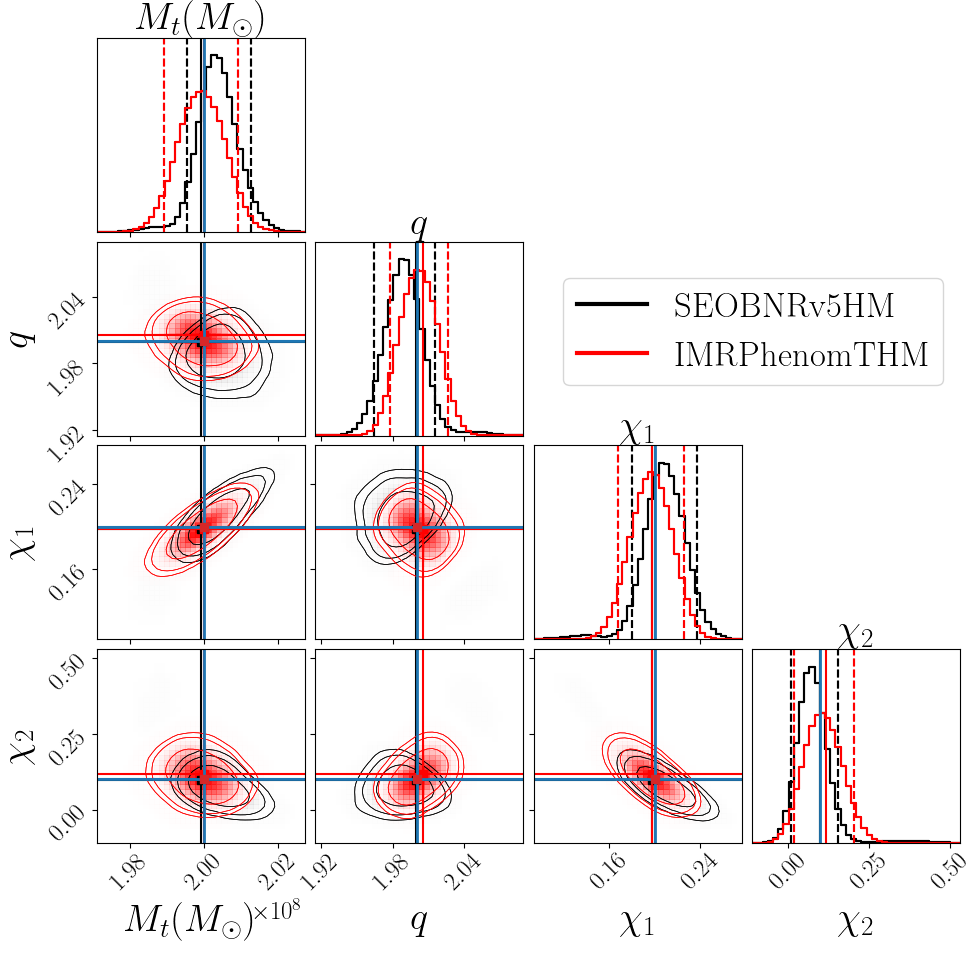}\\
 \centering
 \caption{Comparison of the parameter estimation results obtained using low-tolerance \texttt{SEOBNRv5HM} waveforms (black) and \texttt{IMRPhenomTHM} (red) waveforms for a system with $M_{t,0}=2\times 10^8 \ M_{\odot}$, $\chi_{1,0}=0.2$, $\chi_{2,0}=0.1$, $q_0=2$, $z_0=2.2$, and SNR=241. The latter shows no apparent bias when looking at one-dimensional marginalised posteriors, because the \texttt{IMRPhenomTHM} waveform is a smooth function across the parameter space, and so is the likelihood, and for high SNRs the posterior is fairly Gaussian.}\label{fig:comp_corner_eob_imr}
\end{figure}

For some systems, this apparent bias persists even when using low-tolerance waveforms, as shown in Fig.~\ref{fig:comp_corner_eob_imr}. In order to further validate the argument that this is caused by the nonsmoothness of the waveform across parameter space, we compare to the results obtained using the \texttt{IMRPhenomTHM} waveform model \cite{Estelles:2020twz}. This is an IMR time-domain approximant, built from a phenomenological approach, in the spirit of the frequency-domain approximants of the phenomenological family of templates \cite{Ajith:2007kx,Ajith:2009bn,Santamaria:2010yb,Husa:2015iqa,Khan:2015jqa,Garcia-Quiros:2020qpx}. It is based on post-Newtonian expressions \cite{Blanchet:2013haa} augmented by phenomenological terms fitted against NR simulations and also calibrated to \texttt{SEOBNR} waveforms (where NR data are not available). It includes the same harmonics as \texttt{pSEOBNRv5HM}, except for the subdominant $(3,2)$ and $(4,3)$ harmonics. For nonprecessing systems, the waveform is an analytic smooth function of the waveform parameters. Thus, the likelihood function is smooth across the parameter space, and we observe no apparent bias, even when looking at one-dimensional projections, as can be seen from Fig.~\ref{fig:comp_corner_eob_imr}. We stress that, after the waveform has been generated in the time domain, the steps to compute the likelihood and perform parameter estimation are exactly the same for the two models.

\section{Comparison of measurements to \texttt{IMRPhenomTHM}}\label{app:imrt}

\begin{figure*}[!htbp]
 \includegraphics[width=0.95\textwidth]{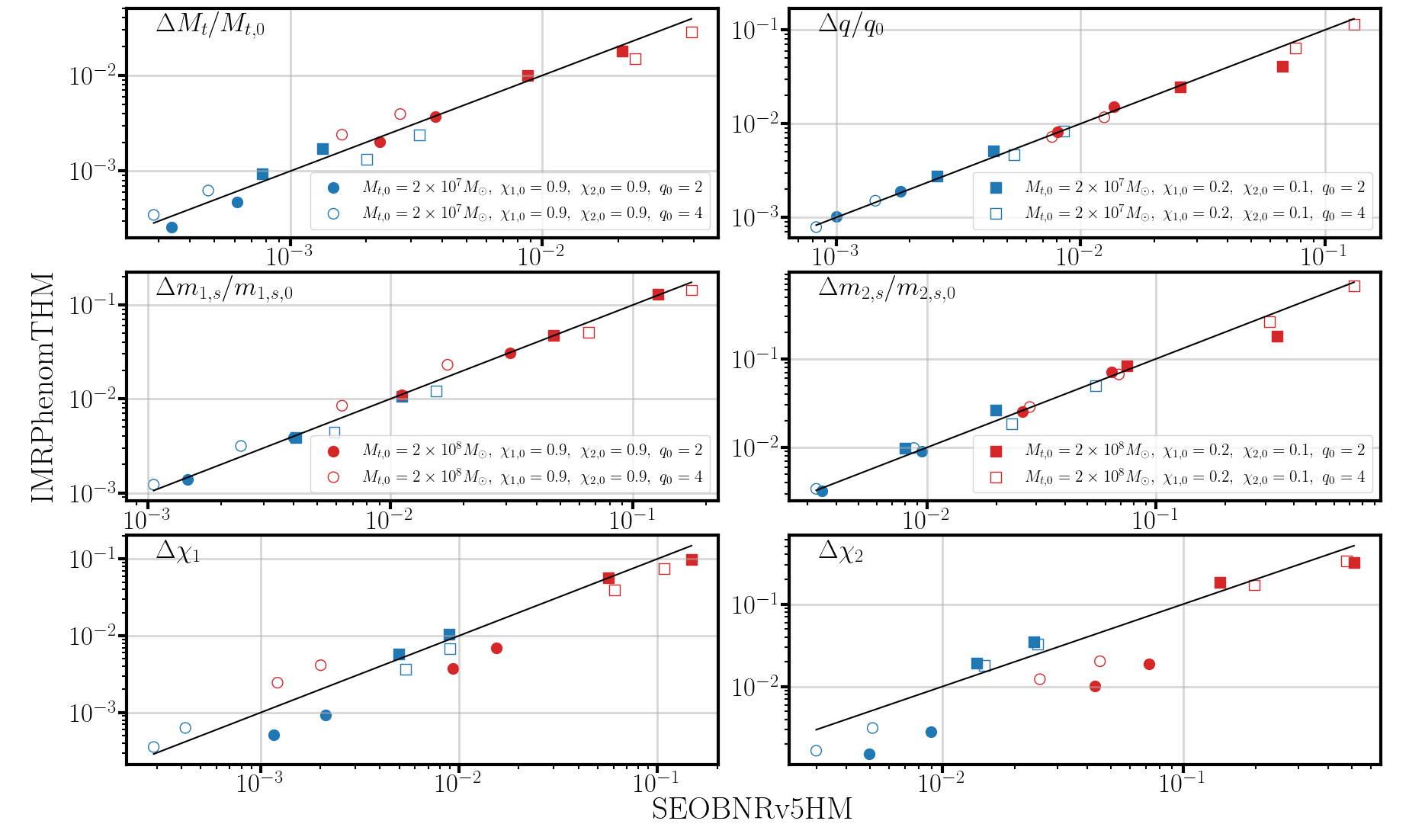}\\
 \caption{Comparison between the measurement errors with \texttt{SEOBNRv5HM} ($x$ axes) and with \texttt{IMRPhenomTHM} ($y$ axes). Black lines show $y=x$. Errors are similar, with a slightly higher discrepancy for the spins, in particular, for high-spin systems (circles). }\label{fig:comp_errors}
\end{figure*}

Here, we run Bayesian analyses for all the binary systems described in Sec.~\ref{sec:astro} using the \texttt{IMRPhenomTHM} waveform, for both the synthetic injection and parameter estimation. We restrict ourselves to the GR case. In Fig.~\ref{fig:comp_errors}, we show how the measurement errors on intrinsic parameters compare when using the \texttt{IMRPhenomTHM} and the \texttt{SEOBNRv5HM} waveforms. Black lines correspond to $y=x$. We see that the estimates are in good agreement, with a slight discrepancy for the error on spins, in particular, in the case of high-spin systems (circles). In this regime, our current waveforms are less accurate, so the agreement between them is worse (see also comparisons between these two waveform models in Ref.~\cite{Pompili:2023tna}).

\section{Errors on intrinsic parameters when allowing for deviations to GR}\label{app:gr_errors}

\begin{figure*}[!htbp]
\centering
 \includegraphics[width=0.95\textwidth]{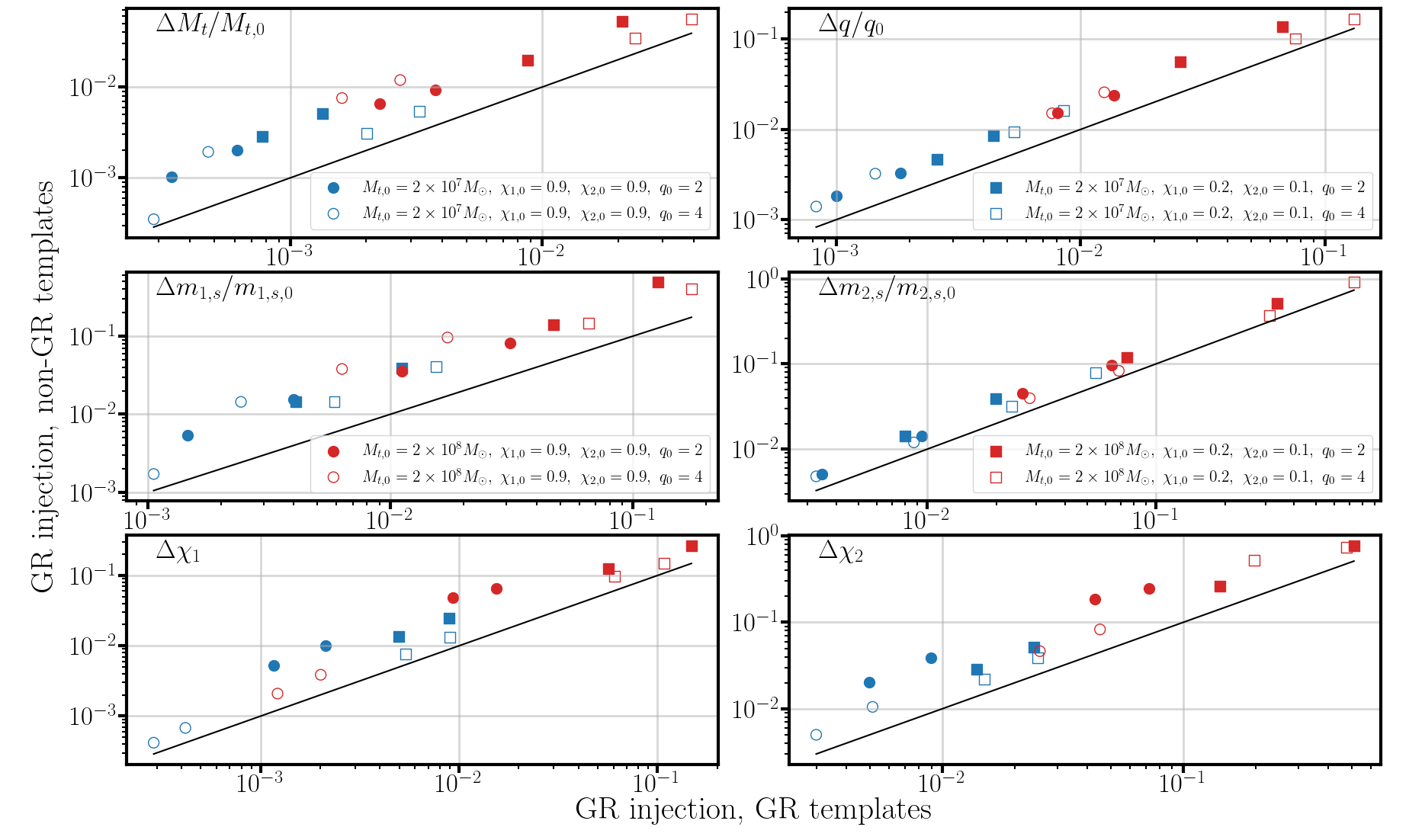}\\
 \centering
 \caption{Comparison of the measurement error on intrinsic parameters when setting deviation parameters to 0 ($x$ axes) and when letting them vary ($y$ axes) for a GR injection. Black lines show $y=x$. Measurements worsen when allowing for deviations from GR to vary but remain comparable to the GR case.}\label{fig:gr_errs_modgr7}
\end{figure*}

We show in Fig.~\ref{fig:gr_errs_modgr7} the error on intrinsic parameters in the case we inject a GR signal and allow for deviations from GR when performing the Bayesian analysis, complementing the results in Sec.~\ref{sec:gr_inj_non_gr_bayes}. Black lines indicate $y=x$. All measurements worsen due to the higher number of parameters but remain comparable to the pure GR case.

\section{Impact of systematics: corner plots}\label{app:pe_nr}

We show here the corner plots comparing the posteriors in the three scenarios considered in Sec.~\ref{sec:syst_meth} for the three choices of total mass.

\begin{figure*}[!htbp]
\centering
 \includegraphics[width=0.9\textwidth]{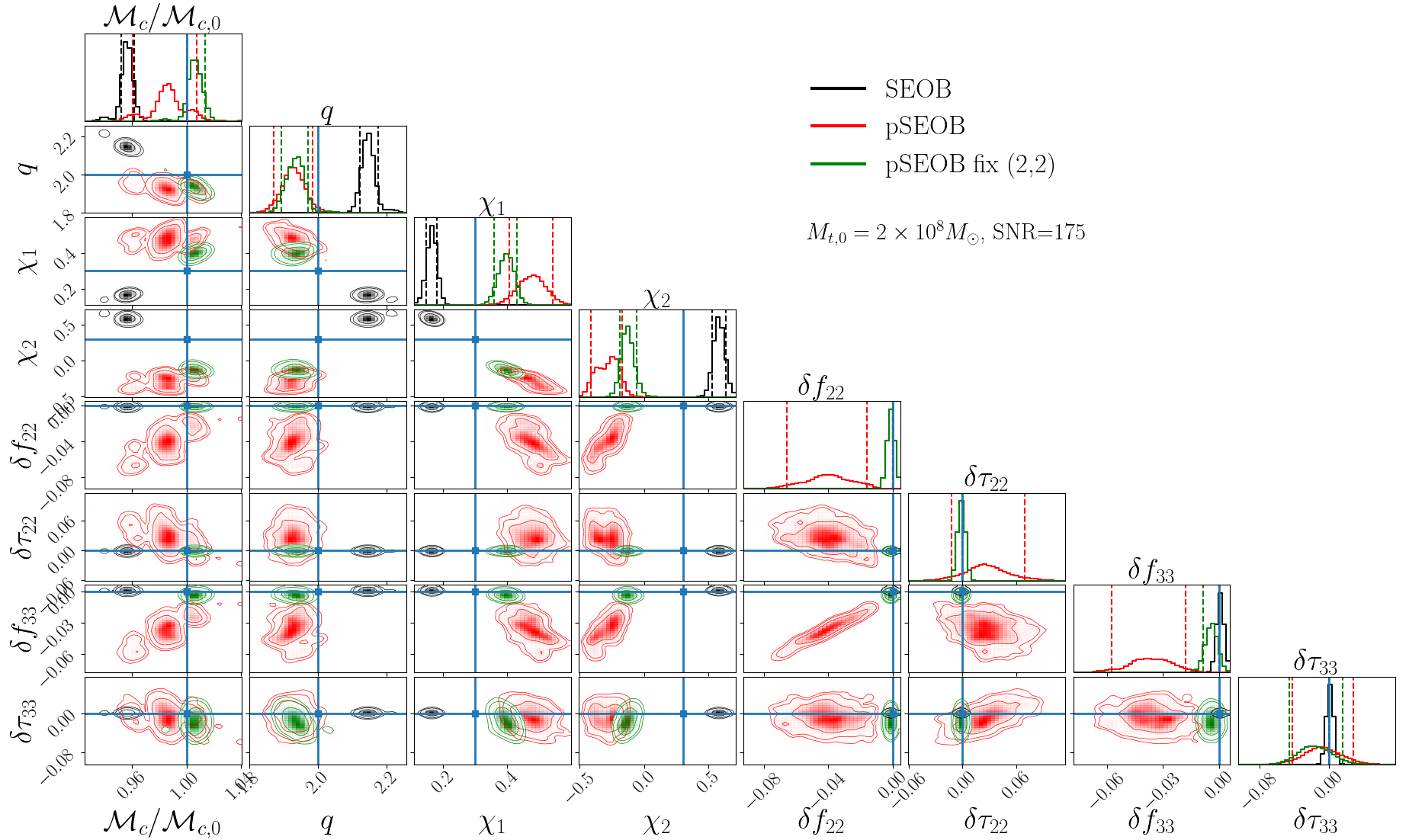}\\
 \centering
 \caption{Corner plot for the parameters estimated in three different scenarios for a total mass of $2\times 10^8  \ M_{\odot}$ including all harmonics of \texttt{SEOBNRv5HM} and fixing the SNR of the injection to be 175. Contours show the $68\%$, $90\%$, and $95\%$ confidence intervals, and dashed lines the 0.05 and 0.95 quantiles.}\label{fig:corner_comp_opts_om8}
\end{figure*}

\begin{figure*}[!htbp]
\centering
 \includegraphics[width=0.9\textwidth]{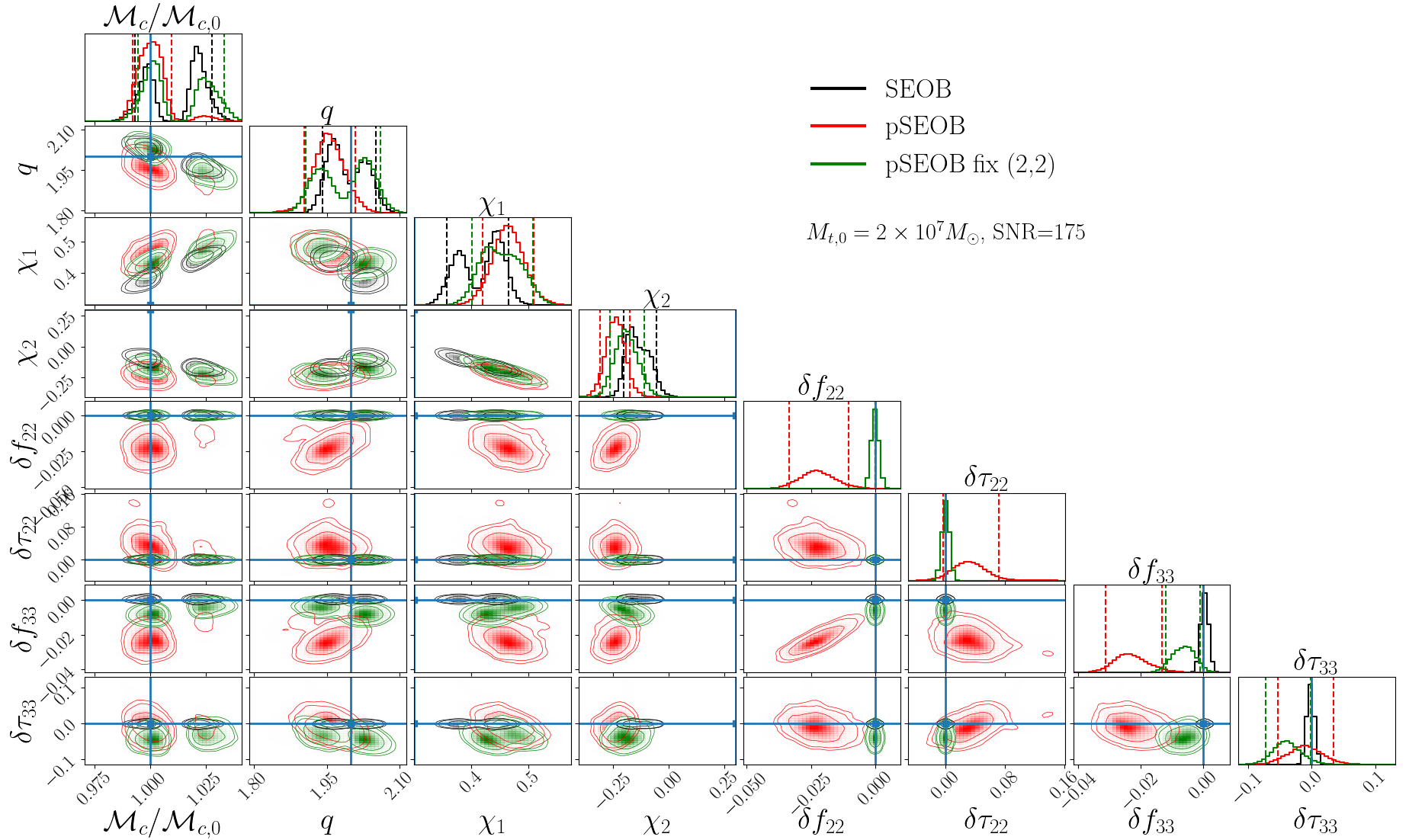}\\
 \centering
 \caption{The same as Fig.~\ref{fig:corner_comp_opts_om8} for $M_t=2\times 10^7 \ M_{\odot}$.}\label{fig:corner_comp_opts_om7}
\end{figure*}

\begin{figure*}[!htbp]
\centering
 \includegraphics[width=\textwidth]{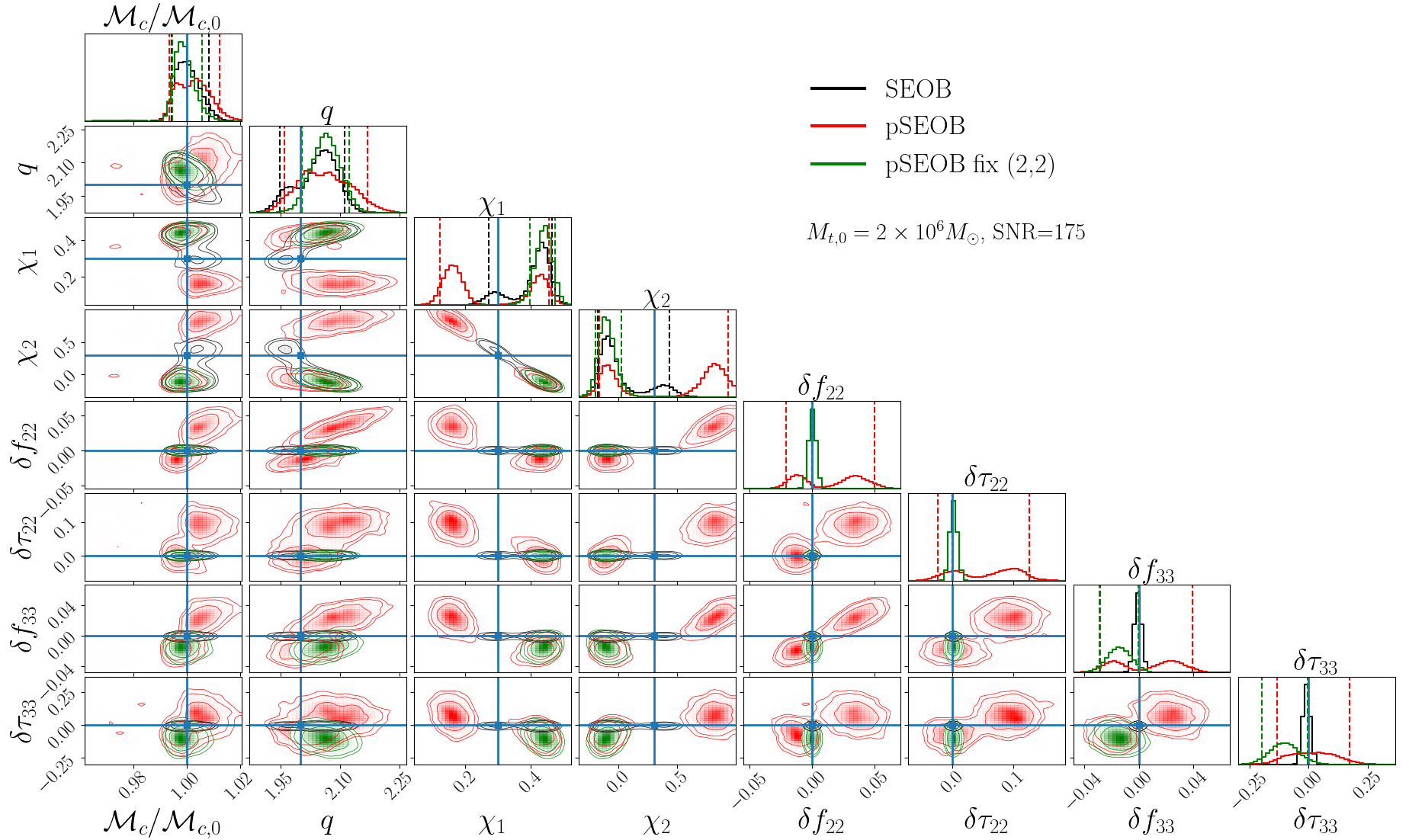}\\
 \centering
 \caption{The same as Fig.~\ref{fig:corner_comp_opts_om8} for $M_t=2\times  \ 10^6M_{\odot}$. }\label{fig:corner_comp_opts_om6}
\end{figure*}

 \FloatBarrier
\bibliography{Ref}

\end{document}